\documentstyle[12pt]{report}
\newtheorem{theorem}{Theorem}
\newtheorem{definition}{Definition}

\begin{document}
\date{July 29, 1998}
{-}
\vskip100pt
\centerline{\Large \bf SPINORS AND FIELD INTERACTIONS IN} \vskip4pt
\centerline{\Large \bf HIGHER ORDER ANISOTROPIC SPACES} \vskip40pt
\centerline{\large \sf Sergiu I. Vacaru}
\vskip18pt
\centerline{\noindent{\em Institute of Applied Physics, Academy of Sciences,}}
\centerline{\noindent{\em 5 Academy str., Chi\c sin\v au 2028,
Republic of Moldova}} \vskip8pt
\centerline{\noindent{ Fax: 011-3732-738149, E-mail: vacaru@lises.as.md}}

\vskip20pt  {\small {\bf Abstract.}

 We formulate the  theory of  field interactions  with higher order
anisot\-ropy. The concepts of higher order anisotropic space
 and locally anisotropic space (in brief, ha--space and la--space) are
 introduced as  general ones for various types of
 higher order  extensions  of Lagrange  and Finsler geometry and  higher
 dimension (Kaluza--Klein type) spaces. The  spinors  on ha--spaces
 are defined in the framework of the geometry of Clifford
 bundles provided with compatible nonlinear and distinguished connections
 and metric structures (d--connection and d--metric).
 The spinor differential geometry of ha--spaces is
 constructed. There are discussed some related issues connected with the
 physical aspects of higher order anisotropic interactions for
 gravitational, gauge, spinor, Dirac spinor and Proca fields. Motion
 equations in higher order generalizations of Finsler spaces, of the
 mentioned type of fields, are defined  by using
 bundles of linear and affine  frames locally adapted to the nonlinear
 connection structure.}
\vskip15pt

\tableofcontents


\section{Introduction}

There is a number of fundamental problems in physics advocating the extension
to locally anisotropic and higher order anisotropic backgrounds of physical
 theories \cite{mat,ma94,am,bej,asa88,mirata,vg,v96jpa}.
In order to
construct physical models on higher order anisotropic spaces it is necessary
a corresponding generalization of the spinor theory. Spinor variables and
interactions of spinor fields on Finsler spaces were used in a heuristic
manner, for instance, in works \cite{asa88,ono}, where the problem of a
rigorous definition of la--spinors for la--spaces was not considered. Here we
note that, in general, the nontrivial nonlinear connection and torsion
structures and possible incompatibility of metric and connections makes the
solution of the mentioned problem very sophisticate. The geometric
definition of la--spinors and a detailed study of the relationship between
Clifford, spinor and nonlinear and distinguished connections structures in
vector bundles, generalized Lagrange and Finsler spaces are presented in
Refs. \cite{vjmp,vb295,vsp96}.

The purpose of this paper is to  develop our results
\cite{vjmp,vb295,vsp96,vg,vlasg} on  the theory of classical
and quantum field interactions on locally anisotropic spaces. Firstly, we
receive an attention to the  necessary geometric
framework,  propose an abstract spinor formalism and formulate the
differential geometry of higher order anisotropic spaces. The next step is
the investigation of higher order anisotropic interactions of fundamental
fields on generic higher order anisotropic spaces (in brief we shall use
instead of higher order anisotropic the abbreviation ha-, for instance,
ha--spaces, ha--interactions and ha--spinors).

In order to develop the higher order anisotropic spinor theory it will be
convenient to extend the Penrose and Rindler abstract index formalism \cite
{pen,penr1,penr2} (see also the Luehr and Rosenbaum index free methods \cite
{lue}) proposed for spinors on locally isotropic spaces. We note that in
order to formulate the locally anisotropic physics usually we have
dimensions $d>4$ for the fundamental, in general higher order anisotropic
space--time, and we must take into account the physical effects of the
nonlinear connection structure. In this case the 2-spinor calculus does
not play a preferential role.

Section 2  contains an introduction into the geometry of higher
order anisotropic spaces, the  distinguishing of geometric objects by
N--connection structures  in such spaces is analyzed, explicit formulas for
coefficients of torsions and curvatures of N- and d--connections are presented
and the field equations for gravitational interactions with higher order
anisotropy are formulated. The distinguished Clifford algebras are introduced
in Section 3 and higher order anisotropic Clifford bundles are defined in
Section 4. We present a study of almost complex structure for the case of
 locally anisotropic spaces modeled in the framework of the almost Hermitian
 model of generalized Lagrange spaces in Section 5. The d--spinor
 techniques is analyzed in Section 6 and the differential
 geometry of higher order anisotropic spinors is formulated in
 Section 7.
 The Section 8 is devoted to geometric aspects of the theory of field
 interactions with higher order anisotropy (the d--tensor and d--spinor form
 of basic field equations for gravitational, gauge and d--spinor fields are
 introduced).

\newpage

\section{ Basic Geometric Objects in Ha--Spaces}

We review some results and methods of the differential geometry of vector
bundles provided with nonlinear and distinguished connections and metric
structures \cite{ma87,ma94,mirata,v96jpa}.\ This subsection
serves the twofold purpose of establishing of abstract index denotations and
starting the geometric backgrounds which are used in the next subsections of
the section.

\subsection{ N-connections and distinguishing of geometric objects}

Let ${\cal E}^{<z>}{\cal =}$ $\left( E^{<z>},p,M,Gr,F^{<z>}\right) $ be a
locally trivial distinguished vector bundle, dv-bundle, where $F^{<z>}={\cal %
R}^{m_1}\oplus ...\oplus {\cal R}^{m_z}$ (a real vector space of dimension $%
m=m_1+...+m_z,\dim F=m,$ ${\cal R\ }$ denotes the real number field) is the
typical fibre, the structural group is chosen to be the group of
automorphisms of ${\cal R}^m$ , i.e. $Gr=GL\left( m,{\cal R}\right) ,\,$ and
$p:E^{<z>}\rightarrow M$ (defined by intermediary projections $%
p_{<z,z-1>}:E^{<z>}\rightarrow E^{<z-1>},p_{<z-1,z-2>}:E^{<z-1>}\rightarrow
E^{<z-2>},...p:E^{<1>}\rightarrow M)$  is a differentiable surjection of
a differentiable manifold $E$ (total space, $\dim E=n+m)$ to a
differentiable manifold $M$ (base
space, $\dim M=n ).$ Local coordinates on ${\cal E}^{<z>}$ are denoted
as
$$
u^{<{\bf \alpha >}}=\left( x^{{\bf i}},y^{<{\bf a>\ }}\right) =\left( x^{%
{\bf i}}\doteq y^{{\bf a}_0},y^{{\bf a}_1},....,y^{{\bf a}_z}\right) =%
$$
$$
(...,y^{{\bf a}_{(p)}},...)=\{y_{(p)}^{{\bf a}_{(p)}}\}=\{y^{{\bf a}%
_{(p)}}\},
$$
or in brief ${\bf u=u_{<z>}=(x,y}_{(1)},...,{\bf y}_{(p)},...,{\bf y}_{(z)})$
where boldfaced indices will be considered as coordinate ones for which the
Einstein summation rule holds (Latin indices ${\bf i,j,k,...=a}_0{\bf ,b}_0%
{\bf ,c}_0{\bf ,...}=1,2,...,n$ will parametrize coordinates of geometrical
objects with respect to a base space $M,$ Latin indices ${\bf a}_p ,
{\bf b}_p,$ ${\bf c}_p,...=$ $1,2,...,m_{(p)}$ will parametrize
fibre coordinates of
geometrical objects and Greek indices ${\bf \alpha ,\beta ,\gamma ,...}$ are
considered as cumulative ones for coordinates of objects defined on the
total space of a v-bundle). We shall correspondingly use abstract indices $%
\alpha =(i,a),$ $\beta =(j,b),\gamma =(k,c),...$ in the Penrose manner \cite
{pen,penr1,penr2} in order to mark geometrical objects and theirs (base,
fibre)-components or, if it will be convenient, we shall consider boldfaced
letters (in the main for pointing to the operator character of tensors and
spinors into consideration) of type ${\bf A\equiv }A=\left(
A^{(h)},A^{(v_1)},...,A^{(v_z)}\right) {\bf ,b=}\left(
b^{(h)},b^{(v_1)},...,b^{(v_z)}\right) ,...,{\bf R},$ ${\bf \omega },$
${\bf \Gamma},...$ for
geometrical objects on ${\cal E}$ and theirs splitting into horizontal (h),
or base, and vertical (v), or fibre, components. For simplicity, we shall
prefer writing out of abstract indices instead of boldface ones if this will
not give rise to ambiguities.

Coordinate trans\-forms $u^{<{\bf \alpha ^{\prime }>\ }}=u^{<{\bf \alpha
^{\prime }>\ }}\left( u^{<{\bf \alpha >}}\right) $ on ${\cal E}^{<z>}$ are
writ\-ten as
$$
\{u^{<\alpha >}=\left( x^{{\bf i}},y^{<{\bf a>}}\right) \}\rightarrow
\{u^{<\alpha ^{\prime }>}=\left( x^{{\bf i^{\prime }\ }},y^{<{\bf a^{\prime
}>}}\right) \}
$$
and written as recurrent maps
$$
x^{{\bf i^{\prime }\ }}=x^{{\bf i^{\prime }\ }}(x^{{\bf i}}),~rank\left(
\frac{\partial x^{{\bf i^{\prime }\ }}}{\partial x^{{\bf i}}}\right) =n,%
\eqno(1)
$$
$$
y_{(1)}^{{\bf a_1^{\prime }\ }}=K_{{\bf a}_1{\bf \ }}^{{\bf a_1^{\prime }}%
}(x^{{\bf i\ }})y_{(1)}^{{\bf a}_1},K_{{\bf a}_1{\bf \ }}^{{\bf a_1^{\prime }%
}}(x^{{\bf i\ }})\in GL\left( m_1,{\cal R}\right) ,
$$
$$
............
$$
$$
y_{(p)}^{{\bf a_p^{\prime }\ }}=K_{{\bf a}_p{\bf \ }}^{{\bf a_p^{\prime }}%
}(u_{(p-1)})y_{(p)}^{{\bf a}_p},K_{{\bf a}_p{\bf \ }}^{{\bf a_p^{\prime }}%
}(u_{(p-1)})\in GL\left( m_p,{\cal R}\right) ,
$$
$$
.............
$$
$$
y_{(z)}^{{\bf a_z^{\prime }\ }}=K_{{\bf a}_z{\bf \ }}^{{\bf a_z^{\prime }}%
}(u_{(z-1)})y_{(z-1)}^{{\bf a}_z},K_{{\bf a}_z{\bf \ }}^{{\bf a_z^{\prime }}%
}(u_{(z-1)})\in GL\left( m_z,{\cal R}\right)
$$
where matrices $K_{{\bf a}_1{\bf \ }}^{{\bf a_1^{\prime }}}(x^{{\bf i\ }%
}),...,K_{{\bf a}_p{\bf \ }}^{{\bf a_p^{\prime }}}(u_{(p-1)}),...,K_{{\bf a}%
_z{\bf \ }}^{{\bf a_z^{\prime }}}(u_{(z-1)})$ are functions of necessary
smoothness class. In brief we write transforms (1) in the form%
$$
x^{{\bf i^{\prime }\ }}=x^{{\bf i^{\prime }\ }}(x^{{\bf i}}),y^{<{\bf %
a^{\prime }>\ }}=K_{<{\bf a>}}^{<{\bf a^{\prime }>}}y^{<{\bf a>}}.
$$
In general form we shall write $K$--matrices $K_{<\alpha {\bf >}}^{<\alpha
{\bf ^{\prime }>}}=\left( K_{{\bf i}}^{{\bf i^{\prime }}},K_{<{\bf a>}}^{<%
{\bf a^{\prime }>}}\right) ,$ where $K_{{\bf i}}^{{\bf i^{\prime }}}=\frac{%
\partial x^{{\bf i^{\prime }\ }}}{\partial x^{{\bf i}}}.$

A local coordinate parametrization of ${\cal E}^{<z>}$ naturally defines a
coordinate basis of the module of d--vector fi\-elds
 $\Xi \left( {\cal E}%
^{<z>}\right) ,$%
$$
\partial _{<\alpha >}=(\partial _i,\partial _{<a>})=(\partial _i,\partial
_{a_1},...,\partial _{a_p},...,\partial _{a_z})=\eqno(2)
$$
$$
\frac \partial {\partial u^{<\alpha >}}=\left( \frac \partial {\partial
x^i},\frac \partial {\partial y^{<a>}}\right) =\left( \frac \partial
{\partial x^i},\frac \partial {\partial y^{a_1}},...\frac \partial {\partial
y^{a_p}},...,\frac \partial {\partial y^{a_z}}\right) ,
$$
and the reciprocal to (2) coordinate basis
$$
d^{<\alpha >}=(d^i,d^{<a>})=(d^i,d^{a_1},...,d^{a_p},...,d^{a_z})=\eqno(3)
$$
$$
du^{<\alpha >}=(dx^i,dy^{<a>})=(dx^i,dy^{a_1},...,dy^{a_p},...,dy^{a_z}),
$$
which is uniquely defined from the equations%
$$
d^{<\alpha >}\circ \partial _{<\beta >}=\delta _{<\beta >}^{<\alpha >},
$$
where $\delta _{<\beta >}^{<\alpha >}$ is the Kronecher symbol and by ''$%
\circ $$"$ we denote the inner (scalar) product in the tangent bundle ${\cal %
TE}^{<z>}.$

The concept of {\bf nonlinear connection,} in brief, N--connection, is
fundamental in the geometry of locally anisotropic and higher order
anisotropic spaces (see a detailed study and basic references in \cite
{ma87,ma94,mirata}). In a dv--bundle ${\cal E}^{<z>}$ it is defined
as a distribution $\{N:E_u\rightarrow H_uE,T_uE=H_uE\oplus V_u^{(1)}E\oplus
...\oplus V_u^{(p)}E...\oplus V_u^{(z)}E\}$ on $E^{<z>}$ being a global
decomposition, as a Whitney sum, into horizontal,${\cal HE,\ }$ and
vertical, ${\cal VE}^{<p>}{\cal ,}p=1,2,...,z$ subbundles of the tangent
bundle ${\cal TE:}$
$$
{\cal TE}=H{\cal E}\oplus V{\cal E}^{<1>}\oplus ...\oplus V{\cal E}%
^{<p>}\oplus ...\oplus V{\cal E}^{<z>}.\eqno(4)
$$

Locally a N-connection in ${\cal E}^{<z>}$ is given by it components $N_{<%
{\bf a}_f>}^{<{\bf a}_p>}({\bf u}),z\geq p>f\geq 0$ (in brief we shall write
$N_{<a_f>}^{<a_p>}(u)$ ) with respect to bases (2) and (3)):

$$
{\bf N}=N_{<a_f>}^{<a_p>}(u)\delta ^{<a_f>}\otimes \delta _{<a_p>},(z\geq
p>f\geq 0),
$$

We note that a linear connection in a dv-bundle ${\cal E}^{<z>}$ can be
considered as a particular case of a N-connection when $%
N_i^{<a>}(u)=K_{<b>i}^{<a>}\left( x\right) y^{<b>},$ where functions $%
K_{<a>i}^{<b>}\left( x\right) $ on the base $M$ are called the Christoffel
coefficients.

To coordinate locally geometric constructions with the global splitting of
${\cal E}^{<z>}$ defined by a N-connection structure, we have to introduce a
lo\-cal\-ly adap\-ted bas\-is ( la--basis, la--frame ):%
$$
\delta _{<\alpha >}=(\delta _i,\delta _{<a>})=(\delta _i,\delta
_{a_1},...,\delta _{a_p},...,\delta _{a_z}),\eqno(5)
$$
with components parametrized as

$$
\delta _i=\partial _i-N_i^{a_1}\partial _{a_1}-...-N_i^{a_z}\partial _{a_z},
$$
$$
\delta _{a_1}=\partial _{a_1}-N_{a_1}^{a_2}\partial
_{a_2}-...-N_{a_1}^{a_z}\partial _{a_z},
$$
$$
................
$$
$$
\delta _{a_p}=\partial _{a_p}-N_{a_p}^{a_{p+1}}\partial
_{a_{p+1}}-...-N_{a_p}^{a_z}\partial _{a_z},
$$
$$
...............
$$
$$
\delta _{a_z}=\partial _{a_z}
$$
and it dual la--basis%
$$
\delta ^{<\alpha >}=(\delta ^i,\delta ^{<a>})=\left( \delta ^i,\delta
^{a_1},...,\delta ^{a_p},...,\delta ^{a_z}\right) ,\eqno(6)
$$
$$
\delta x^i=dx^i,
$$
$$
\delta y^{a_1}=dy^{a_1}+M_i^{a_1}dx^i,
$$
$$
\delta y^{a_2}=dy^{a_2}+M_{a_1}^{a_2}dy^{a_1}+M_i^{a_2}dx^i,
$$
$$
.................
$$
$$
\delta
y^{a_p}=dy^{a_p}+M_{a_{p-1}}^{a_p}dy^{p-1}+M_{a_{p-2}}^{a_p}dy^{a_{p-2}}+...
+M_i^{a_p}dx^i,
$$
$$
...................
$$
$$
\delta
y^{a_z}=dy^{a_z}+
M_{a_{z-1}}^{a_z}dy^{z-1}+M_{a_{z-2}}^{a_z}dy^{a_{z-2}}+...+M_i^{a_z}dx^i .
$$

The{\bf \ nonholonomic coefficients }${\bf w}=\{w_{<\beta ><\gamma
>}^{<\alpha >}\left( u\right) \}$ of the locally adapted to the N--connection
structure frames  are defined as%
$$
\left[ \delta _{<\alpha >},\delta _{<\beta >}\right] =\delta _{<\alpha
>}\delta _{<\beta >}-\delta _{<\beta >}\delta _{<\alpha >}=w_{<\beta
><\gamma >}^{<\alpha >}\left( u\right) \delta _{<\alpha >}.
$$

The {\bf algebra of tensorial distinguished fields} $DT\left( {\cal E}%
^{<z>}\right) $ (d--fields, d--tensors, d--objects) on ${\cal E}^{<z>}$ is
introduced as the tensor algebra\\ ${\cal T}=\{{\cal T}%
_{qs_1...s_p...s_z}^{pr_1...r_p...r_z}\}$ of the dv-bundle ${\cal E}_{\left(
d\right) }^{<z>},$
$$
p_d:{\cal HE}^{<z>}{\cal \oplus V}^1{\cal E}^{<z>}{ \oplus }...{ %
\oplus V}^p{\cal E}^{<z>}{ \oplus }...{ \oplus V}^z{\cal E}^{<z>}%
{\cal \rightarrow E}^{<z>},
$$
${ \ }$ An element ${\bf t}\in {\cal T}_{qs_1...s_z}^{pr_1...r_z},$
d-tensor field of type $\left(
\begin{array}{cccccc}
p & r_1 & ... & r_p & ... & r_z \\
q & s_1 & ... & s_p & ... & s_z
\end{array}
\right) ,$ can be written in local form as%
$$
{\bf t}%
=t_{j_1...j_qb_1^{(1)}...b_{r_1}^{(1)}...b_1^{(p)}...b_{r_p}^{(p)}...
b_1^{(z)}...b_{r_z}^{(z)}}^{i_1...i_pa_1^{(1)}...a_{r_1}^{(1)}...
a_1^{(p)}...a_{r_p}^{(p)}...a_1^{(z)}...a_{r_z}^{(z)}}
\left( u\right) \delta _{i_1}\otimes ...\otimes \delta _{i_p}\otimes
d^{j_1}\otimes ...\otimes d^{j_q}\otimes
$$
$$
\delta _{a_1^{(1)}}\otimes ...\otimes \delta _{a_{r_1}^{(1)}}\otimes \delta
^{b_1^{(1)}}...\otimes \delta ^{b_{s_1}^{(1)}}\otimes ...\otimes \delta
_{a_1^{(p)}}\otimes ...\otimes \delta _{a_{r_p}^{(p)}}\otimes ...\otimes
$$
$$
\delta ^{b_1^{(p)}}...\otimes \delta ^{b_{s_p}^{(p)}}\otimes \delta
_{a_1^{(z)}}\otimes ...\otimes \delta _{a_{rz}^{(z)}}\otimes \delta
^{b_1^{(z)}}...\otimes \delta ^{b_{s_z}^{(z)}}.
$$

We shall respectively use denotations $X{\cal (E}^{<z>})$ (or $X{%
\left( M\right) ),\ }\Lambda ^p\left( {\cal E}^{<z>}\right) $ (or
$\Lambda ^p\left( M\right) ) $ and ${\cal F(E}^{<z>})$ (or $%
{\cal F}$ $\left( M\right) $) for the module of d-vector fields on ${\cal E}%
^{<z>}$ (or $M$ ), the exterior algebra of p-forms on ${\cal E}^{<z>}$
(or $M)$ and the set of real functions on ${\cal E}^{<z>}$(or $%
M). $

In general, d--objects on ${\cal E}^{<z>}$ are introduced as
geometric objects with various group and coordinate transforms coordinated
with the N--connection structure on ${\cal E}^{<z>}.$ For example, a
d--connection $D$ on ${\cal E}^{<z>}$ is defined as a linear
connection $D$ on $E^{<z>}$ conserving under a parallelism the global
decomposition (4) into horizontal and vertical subbundles of ${\cal TE}%
^{<z>}$ .

A N-connection in ${\cal E}^{<z>}$ induces a corresponding decomposition of
d-tensors into sums of horizontal and vertical parts, for example, for every
d-vector $X\in {\cal X(E}^{<z>})$ and 1-form $\widetilde{X}\in \Lambda
^1\left( {\cal E}^{<z>}\right) $ we have respectively
$$
X=hX+v_1X+...+v_zX{\bf \ \quad }\mbox{and \quad }\widetilde{X}=h\widetilde{X}%
+v_1\widetilde{X}+...~v_z\widetilde{X}.\eqno(7)
$$
In consequence, we can associate to every d-covariant derivation along the
d-vector (7), $D_X=X\circ D,$ two new operators of h- and v-covariant
derivations defined respectively as
$$
D_X^{(h)}Y=D_{hX}Y%
$$
and
$$
D_X^{\left( v_1\right) }Y=D_{v_1X}Y{\bf ,...,D_X^{\left( v_z\right)
}Y=D_{v_zX}Y\quad }\forall Y{\bf \in }{\cal X(E}^{<z>}) ,
$$
for which the following conditions hold:%
$$
D_XY{\bf =}D_X^{(h)}Y + D_X^{(v_1)}Y+...+D_X^{(v_z)}Y, \eqno(8)
$$
$$
D_X^{(h)}f=(hX{\bf )}f
$$
and
$$
D_X^{(v_p)}f=(v_pX{\bf )}f,\quad X,Y{\bf \in }{\cal X\left( E\right) ,}f\in
{\cal F}\left( M\right) ,p=1,2,...z.
$$

We define a {\bf metric structure }${\bf G\ }$in the total space $E^{<z>}$
of dv-bundle ${\cal E}^{<z>}{\cal =}$ $\left( E^{<z>},p,M\right) $ over a
connected and paracompact base $M$ as a symmetrical covariant tensor field of
type $\left( 0,2\right) $, $G_{<\alpha ><\beta >,}$ being nondegenerate and
of constant signature on $E^{<z>}.$

Nonlinear connection ${\bf N}$ and metric ${\bf G}$ structures on ${\cal E}%
^{<z>}$ are mutually compatible it there are satisfied the conditions:
$$
{\bf G}\left( \delta _{a_f},\delta _{a_p}\right) =0,\mbox{or equivalently, }%
G_{a_fa_p}\left( u\right) -N_{a_f}^{<b>}\left( u\right) h_{a_f<b>}\left(
u\right) =0,\eqno(9)
$$
where $h_{a_pb_p}={\bf G}\left( \partial _{a_p},\partial _{b_p}\right) $ and
$G_{b_fa_p}={\bf G}\left( \partial _{b_f},\partial _{a_p}\right) ,0\leq
f<p\leq z,\,$ which gives%
$$
N_{c_f}^{b_p}\left( u\right) =h^{<a>b_p}\left( u\right) G_{c_f<a>}\left(
u\right) \eqno(10)
$$
(the matrix $h^{a_pb_p}$ is inverse to $h_{a_pb_p}).$ In consequence one
obtains the following decomposition of metric :%
$$
{\bf G}(X,Y){\bf =hG}(X,Y)+{\bf v}_1{\bf G}(X,Y)+...+{\bf v}_z{\bf G}(X,Y)%
{\bf ,}\eqno(11)
$$
where the d-tensor ${\bf hG}(X,Y){\bf =G}(hX,hY)$ is of type $\left(
\begin{array}{cc}
0 & 0 \\
2 & 0
\end{array}
\right) $ and the d-tensor ${\bf v}_p{\bf G}(X,Y)={\bf G}(v_pX,v_pY)$ is of
type $\left(
\begin{array}{ccccc}
0 & ... & 0(p) & ... & 0 \\
0 & ... & 2 & ... & z
\end{array}
\right) .$ With respect to la--basis (6) the d--metric (11) is written as%
$$
{\bf G}=g_{<\alpha ><\beta >}\left( u\right) \delta ^{<\alpha >}\otimes
\delta ^{<\beta >}=g_{ij}\left( u\right) d^i\otimes d^j+h_{<a><b>}\left(
u\right) \delta ^{<a>}\otimes \delta ^{<b>},\eqno(12)
$$
where $g_{ij}={\bf G}\left( \delta _i,\delta _j\right) .$

A metric structure of type (11) (equivalently, of type (12)) or a metric
on $E^{<z>}$ with components satisfying constraints (9), equivalently
(10)) defines an adapted to the given N-connection inner (d--scalar)
product on the tangent bundle ${\cal TE}^{<z>}{\cal .}$

We shall say that a d-connection $\widehat{D}_X$ is compatible with the
d-scalar product on ${\cal TE}^{<z>}{\cal \ }$ (i.e. is a standard
d-connection) if
$$
\widehat{D}_X\left( {\bf X\cdot Y}\right) =\left( \widehat{D}_X{\bf Y}%
\right) \cdot {\bf Z+Y\cdot }\left( \widehat{D}_X{\bf Z}\right) ,\forall
{\bf X,Y,Z}{\bf \in }{\cal X(E}^{<z>}){\cal .}
$$
An arbitrary d--connection $D_X$ differs from the standard one $\widehat{D}_X$
by an operator $\widehat{P}_X\left( u\right) =\{X^{<\alpha >}\widehat{P}%
_{<\alpha ><\beta >}^{<\gamma >}\left( u\right) \},$ called the deformation
d-tensor with respect to $\widehat{D}_X,$ which is just a d-linear transform
of ${\cal E}_u^{<z>},$ $\forall u\in {\cal E}^{<z>}{\cal .}$ The explicit
form of $\widehat{P}_X$ can be found by using the corresponding axiom
defining linear connections \cite{lue}
$$
\left( D_X-\widehat{D}_X\right) fZ=f\left( D_X-\widehat{D}_X\right) Z{\bf ,}
$$
written with respect to la-bases (5) and (6). From the last expression
we obtain
$$
\widehat{P}_X\left( u\right) =\left[ (D_X-\widehat{D}_X)\delta _{<\alpha
>}\left( u\right) \right] \delta ^{<\alpha >}\left( u\right) ,
$$
therefore
$$
D_XZ{\bf \ }=\widehat{D}_XZ{\bf \ +}\widehat{P}_XZ.\eqno(13)
$$

A d-connection $D_X$ is {\bf metric (}or  compatible with met\-ric ${\bf G%
}$) on ${\cal E}^{<z>}$ if%
$$
D_X{\bf G}=0,\forall X{\bf \in }{\cal X(E}^{<z>}).\eqno(14)
$$

Locally adapted components $\Gamma _{<\beta ><\gamma >}^{<\alpha >}$ of a
d-connection $D_{<\alpha >}=(\delta _{<\alpha >}\circ D)$ are defined by the
equations%
$$
D_{<\alpha >}\delta _{<\beta >}=\Gamma _{<\alpha ><\beta >}^{<\gamma
>}\delta _{<\gamma >},
$$
from which one immediately follows%
$$
\Gamma _{<\alpha ><\beta >}^{<\gamma >}\left( u\right) =\left( D_{<\alpha
>}\delta _{<\beta >}\right) \circ \delta ^{<\gamma >}.\eqno(15)
$$

The operations of h- and v$_{(p)}$-covariant derivations, $%
D_k^{(h)}=\{L_{jk}^i,L_{<b>k\;}^{<a>}\}$ and $D_{c_p}^{(v_p)}=%
\{C_{jc_p}^i,C_{<b>c_p}^i,C_{jc_p}^{<a>},C_{<b>c_p}^{<a>}\}$ (see (8)),
are introduced as corresponding h- and v$_{(p)}$-paramet\-ri\-za\-ti\-ons of
(15):%
$$
L_{jk}^i=\left( D_k\delta _j\right) \circ d^i,\quad L_{<b>k}^{<a>}=\left(
D_k\delta _{<b>}\right) \circ \delta ^{<a>}\eqno(16)
$$
and%
$$
C_{jc_p}^i=\left( D_{c_p}\delta _j\right) \circ \delta ^i,\quad
C_{<b>c_p}^{<a>}=\left( D_{c_p}\delta _{<b>}\right) \circ \delta ^{<a>}%
\eqno(17)
$$
$$
C_{<b>c_p}^i=\left( D_{c_p}\delta _{<b>}\right) \circ \delta ^i,\quad
C_{jc_p}^{<a>}=\left( D_{c_p}\delta _j\right) \circ \delta ^{<a>}.
$$
A set of components (16) and (17), $D\Gamma =\left(
L_{jk}^i,L_{<b>k}^{<a>},C_{j<c>}^i,C_{<b><c>}^{<a>}\right) ,\,$ completely
defines the local action of a d-connection $D$ in ${\cal E}^{<z>}.$ For
instance, taken a d-tensor field of type $\left(
\begin{array}{cccc}
1 & ... & 1(p) & ... \\
1 & ... & 1(p) & ...
\end{array}
\right) ,$ ${\bf t}=t_{jb_p}^{ia_p}\delta _i\otimes \delta _{a_p}\otimes
\delta ^j\otimes \delta ^{b_p},$ and a d-vector ${\bf X}=X^i\delta
_i+X^{<a>}\delta _{<a>}$ we have%
$$
D_X{\bf t=}D_X^{(h)}{\bf t+}D_X^{(v_1)}{\bf %
t+..+.D_X^{(v_p)}t+...+D_X^{(v_z)}t=}
$$
$$
\left( X^kt_{jb_p|k}^{ia_p}+X^{<c>}t_{jb_p\perp <c>}^{ia_p}\right) \delta
_i\otimes \delta _{a_p}\otimes d^j\otimes \delta ^{b_p},
$$
where the h--covariant derivative is written as%
$$
t_{jb_p|k}^{ia_p}=\frac{\delta t_{jb_p}^{ia_p}}{\delta x^k}%
+L_{hk}^it_{jb_p}^{ha_p}+L_{c_pk}^{a_p}t_{jb_p}^{ic_p}-
L_{jk}^ht_{hb_p}^{ia_p}-L_{b_pk}^{c_p}t_{jc_p}^{ia_p}
$$
and the v--covariant derivatives are written as%
$$
t_{jb_p\perp <c>}^{ia_p}=\frac{\partial t_{jb_p}^{ia_p}}{\partial y^{<c>}}%
+C_{h<c>}^it_{jb_p}^{ha_p}+C_{d_p<c>}^{a_p}t_{jb_p}^{id_p}-
C_{j<c>}^ht_{hb_p}^{ia_p}-C_{b_p<c>}^{d_p}t_{jd_p}^{ia_p}.
$$
For a scalar function $f\in {\cal F(E}^{<z>})$ we have%
$$
D_i^{(h)}=\frac{\delta f}{\delta x^i}=\frac{\partial f}{\partial x^i}%
-N_i^{<a>}\frac{\partial f}{\partial y^{<a>}},
$$
$$
D_{a_f}^{(v_f)}=\frac{\delta f}{\delta x^{a_f}}=\frac{\partial f}{\partial
x^{a_f}}-N_{a_f}^{a_p}\frac{\partial f}{\partial y^{a_p}},1\leq f<p\leq z-1,
$$
$$
\mbox{ and }D_{c_z}^{(v_z)}f=\frac{\partial f}{\partial y^{c_z}}.
$$

We emphasize that the geometry of connections in a dv-bundle ${\cal E}^{<z>}$
is very reach. If a triple of fundamental geometric objects
$$(N_{a_f}^{a_p}\left( u\right) ,
 \Gamma _{<\beta ><\gamma >}^{<\alpha >}\left(u\right) ,
 G_{<\alpha ><\beta >}\left( u\right) ) $$ is fixed on ${\cal E}^{<z>},$
a multiconnection structure (with corresponding different
rules of covariant derivation, which are, or not, mutually compatible and
with the same, or not, induced d-scalar products in ${\cal TE}^{<z>}{\cal )}$
is defined on this dv-bundle. For instance, we enumerate some of connections
and covariant derivations which can present interest in investigation of
locally anisotropic gravitational and matter field interactions:

\begin{enumerate}
\item  Every N-connection in ${\cal E}^{<z>},$ with coefficients $%
N_{a_f}^{a_p}\left( u\right) $ being differentiable on y--va\-ri\-ab\-les,
 in\-du\-ces
a structure of linear connection\\ $\widetilde{N}_{<\beta ><\gamma >}^{<\alpha
>},$ where $\widetilde{N}_{b_pc_f}^{a_p}=\frac{\partial N_{c_f}^{a_p}}{%
\partial y^{b_p}}$ and $\widetilde{N}_{b_pc_p}^{a_p}\left( u\right) =0.$ For
some $$Y\left( u\right) =Y^i\left( u\right) \partial _i+Y^{<a>}\left(
u\right) \partial _{<a>}$$ and $$B\left( u\right) =B^{<a>}\left( u\right)
\partial _{<a>}$$ one writes%
$$
D_Y^{(\widetilde{N})}B=\left[ Y^{c_f}\left( \frac{\partial B^{a_p}}{\partial
y^{c_f}}+\widetilde{N}_{b_pi}^{a_p}B^{b_p}\right) +Y^{b_p}\frac{\partial
B^{a_p}}{\partial y^{b_p}}\right] \frac \partial {\partial y^{a_p}}~(0\leq
f<p\leq z).
$$

\item  The d--connection of Berwald type \cite{berw}

$$
\Gamma _{<\beta ><\gamma >}^{(B)<\alpha >}=\left( L_{jk}^i,\frac{\partial
N_k^{<a>}}{\partial y^{<b>}},0,C_{<b><c>}^{<a>}\right) ,\eqno(18)
$$
where
$$
L_{.jk}^i\left( u\right) =\frac 12g^{ir}\left( \frac{\delta g_{jk}}{\partial
x^k}+\frac{\delta g_{kr}}{\partial x^j}-\frac{\delta g_{jk}}{\partial x^r}%
\right) , \eqno(19)
$$

$$
C_{.<b><c>}^{<a>}\left( u\right) =\frac 12h^{<a><d>}\left( \frac{\delta
h_{<b><d>}}{\partial y^{<c>}}+\frac{\delta h_{<c><d>}}{\partial y^{<b>}}-%
\frac{\delta h_{<b><c>}}{\partial y^{<d>}}\right) ,
$$

which is hv-metric, i.e. $D_k^{(B)}g_{ij}=0$ and $D_{<c>}^{(B)}h_{<a><b>}=0.$

\item  The canonical d--connection ${\bf \Gamma ^{(c)}}$ associated to a
metric ${\bf G}$ of type (12) $\Gamma _{<\beta ><\gamma >}^{(c)<\alpha
>}=\left(
L_{jk}^{(c)i},L_{<b>k}^{(c)<a>},C_{j<c>}^{(c)i},C_{<b><c>}^{(c)<a>}\right) ,$
with coefficients%
$$
L_{jk}^{(c)i}=L_{.jk}^i,C_{<b><c>}^{(c)<a>}=C_{.<b><c>}^{<a>}%
\mbox{ (see (19))}
$$
$$
L_{<b>i}^{(c)<a>}=\widetilde{N}_{<b>i}^{<a>}+          \eqno(20)
$$
$$
\frac 12h^{<a><c>}\left( \frac{\delta h_{<b><c>}}{\delta x^i}-\widetilde{N}%
_{<b>i}^{<d>}h_{<d><c>}-\widetilde{N}_{<c>i}^{<d>}h_{<d><b>}\right) ,
$$
$$
C_{j<c>}^{(c)i}=\frac 12g^{ik}\frac{\partial g_{jk}}{\partial y^{<c>}}.%
$$
This is a metric d--connection which satisfies conditions
$$
D_k^{(c)}g_{ij}=0,D_{<c>}^{(c)}g_{ij}=0,D_k^{(c)}h_{<a><b>}=0,
D_{<c>}^{(c)}h_{<a><b>}=0.
$$

\item  We can consider N-adapted Christoffel d--symbols%
$$
\widetilde{\Gamma }_{<\beta ><\gamma >}^{<\alpha >}=$$ $$\frac 12G^{<\alpha
><\tau >}\left( \delta _{<\gamma >}G_{<\tau ><\beta >}+\delta _{<\beta
>}G_{<\tau ><\gamma >}-\delta _{<\tau >}G_{<\beta ><\gamma >}\right) ,%
\eqno(21)
$$
which have the components of d-connection $$\widetilde{\Gamma }_{<\beta
><\gamma >}^{<\alpha >}=\left( L_{jk}^i,0,0,C_{<b><c>}^{<a>}\right) $$ with $%
L_{jk}^i$ and $C_{<b><c>}^{<a>}$ as in (19) if $G_{<\alpha ><\beta >}$ is
taken in the form (12).
\end{enumerate}

Arbitrary linear connections on a dv--bundle ${\cal E}^{<z>}$ can be also
characterized by theirs deformation tensors (see (13)) with respect, for
instance, to d--connect\-i\-on (21):%
$$
\Gamma _{<\beta ><\gamma >}^{(B)<\alpha >}=\widetilde{\Gamma }_{<\beta
><\gamma >}^{<\alpha >}+P_{<\beta ><\gamma >}^{(B)<\alpha >},\Gamma _{<\beta
><\gamma >}^{(c)<\alpha >}=\widetilde{\Gamma }_{<\beta ><\gamma >}^{<\alpha
>}+P_{<\beta ><\gamma >}^{(c)<\alpha >}
$$
or, in general,%
$$
\Gamma _{<\beta ><\gamma >}^{<\alpha >}=\widetilde{\Gamma }_{<\beta ><\gamma
>}^{<\alpha >}+P_{<\beta ><\gamma >}^{<\alpha >},
$$
where $P_{<\beta ><\gamma >}^{(B)<\alpha >},P_{<\beta ><\gamma
>}^{(c)<\alpha >}$ and $P_{<\beta ><\gamma >}^{<\alpha >}$ are respectively
the deformation d--ten\-sors of d-connect\-i\-ons (18),\ (20), or of a
general one.

\subsection{ Torsions and curvatures of N- and d-connections}

The curvature ${\bf \Omega }$$\,$ of a nonlinear connection ${\bf N}$ in a
dv-bundle ${\cal E}^{<z>}$ can be defined as the Nijenhuis tensor field $%
N_v\left( X,Y\right) $ associated to ${\bf N\ }$ \cite{ma87,ma94}:
$$
{\bf \Omega }=N_v={\bf \left[ vX,vY\right] +v\left[ X,Y\right] -v\left[
vX,Y\right] -v\left[ X,vY\right] ,X,Y}\in {\cal X(E}^{<z>}){\cal ,}
$$
where $v=v_1\oplus ...\oplus v_z.$ In local form one has%
$$
{\bf \Omega }=\frac 12\Omega _{b_fc_f}^{a_p}\delta ^{b_f}\bigwedge \delta
^{c_f}\otimes \delta _{a_p},(0\leq f<p\leq z),
$$
where%
$$
\Omega _{b_fc_f}^{a_p}=\frac{\delta N_{c_f}^{a_p}}{\partial y^{b_f}}-\frac{%
\partial N_{b_f}^{a_p}}{\partial y^{c_f}}+N_{b_f}^{<b>}\widetilde{N}%
_{<b>c_f}^{a_p}-N_{c_f}^{<b>}\widetilde{N}_{<b>b_f}^{a_p}.\eqno(22)
$$

The torsion ${\bf T}$ of d--connection ${\bf D\ }$ in ${\cal E}^{<z>}$ is
defined by the equation%
$$
{\bf T\left( X,Y\right) =XY_{\circ }^{\circ }T\doteq }D_X{\bf Y-}D_Y{\bf X\
-\left[ X,Y\right] .}\eqno(23)
$$
One holds the following h- and v$_{(p)}-$--decompositions%
$$
{\bf T\left( X,Y\right) =T\left( hX,hY\right) +T\left( hX,vY\right) +T\left(
vX,hY\right) +T\left( vX,vY\right) .}\eqno(24)
$$
We consider the projections: ${\bf hT\left( X,Y\right) ,v}_{(p)}{\bf T\left(
hX,hY\right) ,hT\left( hX,hY\right) ,...}$ and say that, for instance, ${\bf %
hT\left( hX,hY\right) }$ is the h(hh)-torsion of ${\bf D},$ \\ ${\bf v}_{(p)}%
{\bf T\left( hX,hY\right) \ }$ is the v$_p$(hh)-torsion of ${\bf D}$ and so
on.

The torsion (23) is locally determined by five d-tensor fields, torsions,
defined as
$$
T_{jk}^i={\bf hT}\left( \delta _k,\delta _j\right) \cdot d^i,\quad
T_{jk}^{a_p}={\bf v}_{(p)}{\bf T}\left( \delta _k,\delta _j\right) \cdot
\delta ^{a_p},
$$
$$
P_{jb_p}^i={\bf hT}\left( \delta _{b_p},\delta _j\right) \cdot d^i,\quad
P_{jb_f}^{a_p}={\bf v}_{(p)}{\bf T}\left( \delta _{b_f},\delta _j\right)
\cdot \delta ^{a_p},
$$
$$
S_{b_fc_f}^{a_p}={\bf v}_{(p)}{\bf T}\left( \delta _{c_f},\delta
_{b_f}\right) \cdot \delta ^{a_p}.
$$
Using formulas (5),(6),(22) and (23) we can computer in explicit
form the components of torsions (24) for a d--connection of type (16)
and (17):
$$
T_{.jk}^i=T_{jk}^i=L_{jk}^i-L_{kj}^i,\quad
T_{j<a>}^i=C_{.j<a>}^i,T_{<a>j}^i=-C_{j<a>}^i,\eqno(25)
$$
$$
T_{.j<a>}^i=0,T_{.<b><c>}^{<a>}=S_{.<b><c>}^{<a>}=C_{<b><c>}^{<a>}-
C_{<c><b>}^{<a>},
$$
$$
T_{.b_fc_f}^{a_p}=\frac{\delta N_{c_f}^{a_p}}{\partial y^{b_f}}-\frac{\delta
N_{b_f}^{a_p}}{\partial y^{c_f}},T_{.<b>i}^{<a>}=P_{.<b>i}^{<a>}=\frac{%
\delta N_i^{<a>}}{\partial y^{<b>}}%
-L_{.<b>j}^{<a>},T_{.i<b>}^{<a>}=-P_{.<b>i}^{<a>}.
$$

The curvature ${\bf R}$ of d--connection in ${\cal E}^{<z>}$ is defined by
the equation
$$
{\bf R\left( X,Y\right) Z=XY_{\bullet }^{\bullet }R\bullet Z}=D_XD_Y{\bf Z}%
-D_YD_X{\bf Z-}D_{[X,Y]}{\bf Z.}\eqno(26)
$$
One holds the next properties for the h- and v-decompositions of curvature:%
$$
{\bf v}_{(p)}{\bf R\left( X,Y\right) hZ=0,\ hR\left( X,Y\right) v}_{(p)}{\bf %
Z=0,~v}_{(f)}{\bf R\left( X,Y\right) v}_{(p)}{\bf Z=0,}
$$
$$
{\bf R\left( X,Y\right) Z=hR\left( X,Y\right) hZ+vR\left( X,Y\right) vZ,}
$$
where ${\bf v=v}_1+...+{\bf v}_z.$ From (26) and the equation ${\bf %
R\left( X,Y\right) =-R\left( Y,X\right) }$ we get that the curvature of a
d-con\-nec\-ti\-on ${\bf D}$ in ${\cal E}^{<z>}$ is completely determined by
the following d-tensor fields:%
$$
R_{h.jk}^{.i}=\delta ^i\cdot {\bf R}\left( \delta _k,\delta _j\right) \delta
_h,~R_{<b>.jk}^{.<a>}=\delta ^{<a>}\cdot {\bf R}\left( \delta _k,\delta
_j\right) \delta _{<b>},\eqno(27)
$$
$$
P_{j.k<c>}^{.i}=d^i\cdot {\bf R}\left( \delta _{<c>},\delta _{<k>}\right)
\delta _j,~P_{<b>.<k><c>}^{.<a>}=\delta ^{<a>}\cdot {\bf R}\left( \delta
_{<c>},\delta _{<k>}\right) \delta _{<b>},
$$
$$
S_{j.<b><c>}^{.i}=d^i\cdot {\bf R}\left( \delta _{<c>},\delta _{<b>}\right)
\delta _j,~S_{<b>.<c><d>}^{.<a>}=\delta ^{<a>}\cdot {\bf R}\left( \delta
_{<d>},\delta _{<c>}\right) \delta _{<b>}.
$$
By a direct computation, using (5),(6),(16),(17) and (27) we get :
$$
R_{h.jk}^{.i}=\frac{\delta L_{.hj}^i}{\delta x^h}-\frac{\delta L_{.hk}^i}{%
\delta x^j}+L_{.hj}^mL_{mk}^i-L_{.hk}^mL_{mj}^i+C_{.h<a>}^iR_{.jk}^{<a>},%
\eqno(28)
$$
$$
R_{<b>.jk}^{.<a>}=\frac{\delta L_{.<b>j}^{<a>}}{\delta x^k}-\frac{\delta
L_{.<b>k}^{<a>}}{\delta x^j}%
+L_{.<b>j}^{<c>}L_{.<c>k}^{<a>}-L_{.<b>k}^{<c>}L_{.<c>j}^{<a>}+
C_{.<b><c>}^{<a>}R_{.jk}^{<c>},
$$
$$
P_{j.k<a>}^{.i}=\frac{\delta L_{.jk}^i}{\partial y^{<a>}}%
+C_{.j<b>}^iP_{.k<a>}^{<b>}-
$$
$$
\left( \frac{\partial C_{.j<a>}^i}{\partial x^k}%
+L_{.lk}^iC_{.j<a>}^l-L_{.jk}^lC_{.l<a>}^i-L_{.<a>k}^{<c>}C_{.j<c>}^i\right)
,
$$
$$
P_{<b>.k<a>}^{.<c>}=\frac{\delta L_{.<b>k}^{<c>}}{\partial y^{<a>}}%
+C_{.<b><d>}^{<c>}P_{.k<a>}^{<d>}-
$$
$$
\left( \frac{\partial C_{.<b><a>}^{<c>}}{\partial x^k}+L_{.<d>k}^{<c>%
\,}C_{.<b><a>}^{<d>}-L_{.<b>k}^{<d>}C_{.<d><a>}^{<c>}-
L_{.<a>k}^{<d>}C_{.<b><d>}^{<c>}\right) ,
$$
$$
S_{j.<b><c>}^{.i}=\frac{\delta C_{.j<b>}^i}{\partial y^{<c>}}-\frac{\delta
C_{.j<c>}^i}{\partial y^{<b>}}+C_{.j<b>}^hC_{.h<c>}^i-C_{.j<c>}^hC_{h<b>}^i,
$$
$$
S_{<b>.<c><d>}^{.<a>}=\frac{\delta C_{.<b><c>}^{<a>}}{\partial y^{<d>}}-%
\frac{\delta C_{.<b><d>}^{<a>}}{\partial y^{<c>}}+$$
$$
C_{.<b><c>}^{<e>}C_{.<e><d>}^{<a>}-C_{.<b><d>}^{<e>}C_{.<e><c>}^{<a>}.
$$

We note that torsions (25) and curvatures (28) can be computed by
particular cases of d-connections when d-connections (17), (20) or
(22) are used instead of (16) and (17).

The components of the Ricci d--tensor
$$
R_{<\alpha ><\beta >}=R_{<\alpha >.<\beta ><\tau >}^{.<\tau >}
$$
with respect to locally adapted frame (6) are as follows:%
$$
R_{ij}=R_{i.jk}^{.k},\quad R_{i<a>}=-^2P_{i<a>}=-P_{i.k<a>}^{.k},\eqno(29)
$$
$$
R_{<a>i}=^1P_{<a>i}=P_{<a>.i<b>}^{.<b>},\quad
R_{<a><b>}=S_{<a>.<b><c>}^{.<c>}.
$$
We point out that because, in general, $^1P_{<a>i}\neq ~^2P_{i<a>}$ the
Ricci d--tensor is non symmetric.

Having defined a d--metric of type (12) in ${\cal E}^{<z>}$ we can
introduce the scalar curvature of d--connection ${\bf D}$:
$$
{\overleftarrow{R}}=G^{<\alpha ><\beta >}R_{<\alpha ><\beta >}=R+S,%
$$
where $R=g^{ij}R_{ij}$ and $S=h^{<a><b>}S_{<a><b>}.$

For our further considerations it will be also useful to use an alternative
way of definition torsion (23) and curvature (26) by using the
commutator
$$
\Delta _{<\alpha ><\beta >}\doteq \nabla _{<\alpha >}\nabla _{<\beta
>}-\nabla _{<\beta >}\nabla _{<\alpha >}=2\nabla _{[<\alpha >}\nabla
_{<\beta >]}.
$$
For components (25) of d--torsion we have
$$
\Delta _{<\alpha ><\beta >}f=T_{.<\alpha ><\beta >}^{<\gamma >}\nabla
_{<\gamma >}f\eqno(30)
$$
for every scalar function $f\,\,$ on ${\cal E}^{<z>}{\cal .}$ Curvature can
be introduced as an operator acting on arbitrary d--vector $V^{<\delta >}:$

$$
(\Delta _{<\alpha ><\beta >}-T_{.<\alpha ><\beta >}^{<\gamma >}\nabla
_{<\gamma >})V^{<\delta >}=R_{~<\gamma >.<\alpha ><\beta >}^{.<\delta
>}V^{<\gamma >}\eqno(31)
$$
(in this section we  follow conventions of Miron and
Anastasiei \cite{ma87,ma94} on d--tensors; we can obtain corresponding
Penrose and Rindler abstract index formulas \cite{penr1,penr2} just for a
trivial N-connection structure and by changing denotations for components of
torsion and curvature in this manner:\ $T_{.\alpha \beta }^\gamma
\rightarrow T_{\alpha \beta }^{\quad \gamma }$ and $R_{~\gamma .\alpha \beta
}^{.\delta }\rightarrow R_{\alpha \beta \gamma }^{\qquad \delta }).$

Here we also note that torsion and curvature of a d--connection on ${\cal E}%
^{<z>}$ satisfy generalized for ha--spaces Ricci and Bianchi identities which
in terms of components (30) and (31) are written respectively as%
$$
R_{~[<\gamma >.<\alpha ><\beta >]}^{.<\delta >}+\nabla _{[<\alpha
>}T_{.<\beta ><\gamma >]}^{<\delta >}+T_{.[<\alpha ><\beta >}^{<\nu
>}T_{.<\gamma >]<\nu >}^{<\delta >}=0\eqno(32)
$$
and%
$$
\nabla _{[<\alpha >}R_{|<\nu >|<\beta ><\gamma >]}^{\cdot <\sigma
>}+T_{\cdot [<\alpha ><\beta >}^{<\delta >}R_{|<\nu >|.<\gamma >]<\delta
>}^{\cdot <\sigma >}=0.
$$
Identities (32) can be proved similarly as in \cite{penr1} by taking into
account that indices play a distinguished character.

We can also consider a ha-generalization of the so-called conformal Weyl
tensor (see, for instance, \cite{penr1}) which can be written as a d-tensor
in this form:%
$$
C_{\quad <\alpha ><\beta >}^{<\gamma ><\delta >}=R_{\quad <\alpha ><\beta
>}^{<\gamma ><\delta >}-\frac 4{n+m_1+...+m_z-2}R_{\quad [<\alpha
>}^{[<\gamma >}~\delta _{\quad <\beta >]}^{<\delta >]}+\eqno(33)
$$
$$
\frac 2{(n+m_1+...m_z-1)(n+m_1+...+m_z-2)}{\overleftarrow{R}~\delta _{\quad
[<\alpha >}^{[<\gamma >}~\delta _{\quad <\beta >]}^{<\delta >]}.}
$$
This object is conformally invariant on ha--spaces provided with
d--con\-nec\-ti\-on
generated by d--metric structures.

\subsection{ Field equations for ha--gravity}

The Einstein equations in some models of higher order anisotropic
supergravity have been considered in \cite{v96jpa}. Here we note that the
Einstein equations and conservation laws on v--bundles provided with
N-connection structures were studied in detail in \cite
{ma87,ma94,ana86,ana87,vodg,voa,vcl96}. In Ref. \cite{vg} we proved that the
la-gravity can be formulated in a gauge like manner and analyzed the
conditions when the Einstein la-gravitational field equations are equivalent
to a corresponding form of Yang-Mills equations. Our aim here is to
write the higher order anisotropic gravitational field equations in a form
more convenient for theirs equivalent reformulation in ha--spinor variables.

We define d-tensor $\Phi _{<\alpha ><\beta >}$ as to satisfy conditions
$$
-2\Phi _{<\alpha ><\beta >}\doteq R_{<\alpha ><\beta >}-\frac
1{n+m_1+...+m_z}\overleftarrow{R}g_{<\alpha ><\beta >}
$$
which is the torsionless part of the Ricci tensor for locally isotropic
spaces \cite{penr1,penr2}, i.e. $\Phi _{<\alpha >}^{~~<\alpha >}\doteq 0$.\
The Einstein equations on ha--spaces
$$
\overleftarrow{G}_{<\alpha ><\beta >}+\lambda g_{<\alpha ><\beta >}=\kappa
E_{<\alpha ><\beta >},\eqno(34)
$$
where%
$$
\overleftarrow{G}_{<\alpha ><\beta >}=R_{<\alpha ><\beta >}-\frac 12%
\overleftarrow{R}g_{<\alpha ><\beta >}
$$
is the Einstein d--tensor, $\lambda $ and $\kappa $ are correspondingly the
cosmological and gravitational constants and by $E_{<\alpha ><\beta >}$ is
denoted the locally anisotropic energy--momentum d--tensor, can be rewritten
in equivalent form:%
$$
\Phi _{<\alpha ><\beta >}=-\frac \kappa 2(E_{<\alpha ><\beta >}-\frac
1{n+m_1+...+m_z}E_{<\tau >}^{~<\tau >}~g_{<\alpha ><\beta >}).\eqno(35)
$$

Because ha--spaces generally have nonzero torsions we shall add to (35)
(equivalently to (34)) a system of algebraic d--field equations with the
source $S_{~<\beta ><\gamma >}^{<\alpha >}$ being the locally anisotropic
spin density of matter (if we consider a variant of higher order anisotropic
Einstein--Cartan theory ):
$$
T_{~<\alpha ><\beta >}^{<\gamma >}+2\delta _{~[<\alpha >}^{<\gamma
>}T_{~<\beta >]<\delta >}^{<\delta >}=\kappa S_{~<\alpha ><\beta
>.}^{<\gamma >}\eqno(36)
$$
From (32 ) and (36) one follows the conservation law of higher order
anisot\-rop\-ic spin matter:%
$$
\nabla _{<\gamma >}S_{~<\alpha ><\beta >}^{<\gamma >}-T_{~<\delta ><\gamma
>}^{<\delta >}S_{~<\alpha ><\beta >}^{<\gamma >}=E_{<\beta ><\alpha
>}-E_{<\alpha ><\beta >}.
$$

Finally,  we remark that all presented geometric
constructions contain those elaborated for generalized Lagrange spaces \cite
{ma87,ma94} (for which a tangent bundle $TM$ is considered instead of a
v-bundle ${\cal E}^{<z>}$ ) and for constructions on the so called osculator
bundles with different prolongations and extensions of Finsler and Lagrange
metrics \cite{mirata}. We also note that the Lagrange (Finsler) geometry is
characterized by a metric of type (12) with components parametrized as $%
g_{ij}=\frac 12\frac{\partial ^2{\cal L}}{\partial y^i\partial y^j}$ $\left(
g_{ij}=\frac 12\frac{\partial ^2\Lambda ^2}{\partial y^i\partial y^j}\right)
$ and $h_{ij}=g_{ij},$ where ${\cal L=L}$ $(x,y)$ $\left( \Lambda =\Lambda
\left( x,y\right) \right) $ is a Lagrangian $\left( \mbox{Finsler metric}%
\right) $ on $TM$ (see details in \cite{ma87,ma94,mat,bej}).

\section{Distinguished Clifford Algebras}

The typical fiber of dv-bundle $\xi _d\ ,\ \pi _d:\ HE\oplus V_1E\oplus
...\oplus V_zE\rightarrow E$ is a d-vector space, ${\cal F}=h{\cal F}\oplus
v_1{\cal F\oplus ...}\oplus v_z{\cal F},$ split into horizontal $h{\cal F}$
and verticals $v_p{\cal F,}p=1,...,z$ subspaces, with metric $G(g,h)$
induced by v-bundle metric (12). Clifford algebras (see, for example,
Refs. \cite{kar,tur,penr2}) formulated for d-vector spaces will be called
Clifford d-algebras \cite{vjmp,vb295,vod}. We shall consider
the main properties of Clifford d--algebras. The proof of theorems will be
based on the technique developed in Ref. \cite{kar} correspondingly adapted
to the distinguished character of spaces in consideration.

Let $k$ be a number field (for our purposes $k={\cal R}$ or $k={\cal C},%
{\cal R}$ and ${\cal C},$ are, respectively real and complex number fields)
and define ${\cal F},$ as a d-vector space on $k$ provided with
nondegenerate symmetrical quadratic form (metric)\ $G.$ Let $C$ be an algebra
on $k$ (not necessarily commutative) and $j\ :\ {\cal F}$ $\rightarrow C$ a
homomorphism of underlying vector spaces such that $j(u)^2=\;G(u)\cdot 1\ (1$
is the unity in algebra $C$ and d-vector $u\in {\cal F}).$ We are interested
in definition of the pair $\left( C,j\right) $ satisfying the next
universitality conditions. For every $k$-algebra $A$ and arbitrary
homomorphism $\varphi :{\cal F}\rightarrow A$ of the underlying d-vector
spaces, such that $\left( \varphi (u)\right) ^2\rightarrow G\left( u\right)
\cdot 1,$ there is a unique homomorphism of algebras $\psi \ :\ C\rightarrow
A$ transforming the diagram 1 into a commutative one.
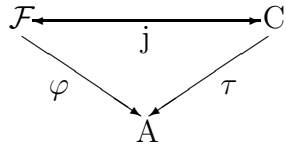
\begin{figure}[htbp]
\begin{center}
\begin{picture}(100,50) \setlength{\unitlength}{1pt}
\thinlines
\put(0,45){${\cal F}$}
\put(96,45){C}
\put(48,2){A}
\put(50,38){j}
\put(50,48){ \vector(-1,0){45}}
\put(50,48){\vector(1,0){45}}
\put(5,42){\vector(3,-2){45}}
\put(98,42){\vector(-3,-2){45}}
\put(15,20){$\varphi$}
\put(80,20){$\tau$}
\end{picture}
\end{center}
\caption{Diagram 1}
\end{figure}

 The algebra solving this problem will be denoted as
$C\left( {\cal F},A\right) $ [equivalently as $C\left( G\right) $ or $%
C\left( {\cal F}\right) ]$ and called as Clifford d--algebra associated with
pair $\left( {\cal F},G\right) .$

\begin{theorem}
The above-presented diagram has a unique solution $\left( C,j\right) $ up to
isomorphism.
\end{theorem}

{\bf Proof:} (We adapt for d-algebras that of Ref. \cite{kar}, p. 127.) For
a universal problem the uniqueness is obvious if we prove the existence of
solution $C\left( G\right) $ . To do this we use tensor algebra ${\cal L}%
^{(F)}=\oplus {\cal L}_{qs}^{pr}\left( {\cal F}\right) $ =$\oplus
_{i=0}^\infty T^i\left( {\cal F}\right) ,$ where $T^0\left( {\cal F}\right)
=k$ and $T^i\left( {\cal F}\right) =k$ and $T^i\left( {\cal F}\right) ={\cal %
F}\otimes ...\otimes {\cal F}$ for $i>0.$ Let $I\left( G\right) $ be the
bilateral ideal generated by elements of form $\epsilon \left( u\right)
=u\otimes u-G\left( u\right) \cdot 1$ where $u\in {\cal F}$ and $1$ is the
unity element of algebra ${\cal L}\left( {\cal F}\right) .$ Every element
from $I\left( G\right) $ can be written as $\sum\nolimits_i\lambda
_i\epsilon \left( u_i\right) \mu _i,$ where $\lambda _{i},\mu _i\in {\cal L}(%
{\cal F})$ and $u_i\in {\cal F}.$ Let $C\left( G\right) $ =${\cal L}({\cal F}%
)/I\left( G\right) $ and define $j:{\cal F}\rightarrow C\left( G\right) $ as
the composition of monomorphism $i:{{\cal F}\rightarrow L}^1 ({\cal F}%
)\subset {\cal L}({\cal F})$ and projection $p:{\cal L}\left( {\cal F}%
\right) \rightarrow C\left( G\right) .$ In this case pair $\left( C\left(
G\right) ,j\right) $ is the solution of our problem. From the general
properties of tensor algebras the homomorphism $\varphi :{\cal F}\rightarrow
A$ can be extended to ${\cal L}({\cal F})$ , i.e., the diagram 2
is commutative, where $\rho $ is a monomorphism of algebras. Because $\left(
\varphi \left( u\right) \right) ^2=G\left( u\right) \cdot 1,$ then $\rho $
vanishes on ideal $I\left( G\right) $ and in this case the necessary
homomorphism $\tau $ is defined. As a consequence of uniqueness of $\rho ,$
the homomorphism $\tau $ is unique.
\begin{figure}[htbp]
\begin{center}
\begin{picture}(100,50) \setlength{\unitlength}{1pt}
\thinlines
\put(0,45){${\cal F}$}
\put(96,45){${\cal L}({\cal F})$}
\put(48,2){A}
\put(50,38){i}
\put(50,48){ \line(-1,0){45}}
\put(50,48){\vector(1,0){45}}
\put(5,42){\vector(3,-2){45}}
\put(98,42){\vector(-3,-2){45}}
\put(15,20){$\varphi$}
\put(80,20){$\rho$}
\end{picture}
\end{center}
\caption{Diagram 2}
\end{figure}
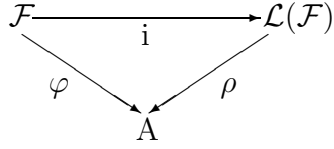

Tensor d-algebra ${\cal L}({\cal F )}$ can be considered as a ${\cal Z}/2$
graded algebra. Really, let us in\-tro\-duce ${\cal L}^{(0)}({\cal F}) =
\sum_{i=1}^\infty T^{2i}\left( {\cal F}\right) $ and ${\cal L}^{(1)}({\cal F}%
) =\sum_{i=1}^\infty T^{2i+1}\left( {\cal F}\right) .$ Setting $I^{(\alpha
)}\left( G\right) =I\left( G\right) \cap {\cal L}^{(\alpha )}({\cal F}).$
Define $C^{(\alpha )}\left( G\right) $ as $p\left( {\cal L}^{(\alpha )} (%
{\cal F})\right) ,$ where $p:{\cal L}\left( {\cal F}\right) \rightarrow
C\left( G\right) $ is the canonical projection. Then $C\left( G\right)
=C^{(0)}\left( G\right) \oplus C^{(1)}\left( G\right) $ and in consequence
we obtain that the Clifford d-algebra is ${\cal Z}/2$ graded.

It is obvious that Clifford d-algebra functorially depends on pair $\left(
{\cal F},G\right) .$ If $f:{\cal F}\rightarrow{\cal F}^{\prime }$ is a
homomorphism of k-vector spaces, such that $G^{\prime }\left( f(u)\right)
=G\left( u\right) ,$ where $G$ and $G^{\prime }$ are, respectively, metrics
on ${\cal F}$ and ${\cal F}^{\prime },$ then $f$ induces an homomorphism of
d-algebras%

$$
C\left( f\right) :C\left( G\right) \rightarrow C\left( G^{\prime }\right)
$$
with identities $C\left( \varphi \cdot f\right) =C\left( \varphi \right)
C\left( f\right) $ and $C\left( Id_{{\cal F}}\right) =Id_{C({\cal F)}}.$

If ${\cal A}^{\alpha}$ and ${\cal B}^{\beta}$ are ${\cal Z}/2$--graded
 d--algebras,
then their graded tensorial product $%
{\cal A}^\alpha \otimes {\cal B}^\beta $ is defined as a d-algebra for
k-vector d-space ${\cal A}^\alpha \otimes {\cal B}^\beta $ with the graded
product induced as $\left( a\otimes b\right) \left( c\otimes d\right)
=\left( -1\right) ^{\alpha \beta }ac\otimes bd,$ where $b\in {\cal B}^\alpha
$ and $c\in {\cal A}^\alpha \quad \left( \alpha ,\beta =0,1\right) .$

Now we reformulate for d--algebras the Chevalley theorem \cite{chev}:

\begin{theorem}
The Clifford d-algebra
$$
C\left( h{\cal F}\oplus v_1{\cal F\oplus ...}\oplus v_z{\cal F}%
,g+h_1+...+h_z\right)
$$
is naturally isomorphic to $C(g)\otimes C\left( h_1\right) \otimes
...\otimes C\left( h_z\right) .$
\end{theorem}

{\bf Proof. }Let $n:h{\cal F}\rightarrow C\left( g\right) $ and $%
n_{(p)}^{\prime }:v_{(p)}{\cal F}\rightarrow C\left( h_{(p)}\right) $ be
canonical maps and map
$$
m:h{\cal F}\oplus v_1{\cal F...}\oplus v_z{\cal F}\rightarrow C(g)\otimes
C\left( h_1\right) \otimes ...\otimes C\left( h_z\right)
$$
is defined as%
$$
m(x,y_{(1)},...,y_{(z)})=
$$
$$
n(x)\otimes 1\otimes ...\otimes 1+1\otimes n^{\prime }(y_{(1)})\otimes
...\otimes 1+1\otimes ...\otimes 1\otimes n^{\prime }(y_{(z)}),
$$
$x\in h{\cal F},y_{(1)}\in v_{(1)}{\cal F,...},y_{(z)}\in v_{(z)}{\cal F.}$
We have
$$
\left( m(x,y_{(1)},...,y_{(z)})\right) ^2=\left[ \left( n\left( x\right)
\right) ^2+\left( n^{\prime }\left( y_{(1)}\right) \right) ^2+...+\left(
n^{\prime }\left( y_{(z)}\right) \right) ^2\right] \cdot 1=
$$
$$
[g\left( x\right) +h\left( y_{(1)}\right) +...+h\left( y_{(z)}\right) ].
$$
\ Taking into account the universality property of Clifford d-algebras we
conclude that $m_1+...+m_z$ induces the homomorphism%
$$
\zeta :C\left( h{\cal F}\oplus v_1{\cal F}\oplus ...\oplus v_z{\cal F}%
,g+h_1+...+h_z\right) \rightarrow
$$
$$
C\left( h{\cal F},g\right) \widehat{\otimes }C\left( v_1{\cal F},h_1\right)
\widehat{\otimes }...C\left( v_z{\cal F},h_z\right) .
$$
We also can define a homomorphism%
$$
\upsilon :C\left( h{\cal F},g\right) \widehat{\otimes }C\left( v_1{\cal F}%
,h_{(1)}\right) \widehat{\otimes }...\widehat{\otimes }C\left( v_z{\cal F}%
,h_{(z)}\right) \rightarrow
$$
$$
C\left( h{\cal F}\oplus v_1{\cal F\oplus ...}\oplus v_z{\cal F}%
,g+h_{(1)}+...+h_{(z)}\right)
$$
by using formula $$\upsilon \left( x\otimes y_{(1)}\otimes ...\otimes
y_{(z)}\right) =\delta \left( x\right) \delta _{(1)}^{\prime }\left(
y_{(1)}\right) ...\delta _{(z)}^{\prime }\left( y_{(z)}\right) ,$$ where
homomorphisms $\delta $ and $\delta _{(1)}^{\prime },...,\delta
_{(z)}^{\prime }$ are, respectively, induced by
im\-bed\-dings of $h{\cal F}$
and $v_1{\cal F}$ into $h{\cal F}\oplus v_1{\cal F\oplus ...}\oplus v_z{\cal %
F}:$%
$$
\delta :C\left( h{\cal F},g\right) \rightarrow C\left( h{\cal F}\oplus v_1%
{\cal F\oplus ...}\oplus v_z{\cal F},g+h_{(1)}+...+h_{(z)}\right) ,
$$
$$
\delta _{(1)}^{\prime }:C\left( v_1{\cal F},h_{(1)}\right) \rightarrow
C\left( h{\cal F}\oplus v_1{\cal F\oplus ...}\oplus v_z{\cal F}%
,g+h_{(1)}+...+h_{(z)}\right) ,
$$
$$
...................................
$$
$$
\delta _{(z)}^{\prime }:C\left( v_z{\cal F},h_{(z)}\right) \rightarrow
C\left( h{\cal F}\oplus v_1{\cal F\oplus ...}\oplus v_z{\cal F}%
,g+h_{(1)}+...+h_{(z)}\right) .
$$

Superpositions of homomorphisms $\zeta $ and $\upsilon $ lead to identities%
$$
\upsilon \zeta =Id_{C\left( h{\cal F},g\right) \widehat{\otimes }C\left( v_1%
{\cal F},h_{(1)}\right) \widehat{\otimes }...\widehat{\otimes }C\left( v_z%
{\cal F},h_{(z)}\right) },\eqno(37)
$$
$$
\zeta \upsilon =Id_{C\left( h{\cal F},g\right) \widehat{\otimes }C\left( v_1%
{\cal F},h_{(1)}\right) \widehat{\otimes }...\widehat{\otimes }C\left( v_z%
{\cal F},h_{(z)}\right) }.
$$
Really, d-algebra $$C\left( h{\cal F}\oplus v_1{\cal F\oplus ...}\oplus v_z%
{\cal F},g+h_{(1)}+...+h_{(z)}\right) $$ is generated by elements of type $%
m(x,y_{(1)},...y_{(z)}).$ Calculating
$$
\upsilon \zeta \left( m\left( x,y_{(1)},...y_{(z)}\right) \right) =\upsilon
(n\left( x\right) \otimes 1\otimes ...\otimes 1+1\otimes n_{(1)}^{\prime
}\left( y_{(1)}\right) \otimes ...\otimes 1+...+
$$
$$
1\otimes ....\otimes n_{(z)}^{\prime }\left( y_{(z)}\right) )=\delta \left(
n\left( x\right) \right) \delta \left( n_{(1)}^{\prime }\left(
y_{(1)}\right) \right) ...\delta \left( n_{(z)}^{\prime }\left(
y_{(z)}\right) \right) =
$$
$$
m\left( x,0,...,0\right) +m(0,y_{(1)},...,0)+...+m(0,0,...,y_{(z)})=m\left(
x,y_{(1)},...,y_{(z)}\right) ,
$$
we prove the first identity in (37).

On the other hand, d-algebra
$$
C\left( h{\cal F},g\right) \widehat{\otimes }C\left( v_1{\cal F}%
,h_{(1)}\right) \widehat{\otimes }...\widehat{\otimes }C\left( v_z{\cal F}%
,h_{(z)}\right)
$$
is generated by elements of type
$$
n\left( x\right) \otimes 1\otimes ...\otimes ,1\otimes n_{(1)}^{\prime
}\left( y_{(1)}\right) \otimes ...\otimes 1,...1\otimes ....\otimes
n_{(z)}^{\prime }\left( y_{(z)}\right) ,
$$
we prove the second identity in (37).

Following from the above--mentioned properties of homomorphisms $\zeta $ and
$\upsilon $ we can assert that the natural isomorphism is explicitly
constructed.$\Box $

In consequence of theorem 2 we conclude that all operations with Clifford
d-algebras can be reduced to calculations for $C\left( h{\cal F},g\right) $
and \\ $C\left( v_{(p)}{\cal F},h_{(p)}\right) $ which are usual Clifford
algebras of dimension $2^n$ and, respectively, $2^{m_p}$ \cite{kar,ati}.

Of special interest is the case when $k={\cal R}$ and ${\cal F}$ is
isomorphic to vector space ${\cal R}^{p+q,a+b}$ provided with quadratic form
$-x_1^2-...-x_p^2+x_{p+q}^2-y_1^2-...-y_a^2+...+y_{a+b}^2.$ In this case,
the Clifford algebra, denoted as $\left( C^{p,q},C^{a,b}\right) ,\,$ is
generated by symbols $%
e_1^{(x)},e_2^{(x)},...,e_{p+q}^{(x)},e_1^{(y)},e_2^{(y)},...,e_{a+b}^{(y)}$
satisfying properties $\left( e_i\right) ^2=-1~\left( 1\leq i\leq p\right)
,\left( e_j\right) ^2=-1~\left( 1\leq j\leq a\right) ,\left( e_k\right)
^2=1~(p+1\leq k\leq p+q),$
 $\left( e_j\right) ^2=1~(n+1\leq s\leq a+b),~e_ie_j=-e_je_i,~i\neq j.\,$
Explicit calculations of $C^{p,q}$ and $C^{a,b}$ are possible by using
isomorphisms \cite{kar,penr2}
$$
C^{p+n,q+n}\simeq C^{p,q}\otimes M_2\left( {\cal R}\right) \otimes
...\otimes M_2\left( {\cal R}\right) \cong C^{p,q}\otimes M_{2^n}\left(
{\cal R}\right) \cong M_{2^n}\left( C^{p,q}\right) ,
$$
where $M_s\left( A\right) $ denotes the ring of quadratic matrices of order $%
s$ with coefficients in ring $A.$ Here we write the simplest isomorphisms $%
C^{1,0}\simeq {\cal C}, C^{0,1}\simeq {\cal R}\oplus {\cal R ,}$ and $%
C^{2,0}={\cal H},$ where by ${\cal H}$ is denoted the body of quaternions.
We summarize this calculus as (as in Ref. \cite{ati})%
$$
C^{0,0}={\cal R}, C^{1,0}={\cal C}, C^{0,1}={\cal R}\oplus {\cal R}, C^{2,0}=%
{\cal H}, C^{0,2}= M_2\left( {\cal R}\right) ,
$$
$$
C^{3,0}={\cal H}\oplus {\cal H} , C^{0,3} = M_2\left( {\cal R}\right),
C^{4,0}=M_2\left( {\cal H}\right) , C^{0,4}=M_2\left( {\cal H}\right) ,
$$
$$
C^{5,0}=M_4\left( {\cal C}\right) ,~C^{0,5}=M_2\left( {\cal H}\right) \oplus
M_2\left( {\cal H}\right) ,~C^{6,0}=M_8\left( {\cal R}\right)
,~C^{0,6}=M_4\left( {\cal H}\right) ,
$$
$$
C^{7,0}=M_8\left( {\cal R}\right) \oplus M_8\left( {\cal R}\right)
,~C^{0,7}=M_8\left( {\cal C}\right) ,~C^{8,0}=M_{16}\left( {\cal R}\right)
,~C^{0,8}=M_{16}\left( {\cal R}\right) .
$$
One of the most important properties of real algebras $C^{0,p}~\left(
C^{0,a}\right) $ and $C^{p,0}~\left( C^{a,0}\right) $ is eightfold
periodicity of $p(a).$

Now, we emphasize that $H^{2n}$-spaces  admit locally a
structure of Clifford algebra on complex vector spaces. Really, by using
almost \ Hermitian structure $J_\alpha ^{\quad \beta }$ and considering
complex space ${\cal C}^n$ with nondegenarate quadratic form $%
\sum_{a=1}^n\left| z_a\right| ^2,~z_a\in {\cal C}^2$ induced locally by
metric (12) (rewritten in complex coordinates $z_a=x_a+iy_a)$ we define
Clifford algebra $\overleftarrow{C}^n=\underbrace{\overleftarrow{C}^1\otimes
...\otimes \overleftarrow{C}^1}_n,$ where $\overleftarrow{C}^1={\cal %
C\otimes }_R{\cal C=C\oplus C}$ or in consequence, $\overleftarrow{C}%
^n\simeq C^{n,0}\otimes _{{\cal R}}{\cal C}\approx C^{0,n}\otimes _{{\cal R}}%
{\cal C}.$ Explicit calculations lead to isomorphisms $\overleftarrow{C}%
^2=C^{0,2}\otimes _{{\cal R}}{\cal C}\approx M_2\left( {\cal R}\right)
\otimes _{{\cal R}}{\cal C}\approx M_2\left( \overleftarrow{C}^n\right)
,~C^{2p}\approx M_{2^p}\left( {\cal C}\right) $ and $\overleftarrow{C}%
^{2p+1}\approx M_{2^p}\left( {\cal C}\right) \oplus M_{2^p}\left( {\cal C}%
\right) ,$ which show that complex Clifford algebras, defined locally for $%
H^{2n}$-spaces, have periodicity 2 on $p.$

Considerations presented in the proof of theorem 2 show that map $j:{\cal F%
}\rightarrow C\left( {\cal F}\right) $ is monomorphic, so we can identify
space ${\cal F}$ with its image in $C\left( {\cal F},G\right) ,$ denoted as $%
u\rightarrow \overline{u},$ if $u\in C^{(0)}\left( {\cal F},G\right) ~\left(
u\in C^{(1)}\left( {\cal F},G\right) \right) ;$ then $u=\overline{u}$ (
respectively, $\overline{u}=-u).$

\begin{definition}
The set of elements $u\in C\left( G\right) ^{*},$ where $C\left( G\right)
^{*}$ denotes the multiplicative group of invertible elements of $C\left(
{\cal F},G\right) $ satisfying $\overline{u}{\cal F}u^{-1}\in {\cal F},$ is
called the twisted Clifford d-group, denoted as $\widetilde{\Gamma }\left(
{\cal F}\right) .$
\end{definition}

Let $\widetilde{\rho }:\widetilde{\Gamma }\left( {\cal F}\right) \rightarrow
GL\left( {\cal F}\right) $ be the homomorphism given by $u\rightarrow \rho
\widetilde{u},$ where $\widetilde{\rho }_u\left( w\right) =\overline{u}%
wu^{-1}.$ We can verify that $\ker \widetilde{\rho }={\cal R}^{*}$is a
subgroup in $\widetilde{\Gamma }\left( {\cal F}\right) .$

Canonical map $j:{\cal F}\rightarrow C\left( {\cal F}\right) $ can be
interpreted as the linear map ${\cal F}\rightarrow C\left( {\cal F}\right)
^0 $ satisfying the universal property of Clifford d-algebras. This leads to
a homomorphism of algebras, $C\left( {\cal F}\right) \rightarrow C\left(
{\cal F}\right) ^t,$ considered by an anti-involution of $C\left( {\cal F}%
\right) $ and denoted as $u\rightarrow ~^tu.$ More exactly, if $u_1...u_n\in
{\cal F,}$ then $t_u=u_n...u_1$ and $^t\overline{u}=\overline{^tu}=\left(
-1\right) ^nu_n...u_1.$

\begin{definition}
The spinor norm of arbitrary $u\in C\left( {\cal F}\right) $ is defined as\\
$S\left( u\right) =~^t\overline{u}\cdot u\in C\left( {\cal F}\right) .$
\end{definition}

It is obvious that if $u,u^{\prime },u^{\prime \prime }\in \widetilde{\Gamma
}\left( {\cal F}\right) ,$ then $S(u,u^{\prime })=S\left( u\right) S\left(
u^{\prime }\right) $ and \\ $S\left( uu^{\prime }u^{\prime \prime }\right)
=S\left( u\right) S\left( u^{\prime }\right) S\left( u^{\prime \prime
}\right) .$ For $u,u^{\prime }\in {\cal F} S\left( u\right) =-G\left(
u\right) $ and $S\left( u,u^{\prime }\right) =S\left( u\right) S\left(
u^{\prime }\right) =S\left( uu^{\prime }\right) .$

Let us introduce the orthogonal group $O\left( G\right) \subset GL\left(
G\right) $ defined by metric $G$ on ${\cal F}$ and denote sets $SO\left(
G\right) =\{u\in O\left( G\right) ,\det \left| u\right| =1\},~Pin\left(
G\right) =\{u\in \widetilde{\Gamma }\left( {\cal F}\right) ,S\left( u\right)
=1\}$ and $Spin\left( G\right) =Pin\left( G\right) \cap C^0\left( {\cal F}%
\right) .$ For ${{\cal F}\cong {\cal R}}^{n+m}$ we write $Spin\left(
n_E\right) .$ By straightforward calculations (see similar considerations in
Ref. \cite{kar}) we can verify the exactness of these sequences:%
$$
1\rightarrow {\cal Z}/2\rightarrow Pin\left( G\right) \rightarrow O\left(
G\right) \rightarrow 1,
$$
$$
1\rightarrow {\cal Z}/2\rightarrow Spin\left( G\right) \rightarrow SO\left(
G\right) \rightarrow 0,
$$
$$
1\rightarrow {\cal Z}/2\rightarrow Spin\left( n_E\right) \rightarrow
SO\left( n_E\right) \rightarrow 1.
$$
We conclude this subsection by emphasizing that the spinor norm was defined
with respect to a quadratic form induced by a metric in dv-bundle ${\cal E}%
^{<z>}$. This approach differs from those presented in Refs. \cite{asa88}
and \cite{ono} where the existence of spinor spaces only with antisymmetric
metrics on Finsler like spaces is postulated. The introduction of $Pin$
structures is closely related with works \cite{cgt,tt} in our case considered
for spaces with local anisotropy.

\section{Higher Order An\-i\-sot\-rop\-ic Clifford Bundles}

We shall consider two variants of generalization of spinor constructions
defined for d-vector spaces to the case of distinguished vector bundle
spaces enabled with the structure of N-connection. The first is to use the
extension to the category of vector bundles. The second is to define the
Clifford fibration associated with compatible linear d-connection and metric
$G$ on a vector bundle. We shall analyze both variants.

\subsection{Clifford d--module structure in dv--bundles}

Because functor ${\cal F}\to C({\cal F})$ is smooth we can extend it to the
category of vector bundles of type $\xi ^{<z>}=\{\pi _d:HE^{<z>}\oplus
V_1E^{<z>}\oplus ...\oplus V_zE^{<z>}\rightarrow E^{<z>}\}.$ Recall that by $%
{\cal F}$ we denote the typical fiber of such bundles. For $\xi ^{<z>}$ we
obtain a bundle of algebras, denoted as $C\left( \xi ^{<z>}\right) ,\,$ such
that $C\left( \xi ^{<z>}\right) _u=C\left( {\cal F}_u\right) .$
Multiplication in every fibre defines a continuous map $C\left( \xi
^{<z>}\right) \times C\left( \xi ^{<z>}\right) \rightarrow C\left( \xi
^{<z>}\right) .$ If $\xi ^{<z>}$ is a vector bundle on number field $k$%
,\thinspace \thinspace the structure of the $C\left( \xi ^{<z>}\right) $%
-module, the d-module, the d-module, on $\xi ^{<z>}$ is given by the
continuous map $C\left( \xi ^{<z>}\right) \times _E\xi ^{<z>}\rightarrow \xi
^{<z>}$ with every fiber ${\cal F}_u$ provided with the structure of the $%
C\left( {\cal F}_u\right) -$module, correlated with its $k$-module
structure, Because ${\cal F}\subset C\left( {\cal F}\right) ,$ we have a
fiber to fiber map ${\cal F}\times _E\xi ^{<z>}\rightarrow \xi ^{<z>},$
inducing on every fiber the map ${\cal F}_u\times _E\xi
_{(u)}^{<z>}\rightarrow \xi _{(u)}^{<z>}$ (${\cal R}$-linear on the first
factor and $k$-linear on the second one ). Inversely, every such bilinear
map defines on $\xi ^{<z>}$ the structure of the $C\left( \xi ^{<z>}\right) $%
-module by virtue of universal properties of Clifford d-algebras.
Equivalently, the above-mentioned bilinear map defines a morphism of
v-bundles $m:\xi ^{<z>}\rightarrow HOM\left( \xi ^{<z>},\xi ^{<z>}\right)
\quad [HOM\left( \xi ^{<z>},\xi ^{<z>}\right) $ denotes the bundles of
homomorphisms] when $\left( m\left( u\right) \right) ^2=G\left( u\right) $
on every point.

Vector bundles $\xi ^{<z>}$ provided with $C\left( \xi ^{<z>}\right) $%
-structures are objects of the category with morphisms being morphisms of
dv-bundles, which induce on every point $u\in \xi ^{<z>}$ morphisms of $%
C\left( {\cal F}_u\right) -$modules. This is a Banach category contained in
the category of finite-dimensional d-vector spaces on filed $k.$ We shall
not use category formalism in this work, but point to its advantages in
further formulation of new directions of K-theory (see , for example, an
introduction in Ref. \cite{kar}) concerned with la-spaces.

Let us denote by $H^s\left( {\cal E}^{<z>},GL_{n_E}\left( {\cal R}\right)
\right) ,\,$ where $n_E=n+m_1+...+m_z,\,$ the s-dimensional cohomology group
of the algebraic sheaf of germs of continuous maps of dv-bundle ${\cal E}%
^{<z>}$ with group $GL_{n_E}\left( {\cal R}\right) $ the group of
automorphisms of ${\cal R}^{n_E}\,$ (for the language of algebraic topology
see, for example, Refs. \cite{kar} and \cite{god}). We shall also use the
group $SL_{n_E}\left( {\cal R}\right) =\{A\subset GL_{n_E}\left( {\cal R}%
\right) ,\det A=1\}.\,$ Here we point out that cohomologies $H^s(M,Gr)$
characterize the class of a principal bundle $\pi :P\rightarrow M$ on $M$
with structural group $Gr.$ Taking into account that we deal with bundles
distinguished by an N-connection we introduce into consideration
cohomologies $H^s\left( {\cal E}^{<z>},GL_{n_E}\left( {\cal R}\right)
\right) $ as distinguished classes (d-classes) of bundles ${\cal E}^{<z>}$
provided with a global N-connection structure.

For a real vector bundle $\xi ^{<z>}$ on compact base ${\cal E}^{<z>}$ we
can define the orientation on $\xi ^{<z>}$ as an element $\alpha _d\in
H^1\left( {\cal E}^{<z>},GL_{n_E}\left( {\cal R}\right) \right) $ whose
image on map%
$$
H^1\left( {\cal E}^{<z>},SL_{n_E}\left( {\cal R}\right) \right) \rightarrow
H^1\left( {\cal E}^{<z>},GL_{n_E}\left( {\cal R}\right) \right)
$$
is the d-class of bundle ${\cal E}^{<z>}.$

\begin{definition}
The spinor structure on $\xi ^{<z>}$ is defined as an element\\ $\beta _d\in
H^1\left( {\cal E}^{<z>},Spin\left( n_E\right) \right) $ whose image in the
composition%
$$
H^1\left( {\cal E}^{<z>},Spin\left( n_E\right) \right) \rightarrow H^1\left(
{\cal E}^{<z>},SO\left( n_E\right) \right) \rightarrow H^1\left( {\cal E}%
^{<z>},GL_{n_E}\left( {\cal R}\right) \right)
$$
is the d-class of ${\cal E}^{<z>}.$
\end{definition}

The above definition of spinor structures can be reformulated in terms of
principal bundles. Let $\xi ^{<z>}$ be a real vector bundle of rank n+m on a
compact base ${\cal E}^{<z>}.$ If there is a principal bundle $P_d$ with
structural group $SO( n_E ) $   [ or $Spin( n_E ) ],$ this bundle $%
\xi ^{<z>}$ can be provided with orientation (or spinor) structure. The
bundle $P_d$ is associated with element $\alpha _d\in H^1\left( {\cal E}%
^{<z>},SO(n_{<z>})\right) $ [or $\beta _d\in H^1\left( {\cal E}%
^{<z>},Spin\left( n_E\right) \right) .$

We remark that a real bundle is oriented if and only if its first
Stiefel-Whitney d--class vanishes,
$$
w_1\left( \xi _d\right) \in H^1\left( \xi ,{\cal Z}/2\right) =0,
$$
where $H^1\left( {\cal E}^{<z>},{\cal Z}/2\right) $ is the first group of
Chech cohomology with coefficients in ${\cal Z}/2,$ Considering the second
Stiefel--Whitney class $w_2\left( \xi ^{<z>}\right) \in H^2\left( {\cal E}%
^{<z>},{\cal Z}/2\right) $ it is well known that vector bundle $\xi ^{<z>}$
admits the spinor structure if and only if $w_2\left( \xi ^{<z>}\right) =0.$
Finally,  we emphasize that taking into account that base
space ${\cal E}^{<z>}$ is also a v-bundle, $p:E^{<z>}\rightarrow M,$ we have
to make explicit calculations in order to express cohomologies $H^s\left(
{\cal E}^{<z>},GL_{n+m}\right) \,$ and $H^s\left( {\cal E}^{<z>},SO\left(
n+m\right) \right) $ through cohomologies $H^s\left( M,GL_n\right)
,H^s\left( M,SO\left( m_1\right) \right) ,$ $...H^s\left( M,SO\left(
m_z\right) \right) ,$ , which depends on global topological structures of
spaces $M$ and ${\cal E}^{<z>}$ $.$ For general bundle and base spaces this
requires a cumbersome cohomological calculus.

\subsection{Clifford fibration}

Another way of defining the spinor structure is to use Clifford fibrations.
Consider the principal bundle with the structural group $Gr$ being a
subgroup of orthogonal group $O\left( G\right) ,$ where $G$ is a quadratic
nondegenerate form (see(12)) defined on the base (also being a bundle
space) space ${\cal E}^{<z>}.$ The fibration associated to principal
fibration $P\left( {\cal E}^{<z>},Gr\right) $ with a typical fiber having
Clifford algebra $C\left( G\right) $ is, by definition, the Clifford
fibration $PC\left( {\cal E}^{<z>},Gr\right) .$ We can always define a
metric on the Clifford fibration if every fiber is isometric to $PC\left(
{\cal E}^{<z>},G\right) $ (this result is proved for arbitrary quadratic
forms $G$ on pseudo-Riemannian bases \cite{tur}). If, additionally, $%
Gr\subset SO\left( G\right) $ a global section can be defined on $PC\left(
G\right) .$

Let ${\cal P}\left( {\cal E}^{<z>},Gr\right) $ be the set of principal
bundles with differentiable base ${\cal E}^{<z>}$ and structural group $Gr.$
If $g:Gr\rightarrow Gr^{\prime }$ is an homomorphism of Lie groups and $%
P\left( {\cal E}^{<z>},Gr\right) \subset {\cal P}\left( {\cal E}%
^{<z>},Gr\right) $ (for simplicity in this subsection we shall denote mentioned
bundles and sets of bundles as $P,P^{\prime }$ and respectively, ${\cal P},%
{\cal P}^{\prime }),$ we can always construct a principal bundle with the
property that there is as homomorphism $f:P^{\prime }\rightarrow P$ of
principal bundles which can be projected to the identity map of ${\cal E}%
^{<z>}$ and corresponds to isomorphism $g:Gr\rightarrow Gr^{\prime }.$ If
the inverse statement also holds, the bundle $P^{\prime }$ is called as the
extension of $P$ associated to $g$ and $f$ is called the extension
homomorphism denoted as $\widetilde{g.}$

Now we can define distinguished spinor structures on bundle spaces (compare
with definition 3 ).

\begin{definition}
Let $P\in {\cal P}\left( {\cal E}^{<z>},O\left( G\right) \right) $ be a
principal bundle. A distinguished spinor structure of $P,$ equivalently a
ds-structure of ${\cal E}^{<z>}$ is an extension $\widetilde{P}$ of $P$
associated to homomorphism $h:PinG\rightarrow O\left( G\right) $ where $%
O\left( G\right) $ is the group of orthogonal rotations, generated by metric
$G,$ in bundle ${\cal E}^{<z>}.$
\end{definition}

So, if $\widetilde{P}$ is a spinor structure of the space ${\cal E}^{<z>},$
then\\ $\widetilde{P}\in {\cal P}\left( {\cal E}^{<z>},PinG\right) .$

The definition of spinor structures on varieties was given in Ref.\cite{cru1}.
In Refs. \cite{cru1} and \cite{cru2} it is proved that a necessary and
sufficient condition for a space time to be orientable is to admit a global
field of orthonormalized frames. We mention that spinor structures can be
also defined on varieties modeled on Banach spaces \cite{ana77}. As we have
shown  similar constructions are possible for the cases
when space time has the structure of a v-bundle with an N-connection.

\begin{definition}
A special distinguished spinor structure, ds-structure, of principal bundle $%
P=P\left( {\cal E}^{<z>},SO\left( G\right) \right) $ is a principal bundle\\ $%
\widetilde{P}=\widetilde{P}\left( {\cal E}^{<z>},SpinG\right) $ for which a
homomorphism of principal bundles $\widetilde{p}:\widetilde{P}\rightarrow P,$
projected on the identity map of ${\cal E}^{<z>}$ and corresponding to
representation%
$$
R:SpinG\rightarrow SO\left( G\right) ,
$$
is defined.
\end{definition}

In the case when the base space variety is oriented, there is a natural
bijection between tangent spinor structures with a common base. For special
ds--structures we can define, as for any spinor structure, the concepts of
spin tensors, spinor connections, and spinor covariant derivations.

\section{Almost Complex Spinor Structures}

Almost complex structures are an important characteristic of $H^{2n}$-spaces
and of osculator bundles $Osc^{k=2k_1}(M),$ where $k_1=1,2,...$ . For
simplicity in this subsection we restrict our analysis to the case of $H^{2n}$%
-spaces. We can rewrite the almost Hermitian metric \cite{ma87,ma94}, $%
H^{2n} $-metric ( see considerations for
metrics and conditions of type (12) and correspondingly (14) ), in
complex form \cite{vjmp}:

$$
G=H_{ab}\left( z,\xi \right) dz^a\otimes dz^b,\eqno(38)
$$
where
$$
z^a=x^a+iy^a,~\overline{z^a}=x^a-iy^a,~H_{ab}\left( z,\overline{z}\right)
=g_{ab}\left( x,y\right) \mid _{y=y\left( z,\overline{z}\right) }^{x=x\left(
z,\overline{z}\right) },
$$
and define almost complex spinor structures. For given metric (38) on $%
H^{2n}$-space there is always a principal bundle $P^U$ with unitary
structural group U(n) which allows us to transform $H^{2n}$-space into
v-bundle $\xi ^U\approx P^U\times _{U\left( n\right) }{\cal R}^{2n}.$ This
statement will be proved after we introduce complex
spinor structures on oriented real vector bundles \cite{kar}.

Let us consider momentarily $k={\cal C}$ and introduce into consideration
[instead of the group $Spin(n)]$ the group $Spin^c\times _{{\cal Z}%
/2}U\left( 1\right) $ being the factor group of the product $Spin(n)\times
U\left( 1\right) $ with the respect to equivalence%
$$
\left( y,z\right) \sim \left( -y,-a\right) ,\quad y\in Spin(m).
$$
This way we define the short exact sequence%
$$
1\rightarrow U\left( 1\right) \rightarrow Spin^c\left( n\right) \stackrel{S^c%
}{\to }SO\left( n\right) \rightarrow 1,        \eqno(39)
$$
where $\rho ^c\left( y,a\right) =\rho ^c\left( y\right) .$ If $\lambda $ is
oriented , real, and rank $n,$ $\gamma $-bundle $\pi :E_\lambda \rightarrow
M^n,$ with base $M^n,$ the complex spinor structure, spin structure, on
$\lambda $ is given by the principal bundle $P$ with structural group $%
Spin^c\left( m\right) $ and isomorphism $\lambda \approx P\times
_{Spin^c\left( n\right) }{\cal R}^n$ (see (39)).
For such bundles the categorial equivalence can be defined as
$$
\epsilon ^c:{\cal E}_{{\cal C}}^T\left( M^n\right) \rightarrow {\cal E}_{%
{\cal C}}^\lambda \left( M^n\right) ,\eqno(40)
$$
where $\epsilon ^c\left( E^c\right) =P\bigtriangleup _{Spin^c\left( n\right)
}E^c$ is the category of trivial complex bundles on $M^n,{\cal E}_{{\cal C}%
}^\lambda \left( M^n\right) $ is the category of complex v-bundles on $M^n$
with action of Clifford bundle $C\left( \lambda \right) ,P\bigtriangleup
_{Spin^c(n)}$ and $E^c$ is the factor space of the bundle product $P\times
_ME^c$ with respect to the equivalence $\left( p,e\right) \sim \left( p%
\widehat{g}^{-1},\widehat{g}e\right) ,p\in P,e\in E^c,$ where $\widehat{g}%
\in Spin^c\left( n\right) $ acts on $E$ by via the imbedding $Spin\left(
n\right) \subset C^{0,n}$ and the natural action $U\left( 1\right) \subset
{\cal C}$ on complex v-bundle $\xi ^c,E^c=tot\xi ^c,$ for bundle $\pi
^c:E^c\rightarrow M^n.$

Now we return to the bundle $\xi ={\cal E}^{<1>}.$ A real v-bundle (not
being a spinor bundle) admits a complex spinor structure if and only if
there exist a homomorphism $\sigma :U\left( n\right) \rightarrow
Spin^c\left( 2n\right) $ making the diagram 3 commutative. The explicit
construction of $\sigma $ for arbitrary $\gamma $-bundle is given in Refs.
\cite{kar} and \cite{ati}. For $H^{2n}$-spaces it is obvious that a diagram
similar to (40) can be defined for the tangent bundle $TM^n,$ which
directly points to the possibility of defining the $^cSpin$-structure on $%
H^{2n}$-spaces.
\begin{figure}[htbp]
\begin{center}
\begin{picture}(255,50) \setlength{\unitlength}{1pt}
\thinlines
\put(50,45){$U(n)$}
\put(166,45){$SO(2n)$}
\put(103,2){${Spin}^c (2n)$}
\put(120,38){i}
\put(120,48){ \line(-1,0){45}}
\put(120,48){\vector(1,0){45}}
\put(75,42){\vector(3,-2){45}}
\put(128,13) {\vector(3,2){45}}
\put(85,20){$\sigma$}
\put(150,20){${\rho}^c$}
\end{picture}
\end{center}
\caption{Diagram 3}
\end{figure}
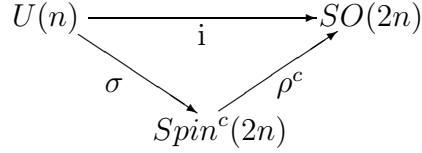

Let $\lambda $ be a complex, rank\thinspace $n,$ spinor bundle with
$$
\tau :Spin^c\left( n\right) \times _{{\cal Z}/2}U\left( 1\right) \rightarrow
U\left( 1\right) \eqno(41)
$$
the homomorphism defined by formula $\tau \left( \lambda ,\delta \right)
=\delta ^2.$ For $P_s$ being the principal bundle with fiber $Spin^c\left(
n\right) $ we introduce the complex linear bundle $L\left( \lambda ^c\right)
=P_S\times _{Spin^c(n)}{\cal C}$ defined as the factor space of $P_S\times
{\cal C}$ on equivalence relation%

$$
\left( pt,z\right) \sim \left( p,l\left( t\right) ^{-1}z\right) ,
$$
where $t\in Spin^c\left( n\right) .$ This linear bundle is associated to
complex spinor structure on $\lambda ^c.$

If $\lambda ^c$ and $\lambda ^{c^{\prime }}$ are complex spinor bundles, the
Whitney sum $\lambda ^c\oplus \lambda ^{c^{\prime }}$ is naturally provided
with the structure of the complex spinor bundle. This follows from the
holomorphism%
$$
\omega ^{\prime }:Spin^c\left( n\right) \times Spin^c\left( n^{\prime
}\right) \rightarrow Spin^c\left( n+n^{\prime }\right) ,\eqno(42)
$$
given by formula $\left[ \left( \beta ,z\right) ,\left( \beta ^{\prime
},z^{\prime }\right) \right] \rightarrow \left[ \omega \left( \beta ,\beta
^{\prime }\right) ,zz^{\prime }\right] ,$ where $\omega $ is the
homomorphism making the diagram 4 commutative.%
Here, $z,z^{\prime }\in U\left( 1\right) .$ It is obvious that $L\left(
\lambda ^c\oplus \lambda ^{c^{\prime }}\right) $ is isomorphic to $L\left(
\lambda ^c\right) \otimes L\left( \lambda ^{c^{\prime }}\right) .$
\begin{figure}[htbp]
\begin{center}
\begin{picture}(190,50) \setlength{\unitlength}{1pt}
\thinlines
\put(0,45){$Spin(n)\times Spin(n')$}
\put(160,45){$Spin(n+n')$}
\put(15,2){$O(n)\times O(n')$}
\put(168,02){$O(n+n')$}
\put(100,45){ \vector(1,0){55}}
\put(48,40){\vector(0,-1){25}}
\put(90,5){\vector(1,0){70}}
\put(185,40) {\vector(0,-1){25}}
\end{picture}
\end{center}
\caption{Diagram 4}
\end{figure}
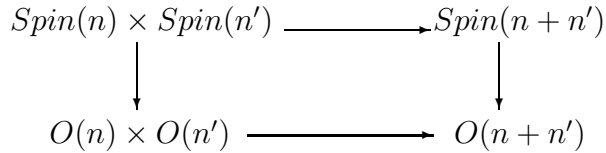

We conclude this subsection by formulating our main result on complex spinor
structures for $H^{2n}$-spaces:

\begin{theorem}
Let $\lambda ^c$ be a complex spinor bundle of rank $n$ and $H^{2n}$-space
considered as a real vector bundle $\lambda ^c\oplus \lambda ^{c^{\prime }}$
provided with almost complex structure $J_{\quad \beta }^\alpha ;$
multiplication on $i$ is given by $\left(
\begin{array}{cc}
0 & -\delta _j^i \\
\delta _j^i & 0
\end{array}
\right) $. Then, the diagram 5 is commutative up to isomorphisms $\epsilon ^c
$ and $\widetilde{\epsilon }^c$ defined as in (40), ${\cal H}$ is functor $%
E^c\rightarrow E^c\otimes L\left( \lambda ^c\right) $ and ${\cal E}_{{\cal C}%
}^{0,2n}\left( M^n\right) $ is defined by functor ${\cal E}_{{\cal C}}\left(
M^n\right) \rightarrow {\cal E}_{{\cal C}}^{0,2n}\left( M^n\right) $ given
as correspondence $E^c\rightarrow \Lambda \left( {\cal C}^n\right) \otimes
E^c$ (which is a categorial equivalence), $\Lambda \left( {\cal C}^n\right) $
is the exterior algebra on ${\cal C}^n.$ $W$ is the real bundle $\lambda
^c\oplus \lambda ^{c^{\prime }}$ provided with complex structure.
\end{theorem}
\begin{figure}[htbp]
\begin{center}
\begin{picture}(150,50) \setlength{\unitlength}{1pt}
\thinlines
\put(0,45){${\cal E}_{\cal C}^{0,2n} (M^{2n})$}
\put(145,45){${\cal E}_{\cal C}^{{\lambda}^c \oplus {\lambda}^c } (M^n )$}
\put(81,0){${\cal E}_{\cal C}^W (M^n )$}
\put(99,38){${\epsilon}^c$}
\put(99,48){ \line(-1,0){45}}
\put(99,48){\vector(1,0){45}}
\put(54,42){\vector(3,-2){45}}
\put(147,42){\vector(-3,-2){45}}
\put(64,20){${\tilde{\varepsilon}}^c$}
\put(129,20){$\cal H$}
\end{picture}
\end{center}
\caption{Diagram 5}
\end{figure}

{\bf Proof: }We use composition of homomorphisms%
$$
\mu :Spin^c\left( 2n\right) \stackrel{\pi }{\to }SO\left( n\right) \stackrel{%
r}{\to }U\left( n\right) \stackrel{\sigma }{\to }Spin^c\left( 2n\right)
\times _{{\cal Z}/2}U\left( 1\right) ,
$$
commutative diagram 6 and introduce composition of homomorphisms%
$$
\mu :Spin^c\left( n\right) \stackrel{\Delta }{\to }Spin^c\left( n\right)
\times Spin^c\left( n\right) \stackrel{{\omega }^c}{\to }Spin^c\left(
n\right) ,
$$
where $\Delta $ is the diagonal homomorphism and $\omega ^c$ is defined as
in (42). Using homomorphisms (41) and (42) we obtain formula $\mu
\left( t\right) =\mu \left( t\right) r\left( t\right) .$
\begin{figure}[htbp]
\begin{center}
\begin{picture}(140,70) \setlength{\unitlength}{1pt}
\thinlines
\put(0,60){$Spin(2n)$}
\put(100,60){${Spin}^c (2n)$}
\put(6,0){$SO(n)$}
\put(106,0){$SO(2n)$}
\put(36,2){ \vector(1,0){65}}
\put(20,13){\vector(0,1){40}}
\put(126,13) {\vector(0,1){40}}
\put(8,30){$\beta$}
\put(70,60){$\subset$}
\end{picture}
\end{center}
\caption{Diagram 6}
\end{figure}

Now consider bundle $P\times _{Spin^c\left( n\right) }Spin^c\left( 2n\right)
$ as the principal $Spin^c\left( 2n\right) $-bundle, associated to $M\oplus
M $ being the factor space of the product $P\times Spin^c\left( 2n\right) $
on the equivalence relation $\left( p,t,h\right) \sim \left( p,\mu \left(
t\right) ^{-1}h\right) .$ In this case the categorial equivalence (40) can
be rewritten as
$$
\epsilon ^c\left( E^c\right) =P\times _{Spin^c\left( n\right) }Spin^c\left(
2n\right) \Delta _{Spin^c\left( 2n\right) }E^c
$$
and seen as factor space of $P\times Spin^c\left( 2n\right) \times _ME^c$ on
equivalence relation
$$
\left( pt,h,e\right) \sim \left( p,\mu \left( t\right) ^{-1}h,e\right)
\mbox{and}\left( p,h_1,h_2,e\right) \sim \left( p,h_1,h_2^{-1}e\right)
$$
(projections of elements $p$ and $e$ coincides on base $M).$ Every element
of $\epsilon ^c\left( E^c\right) $ can be represented as $P\Delta
_{Spin^c\left( n\right) }E^c,$ i.e., as a factor space $P\Delta E^c$ on
equivalence relation $\left( pt,e\right) \sim \left( p,\mu ^c\left( t\right)
,e\right) ,$ when $t\in Spin^c\left( n\right) .$
The complex line bundle $L\left( \lambda ^c\right) $ can be interpreted as
the factor space of\\ $P\times _{Spin^c\left( n\right) }{\cal C}$ on
equivalence relation $\left( pt,\delta \right) \sim \left( p,r\left(
t\right) ^{-1}\delta \right) .$

Putting $\left( p,e\right) \otimes \left( p,\delta \right) \left( p,\delta
e\right) $ we introduce morphism%
$$
\epsilon ^c\left( E\right) \times L\left( \lambda ^c\right) \rightarrow
\epsilon ^c\left( \lambda ^c\right)
$$
with properties $\left( pt,e\right) \otimes \left( pt,\delta \right)
\rightarrow \left( pt,\delta e\right) =\left( p,\mu ^c\left( t\right)
^{-1}\delta e\right) ,$

$\left( p,\mu ^c\left( t\right) ^{-1}e\right) \otimes \left( p,l\left(
t\right) ^{-1}e\right) \rightarrow \left( p,\mu ^c\left( t\right) r\left(
t\right) ^{-1}\delta e\right) $ pointing to the fact that we have defined
the isomorphism correctly and that it is an isomorphism on every fiber. $%
\Box $

\section{ D--Spinor Techniques}

The purpose of this subsection is to show how a corresponding abstract spinor
technique entailing notational and calculations advantages can be developed
for arbitrary splits of dimensions of a d-vector space ${\cal F}=h{\cal F}%
\oplus v_1{\cal F\oplus ...}\oplus v_z{\cal F}$, where $\dim h{\cal F}=n$
and $\dim v_p{\cal F}=m_p.$ For convenience we shall also present some
necessary coordinate expressions.

The problem of a rigorous definition of spinors on la-spaces (la-spinors,
d-spinors) was posed and solved \cite{vjmp,vb295,vod} in the
framework of the formalism of Clifford and spinor structures on v-bundles
provided with compatible nonlinear and distinguished connections and metric.
We introduced d-spinors as corresponding objects of the Clifford d-algebra $%
{\cal C}\left( {\cal F},G\right) $, defined for a d-vector space ${\cal F}$
in a standard manner (see, for instance, \cite{kar}) and proved that
operations with ${\cal C}\left( {\cal F},G\right) \ $ can be reduced to
calculations for ${\cal C}\left( h{\cal F},g\right) ,{\cal C}\left( v_1{\cal %
F},h_1\right) ,...$ and ${\cal C}\left( v_z{\cal F},h_z\right) ,$ which are
usual Clifford algebras of respective dimensions 2$^n,2^{m_1},...$ and 2$%
^{m_z}$ (if it is necessary we can use quadratic forms $g$ and $h_p$
correspondingly induced on $h{\cal F}$ and $v_p{\cal F}$ by a metric ${\bf G}
$ (12)). Considering the orthogonal subgroup $O{\bf \left( G\right) }%
\subset GL{\bf \left( G\right) }$ defined by a metric ${\bf G}$ we can
define the d-spinor norm and parametrize d-spinors by ordered pairs of
elements of Clifford algebras ${\cal C}\left( h{\cal F},g\right) $ and $%
{\cal C}\left( v_p{\cal F},h_p\right) ,p=1,2,...z.$ We emphasize that the
splitting of a Clifford d-algebra associated to a dv-bundle ${\cal E}^{<z>}$
is a straightforward consequence of the global decomposition (3) defining
a N-connection structure in ${\cal E}^{<z>}{\cal .}$

In this subsection we shall omit detailed proofs which in most cases are
mechanical but rather tedious. We can apply the methods developed in \cite
{pen,penr1,penr2,lue} in a straightforward manner on h- and v-subbundles in
order to verify the correctness of affirmations.

\subsection{Clifford d--algebra, d--spinors and d--twistors}

In order to relate the succeeding constructions with Clifford d-algebras
\cite{vjmp,vb295} we consider a la-frame decomposition of the metric (12):%
$$
G_{<\alpha ><\beta >}\left( u\right) =l_{<\alpha >}^{<\widehat{\alpha }%
>}\left( u\right) l_{<\beta >}^{<\widehat{\beta }>}\left( u\right) G_{<%
\widehat{\alpha }><\widehat{\beta }>},
$$
where the frame d-vectors and constant metric matrices are distinguished as

$$
l_{<\alpha >}^{<\widehat{\alpha }>}\left( u\right) =\left(
\begin{array}{cccc}
l_j^{\widehat{j}}\left( u\right) & 0 & ... & 0 \\
0 & l_{a_1}^{\widehat{a}_1}\left( u\right) & ... & 0 \\
... & ... & ... & ... \\
0 & 0 & .. & l_{a_z}^{\widehat{a}_z}\left( u\right)
\end{array}
\right) ,
$$
$$
G_{<\widehat{\alpha }><\widehat{\beta }>}=\left(
\begin{array}{cccc}
g_{\widehat{i}\widehat{j}} & 0 & ... & 0 \\
0 & h_{\widehat{a}_1\widehat{b}_1} & ... & 0 \\
... & ... & ... & ... \\
0 & 0 & 0 & h_{\widehat{a}_z\widehat{b}_z}
\end{array}
\right) ,
$$
$g_{\widehat{i}\widehat{j}}$ and $h_{\widehat{a}_1\widehat{b}_1},...,h_{%
\widehat{a}_z\widehat{b}_z}$ are diagonal matrices with $g_{\widehat{i}%
\widehat{i}}=$ $h_{\widehat{a}_1\widehat{a}_1}=...=h_{\widehat{a}_z\widehat{b%
}_z}=\pm 1.$

To generate Clifford d-algebras we start with matrix equations%
$$
\sigma _{<\widehat{\alpha }>}\sigma _{<\widehat{\beta }>}+\sigma _{<\widehat{%
\beta }>}\sigma _{<\widehat{\alpha }>}=-G_{<\widehat{\alpha }><\widehat{%
\beta }>}I,\eqno(43)
$$
where $I$ is the identity matrix, matrices $\sigma _{<\widehat{\alpha }%
>}\,(\sigma $-objects) act on a d-vector space ${\cal F}=h{\cal F}\oplus v_1%
{\cal F}\oplus ...\oplus v_z{\cal F}$ and theirs components are
distinguished as
$$
\sigma _{<\widehat{\alpha }>}\,=\left\{ (\sigma _{<\widehat{\alpha }>})_{%
\underline{\beta }}^{\cdot \underline{\gamma }}=\left(
\begin{array}{cccc}
(\sigma _{\widehat{i}})_{\underline{j}}^{\cdot \underline{k}} & 0 & ... & 0
\\
0 & (\sigma _{\widehat{a}_1})_{\underline{b}_1}^{\cdot \underline{c}_1} &
... & 0 \\
... & ... & ... & ... \\
0 & 0 & ... & (\sigma _{\widehat{a}_z})_{\underline{b}_z}^{\cdot \underline{c%
}_z}
\end{array}
\right) \right\} ,\eqno(44)
$$
indices \underline{$\beta $},\underline{$\gamma $},... refer to spin spaces
of type ${\cal S}=S_{(h)}\oplus S_{(v_1)}\oplus ...\oplus S_{(v_z)}$ and
underlined Latin indices \underline{$j$},$\underline{k},...$ and $\underline{%
b}_1,\underline{c}_1,...,\underline{b}_z,\underline{c}_z...$ refer
respectively to h-spin space ${\cal S}_{(h)}$ and v$_p$-spin space ${\cal S}%
_{(v_p)},(p=1,2,...,z)\ $which are correspondingly associated to a h- and v$%
_p$-decomposition of a dv-bundle ${\cal E}^{<z>}.$ The irreducible algebra
of matrices $\sigma _{<\widehat{\alpha }>}$ of minimal dimension $N\times N,$
where $N=N_{(n)}+N_{(m_1)}+...+N_{(m_z)},$ $\dim {\cal S}_{(h)}$=$N_{(n)}$
and $\dim {\cal S}_{(v_p)}$=$N_{(m_p)},$ has these dimensions%
$$
{N_{(n)}=\left\{
\begin{array}{rl}
{\ 2^{(n-1)/2},} & n=2k+1 \\
{2^{n/2},\ } & n=2k;
\end{array}
\right. }$$
 and
 $$
{N_{(m_p)}=\left\{
\begin{array}{rl}
{\ 2^{(m_p-1)/2},} & m_p=2k_p+1 \\
{2^{m_p},\ } & m_p=2k_p,
\end{array}
\right. } $$
where $k=1,2,...,k_p=1,2,....$

The Clifford d-algebra is generated by sums on $n+1$ elements of form%
$$
A_1I+B^{\widehat{i}}\sigma _{\widehat{i}}+C^{\widehat{i}\widehat{j}}\sigma _{%
\widehat{i}\widehat{j}}+D^{\widehat{i}\widehat{j}\widehat{k}}\sigma _{%
\widehat{i}\widehat{j}\widehat{k}}+...\eqno(45)
$$
and sums of $m_p+1$ elements of form%
$$
A_{2(p)}I+B^{\widehat{a}_p}\sigma _{\widehat{a}_p}+C^{\widehat{a}_p\widehat{b%
}_p}\sigma _{\widehat{a}_p\widehat{b}_p}+D^{\widehat{a}_p\widehat{b}_p%
\widehat{c}_p}\sigma _{\widehat{a}_p\widehat{b}_p\widehat{c}_p}+...
$$
with antisymmetric coefficients\\ $C^{\widehat{i}\widehat{j}}=C^{[\widehat{i}%
\widehat{j}]},C^{\widehat{a}_p\widehat{b}_p}=C^{[\widehat{a}_p\widehat{b}%
_p]},D^{\widehat{i}\widehat{j}\widehat{k}}=D^{[\widehat{i}\widehat{j}%
\widehat{k}]},D^{\widehat{a}_p\widehat{b}_p\widehat{c}_p}=D^{[\widehat{a}_p%
\widehat{b}_p\widehat{c}_p]},...$ and matrices $\sigma _{\widehat{i}\widehat{%
j}}=\sigma _{[\widehat{i}}\sigma _{\widehat{j}]},\sigma _{\widehat{a}_p%
\widehat{b}_p}=\sigma _{[\widehat{a}_p}\sigma _{\widehat{b}_p]},\sigma _{%
\widehat{i}\widehat{j}\widehat{k}}=\sigma _{[\widehat{i}}\sigma _{\widehat{j}%
}\sigma _{\widehat{k}]},...$ . Really, we have 2$^{n+1}$ coefficients $%
\left( A_1,C^{\widehat{i}\widehat{j}},D^{\widehat{i}\widehat{j}\widehat{k}%
},...\right) $ and 2$^{m_p+1}$ coefficients $( A_{2(p)},C^{\widehat{a}%
_p\widehat{b}_p},D^{\widehat{a}_p\widehat{b}_p\widehat{c}_p},...) $ of
the Clifford algebra on ${\cal F.}$

For simplicity,  we shall present the necessary geometric
constructions only for h-spin spaces ${\cal S}_{(h)}$ of dimension $N_{(n)}.$
Considerations for a v-spin space ${\cal S}_{(v)}$ are similar but with
proper characteristics for a dimension $N_{(m)}.$

In order to define the scalar (spinor) product on ${\cal S}_{(h)}$ we
introduce into consideration this finite sum (because of a finite number of
elements $\sigma _{[\widehat{i}\widehat{j}...\widehat{k}]}):$
$$
^{(\pm )}E_{\underline{k}\underline{m}}^{\underline{i}\underline{j}}=\delta
_{\underline{k}}^{\underline{i}}\delta _{\underline{m}}^{\underline{j}%
}+\frac 2{1!}(\sigma _{\widehat{i}})_{\underline{k}}^{.\underline{i}}(\sigma
^{\widehat{i}})_{\underline{m}}^{.\underline{j}}+\frac{2^2}{2!}(\sigma _{%
\widehat{i}\widehat{j}})_{\underline{k}}^{.\underline{i}}(\sigma ^{\widehat{i%
}\widehat{j}})_{\underline{m}}^{.\underline{j}}+\frac{2^3}{3!}(\sigma _{%
\widehat{i}\widehat{j}\widehat{k}})_{\underline{k}}^{.\underline{i}}(\sigma
^{\widehat{i}\widehat{j}\widehat{k}})_{\underline{m}}^{.\underline{j}}+...%
\eqno(46)
$$
which can be factorized as
$$
^{(\pm )}E_{\underline{k}\underline{m}}^{\underline{i}\underline{j}}=N_{(n)}{%
\ }^{(\pm )}\epsilon _{\underline{k}\underline{m}}{\ }^{(\pm )}\epsilon ^{%
\underline{i}\underline{j}}\mbox{ for }n=2k\eqno(47)
$$
and%
$$
^{(+)}E_{\underline{k}\underline{m}}^{\underline{i}\underline{j}%
}=2N_{(n)}\epsilon _{\underline{k}\underline{m}}\epsilon ^{\underline{i}%
\underline{j}},{\ }^{(-)}E_{\underline{k}\underline{m}}^{\underline{i}%
\underline{j}}=0\mbox{ for }n=3(mod4),\eqno(48)
$$
$$
^{(+)}E_{\underline{k}\underline{m}}^{\underline{i}\underline{j}}=0,{\ }%
^{(-)}E_{\underline{k}\underline{m}}^{\underline{i}\underline{j}%
}=2N_{(n)}\epsilon _{\underline{k}\underline{m}}\epsilon ^{\underline{i}%
\underline{j}}\mbox{ for }n=1(mod4).
$$

Antisymmetry of $\sigma _{\widehat{i}\widehat{j}\widehat{k}...}$ and the
construction of the objects (45), (46), (47) and (48) define the
properties of $\epsilon $-objects $^{(\pm )}\epsilon _{\underline{k}%
\underline{m}}$ and $\epsilon _{\underline{k}\underline{m}}$ which have an
eight-fold periodicity on $n$ (see details in \cite{penr2} and, with respect
to la-spaces, \cite{vjmp}).

For even values of $n$ it is possible the decomposition of every h-spin
space ${\cal S}_{(h)}$into irreducible h-spin spaces ${\bf S}_{(h)}$ and $%
{\bf S}_{(h)}^{\prime }$ (one considers splitting of h-indices, for
instance, \underline{$l$}$=L\oplus L^{\prime },\underline{m}=M\oplus
M^{\prime },...;$ for v$_p$-indices we shall write $\underline{a}%
_p=A_p\oplus A_p^{\prime },\underline{b}_p=B_p\oplus B_p^{\prime },...)$ and
defines new $\epsilon $-objects
$$
\epsilon ^{\underline{l}\underline{m}}=\frac 12\left( ^{(+)}\epsilon ^{%
\underline{l}\underline{m}}+^{(-)}\epsilon ^{\underline{l}\underline{m}%
}\right) \mbox{ and }\widetilde{\epsilon }^{\underline{l}\underline{m}%
}=\frac 12\left( ^{(+)}\epsilon ^{\underline{l}\underline{m}}-^{(-)}\epsilon
^{\underline{l}\underline{m}}\right) \eqno(49)
$$
We shall omit similar formulas for $\epsilon $-objects with lower indices.

We can verify, by using expressions (48) and straightforward calculations,
these para\-met\-ri\-za\-ti\-ons on symmetry properties of $\epsilon $%
-objects (49)
$$
\epsilon ^{\underline{l}\underline{m}}=\left(
\begin{array}{cc}
\epsilon ^{LM}=\epsilon ^{ML} & 0 \\
0 & 0
\end{array}
\right) \mbox{ and }\widetilde{\epsilon }^{\underline{l}\underline{m}%
}=\left(
\begin{array}{cc}
0 & 0 \\
0 & \widetilde{\epsilon }^{LM}=\widetilde{\epsilon }^{ML}
\end{array}
\right) \mbox{ for }n=0(mod8);\eqno(50)
$$
$$
\epsilon ^{\underline{l}\underline{m}}=-\frac 12{}^{(-)}\epsilon ^{%
\underline{l}\underline{m}}=\epsilon ^{\underline{m}\underline{l}},%
\mbox{ where }^{(+)}\epsilon ^{\underline{l}\underline{m}}=0,\mbox{ and }
$$
$$
\widetilde{\epsilon }^{\underline{l}\underline{m}}=-\frac 12{}^{(-)}\epsilon
^{\underline{l}\underline{m}}=\widetilde{\epsilon }^{\underline{m}\underline{%
l}}\mbox{
for }n=1(mod8);
$$
$$
\epsilon ^{\underline{l}\underline{m}}=\left(
\begin{array}{cc}
0 & 0 \\
\epsilon ^{L^{\prime }M} & 0
\end{array}
\right) \mbox{ and }\widetilde{\epsilon }^{\underline{l}\underline{m}%
}=\left(
\begin{array}{cc}
0 & \widetilde{\epsilon }^{LM^{\prime }}=-\epsilon ^{M^{\prime }L} \\ 0 & 0
\end{array}
\right) \mbox{ for }n=2(mod8);
$$
$$
\epsilon ^{\underline{l}\underline{m}}=-\frac 12{}^{(+)}\epsilon ^{%
\underline{l}\underline{m}}=-\epsilon ^{\underline{m}\underline{l}},%
\mbox{ where }^{(-)}\epsilon ^{\underline{l}\underline{m}}=0,\mbox{ and }
$$
$$
\widetilde{\epsilon }^{\underline{l}\underline{m}}=\frac 12{}^{(+)}\epsilon
^{\underline{l}\underline{m}}=-\widetilde{\epsilon }^{\underline{m}%
\underline{l}}\mbox{
for }n=3(mod8);
$$
$$
\epsilon ^{\underline{l}\underline{m}}=\left(
\begin{array}{cc}
\epsilon ^{LM}=-\epsilon ^{ML} & 0 \\
0 & 0
\end{array}
\right) \mbox{ and }\widetilde{\epsilon }^{\underline{l}\underline{m}%
}=\left(
\begin{array}{cc}
0 & 0 \\
0 & \widetilde{\epsilon }^{LM}=-\widetilde{\epsilon }^{ML}
\end{array}
\right)$$  for $n=4(mod8);$
$$
\epsilon ^{\underline{l}\underline{m}}=-\frac 12{}^{(-)}\epsilon ^{%
\underline{l}\underline{m}}=-\epsilon ^{\underline{m}\underline{l}},%
\mbox{ where }^{(+)}\epsilon ^{\underline{l}\underline{m}}=0,\mbox{ and }
$$
$$
\widetilde{\epsilon }^{\underline{l}\underline{m}}=-\frac 12{}^{(-)}\epsilon
^{\underline{l}\underline{m}}=-\widetilde{\epsilon }^{\underline{m}%
\underline{l}}\mbox{ for }n=5(mod8);
$$
$$
\epsilon ^{\underline{l}\underline{m}}=\left(
\begin{array}{cc}
0 & 0 \\
\epsilon ^{L^{\prime }M} & 0
\end{array}
\right) \mbox{ and }\widetilde{\epsilon }^{\underline{l}\underline{m}%
}=\left(
\begin{array}{cc}
0 & \widetilde{\epsilon }^{LM^{\prime }}=\epsilon ^{M^{\prime }L} \\ 0 & 0
\end{array}
\right) \mbox{ for }n=6(mod8);
$$
$$
\epsilon ^{\underline{l}\underline{m}}=\frac 12{}^{(-)}\epsilon ^{\underline{%
l}\underline{m}}=\epsilon ^{\underline{m}\underline{l}},\mbox{ where }%
{}^{(+)}\epsilon ^{\underline{l}\underline{m}}=0,\mbox{ and }
$$
$$
\widetilde{\epsilon }^{\underline{l}\underline{m}}=-\frac 12{}^{(-)}\epsilon
^{\underline{l}\underline{m}}=\widetilde{\epsilon }^{\underline{m}\underline{%
l}}\mbox{
for }n=7(mod8).
$$

Let denote reduced and irreducible h-spinor spaces in a form pointing to the
symmetry of spinor inner products in dependence of values $n=8k+l$ ($%
k=0,1,2,...;l=1,2,...7)$ of the dimension of the horizontal subbundle (we
shall write respectively $\bigtriangleup $ and $\circ $ for antisymmetric
and symmetric inner products of reduced spinors and $\diamondsuit
=(\bigtriangleup ,\circ )$ and $\widetilde{\diamondsuit }=(\circ
,\bigtriangleup )$ for corresponding parametrizations of inner products, in
brief {\it i.p.}, of irreducible spinors; properties of scalar products of
spinors are defined by $\epsilon $-objects (50); we shall use $\Diamond $
for a general {\it i.p.} when the symmetry is not pointed out):%
$$
{\cal S}_{(h)}{\ }\left( 8k\right) ={\bf S}_{\circ }\oplus {\bf S}_{\circ
}^{\prime};\quad \eqno(51)
$$
$$
{\cal S}_{(h)}{\ }\left( 8k+1\right) ={\cal S}_{\circ }^{(-)}\
\mbox{({\it i.p.} is defined by an }^{(-)}\epsilon \mbox{-object);}
$$
$$
{\cal S}_{(h)}{\ }\left( 8k+2\right) =\{
\begin{array}{c}
{\cal S}_{\Diamond }=({\bf S}_{\Diamond },{\bf S}_{\Diamond }),\mbox{ or} \\
{\cal S}_{\Diamond }^{\prime }=({\bf S}_{\widetilde{\Diamond }}^{\prime },%
{\bf S}_{\widetilde{\Diamond }}^{\prime });
\end{array}
\qquad
$$
$$
{\cal S}_{(h)}\left( 8k+3\right) ={\cal S}_{\bigtriangleup }^{(+)}\
\mbox{({\it i.p.} is defined by an }^{(+)}\epsilon \mbox{-object);}
$$
$$
{\cal S}_{(h)}\left( 8k+4\right) ={\bf S}_{\bigtriangleup }\oplus {\bf S}%
_{\bigtriangleup }^{\prime };\quad
$$
$$
{\cal S}_{(h)}\left( 8k+5\right) ={\cal S}_{\bigtriangleup }^{(-)}\
\mbox{({\it i.p. }is defined
by an }^{(-)}\epsilon \mbox{-object),}
$$
$$
{\cal S}_{(h)}\left( 8k+6\right) =\{
\begin{array}{c}
{\cal S}_{\Diamond }=({\bf S}_{\Diamond },{\bf S}_{\Diamond }),\mbox{ or} \\
{\cal S}_{\Diamond }^{\prime }=({\bf S}_{\widetilde{\Diamond }}^{\prime },%
{\bf S}_{\widetilde{\Diamond }}^{\prime });
\end{array}
$$
\qquad
$$
{\cal S}_{(h)}\left( 8k+7\right) ={\cal S}_{\circ }^{(+)}\
\mbox{({\it i.p. } is defined by an }^{(+)}\epsilon \mbox{-object)}.
$$
We note that by using corresponding $\epsilon $-objects we can lower and
rise indices of reduced and irreducible spinors (for $n=2,6(mod4)$ we can
exclude primed indices, or inversely, see details in \cite{pen,penr1,penr2}).

The similar v-spinor spaces are denoted by the same symbols as in (51)
provided with a left lower mark ''$|"$ and parametrized with respect to the
values $m=8k^{\prime }+l$ (k'=0,1,...; l=1,2,...,7) of the dimension of the
vertical subbundle, for example, as
$$
{\cal S}_{(v_p)}(8k^{\prime })={\bf S}_{|\circ }\oplus {\bf S}_{|\circ
}^{\prime },{\cal S}_{(v_p)}\left( 8k+1\right) ={\cal S}_{|\circ }^{(-)},...%
\eqno(52)
$$
We use '' $\widetilde{}$ ''-overlined symbols,
$$
{\widetilde{{\cal S}}}_{(h)}\left( 8k\right) ={\widetilde{{\bf S}}}_{\circ
}\oplus \widetilde{S}_{\circ }^{\prime },{\widetilde{{\cal S}}}_{(h)}\left(
8k+1\right) ={\widetilde{{\cal S}}}_{\circ }^{(-)},...\eqno(53)
$$
and
$$
{\widetilde{{\cal S}}}_{(v_p)}(8k^{\prime })={\widetilde{{\bf S}}}_{|\circ
}\oplus {\widetilde{S}}_{|\circ }^{\prime },{\widetilde{{\cal S}}}%
_{(v_p)}\left( 8k^{\prime }+1\right) ={\widetilde{{\cal S}}}_{|\circ
}^{(-)},...\eqno(54)
$$
respectively for the dual to (50) and (51) spinor spaces.

The spinor spaces (50),(52), (53) and (54) are
 called the prime spinor spaces, in brief
p-spinors. They are considered as building blocks of distinguished, for
simplicity we consider $\left( n,m_1\right) $--spinor spaces constructed in
this manner:%
$$
{\cal S}(_{\circ \circ ,\circ \circ })={\bf S_{\circ }\oplus S_{\circ
}^{\prime }\oplus S_{|\circ }\oplus S_{|\circ }^{\prime },}{\cal S}(_{\circ
\circ ,\circ }\mid ^{\circ })={\bf S_{\circ }\oplus S_{\circ }^{\prime
}\oplus S_{|\circ }\oplus \widetilde{S}_{|\circ }^{\prime },}\eqno(55)
$$
$$
{\cal S}(_{\circ \circ ,}\mid ^{\circ \circ })={\bf S_{\circ }\oplus
S_{\circ }^{\prime }\oplus \widetilde{S}_{|\circ }\oplus \widetilde{S}%
_{|\circ }^{\prime },}{\cal S}(_{\circ }\mid ^{\circ \circ \circ })={\bf %
S_{\circ }\oplus \widetilde{S}_{\circ }^{\prime }\oplus \widetilde{S}%
_{|\circ }\oplus \widetilde{S}_{|\circ }^{\prime },}
$$
$$
...............................................
$$
$$
{\cal S}(_{\triangle },_{\triangle })={\cal S}_{\triangle }^{(+)}\oplus
S_{|\bigtriangleup }^{(+)},S(_{\triangle },^{\triangle })={\cal S}%
_{\triangle }^{(+)}\oplus \widetilde{S}_{|\triangle }^{(+)},
$$
$$
................................
$$
$$
{\cal S}(_{\triangle }|^{\circ },_\diamondsuit )={\bf S}_{\triangle }\oplus
\widetilde{S_{\circ }}^{\prime }\oplus {\cal S}_{|\diamondsuit },{\cal S}%
(_{\triangle }|^{\circ },^\diamondsuit )={\bf S}_{\triangle }\oplus
\widetilde{S_{\circ }}^{\prime }\oplus {\cal \widetilde{S}}_{|}^\diamondsuit
,
$$
$$
................................
$$
Considering the operation of dualization of prime components in (55) we
can generate different isomorphic variants of distinguished $\left(
n,m_1\right) $-spinor spaces. If we add anisotropic ''shalls'' with $%
m_2,...,m_z,$ we have to extend correspondingly spaces (55), for instance,%
$$
{\cal S}(_{\circ \circ ,\circ \circ (1)},...,_{\infty (p)},...,_{\infty
(z)})={\bf S_{\circ }\oplus S_{\circ }^{\prime }\oplus S_{|(1)\circ }\oplus
S_{|(1)\circ }^{\prime }\oplus ...}
$$
$$
{\bf \oplus S_{|(p)\circ }\oplus S_{|(p)\circ }^{\prime }\oplus ...\oplus
S_{|(z)\circ }\oplus S_{|(z)\circ ,}^{\prime }}
$$
and so on.

We define a d-spinor space ${\cal S}_{(n,m_1)}\ $ as a direct sum of a
horizontal and a vertical spinor spaces of type (55), for instance,
$$
{\cal S}_{(8k,8k^{\prime })}={\bf S}_{\circ }\oplus {\bf S}_{\circ }^{\prime
}\oplus {\bf S}_{|\circ }\oplus {\bf S}_{|\circ }^{\prime },{\cal S}%
_{(8k,8k^{\prime }+1)}\ ={\bf S}_{\circ }\oplus {\bf S}_{\circ }^{\prime
}\oplus {\cal S}_{|\circ }^{(-)},...,
$$
$$
{\cal S}_{(8k+4,8k^{\prime }+5)}={\bf S}_{\triangle }\oplus {\bf S}%
_{\triangle }^{\prime }\oplus {\cal S}_{|\triangle }^{(-)},...
$$
The scalar product on a ${\cal S}_{(n,m_1)}\ $ is induced by (corresponding
to fixed values of $n$ and $m_1$ ) $\epsilon $-objects (50) considered for
h- and v$_1$-components. We present also an example for ${\cal S}%
_{(n,m_1+...+m_z)}:$%
$$
{\cal S}_{(8k+4,8k_{(1)}+5,...,8k_{(p)}+4,...8k_{(z)})}=
$$
$$
{\bf S}_{\triangle }\oplus {\bf S}_{\triangle }^{\prime }\oplus {\cal S}%
_{|(1)\triangle }^{(-)}\oplus ...\oplus {\bf S}_{|(p)\triangle }\oplus {\bf S%
}_{|(p)\triangle }^{\prime }\oplus ...\oplus {\bf S}_{|(z)\circ }\oplus {\bf %
S}_{|(z)\circ }^{\prime }.
$$

Having introduced d-spinors for dimensions $\left( n,m_1+...+m_z\right) $ we
can write out the generalization for ha--spaces of twistor equations \cite
{penr1} by using the distinguished $\sigma $-objects (44):%
$$
(\sigma _{(<\widehat{\alpha }>})_{|\underline{\beta }|}^{..\underline{\gamma
}}\quad \frac{\delta \omega ^{\underline{\beta }}}{\delta u^{<\widehat{\beta
}>)}}=\frac 1{n+m_1+...+m_z}\quad G_{<\widehat{\alpha }><\widehat{\beta }%
>}(\sigma ^{\widehat{\epsilon }})_{\underline{\beta }}^{..\underline{\gamma }%
}\quad \frac{\delta \omega ^{\underline{\beta }}}{\delta u^{\widehat{%
\epsilon }}},\eqno(56)
$$
where $\left| \underline{\beta }\right| $ denotes that we do not consider
symmetrization on this index. The general solution of (56) on the d-vector
space ${\cal F}$ looks like as
$$
\omega ^{\underline{\beta }}=\Omega ^{\underline{\beta }}+u^{<\widehat{%
\alpha }>}(\sigma _{<\widehat{\alpha }>})_{\underline{\epsilon }}^{..%
\underline{\beta }}\Pi ^{\underline{\epsilon }},\eqno(57)
$$
where $\Omega ^{\underline{\beta }}$ and $\Pi ^{\underline{\epsilon }}$ are
constant d-spinors. For fixed values of dimensions $n$ and $m=m_1+...m_z$ we
mast analyze the reduced and irreducible components of h- and v$_p$-parts of
equations (56) and their solutions (57) in order to find the symmetry
properties of a d-twistor ${\bf Z^\alpha \ }$ defined as a pair of d-spinors%
$$
{\bf Z}^\alpha =(\omega ^{\underline{\alpha }},\pi _{\underline{\beta }%
}^{\prime }),
$$
where $\pi _{\underline{\beta }^{\prime }}=\pi _{\underline{\beta }^{\prime
}}^{(0)}\in {\widetilde{{\cal S}}}_{(n,m_1,...,m_z)}$ is a constant dual
d-spinor. The problem of definition of spinors and twistors on ha-spaces was
firstly considered in \cite{vod} (see also \cite{v87,vb12}) in connection
with the possibility to extend the equations (56) and theirs solutions
(57), by using nearly autoparallel maps, on curved, locally isotropic or
anisotropic, spaces. We note that the definition of twistors have been
extended to higher order anisotropic spaces with trivial N-- and
d--connections.

\subsection{ Mutual transforms of d-tensors and d-spinors}

The spinor algebra for spaces of higher dimensions can not be considered as
a real alternative to the tensor algebra as for locally isotropic spaces of
dimensions $n=3,4$ \cite{pen,penr1,penr2}. The same holds true for ha-spaces
and we emphasize that it is not quite convenient to perform a spinor
calculus for dimensions $n,m>>4$. Nevertheless, the concept of spinors is
important for every type of spaces, we can deeply understand the fundamental
properties of geometrical objects on ha-spaces, and we shall consider in this
subsection some questions concerning transforms of d-tensor objects into
d-spinor ones.

\subsection{ Transformation of d-tensors into d-spinors}

In order to pass from d-tensors to d-spinors we must use $\sigma $-objects
(44) written in reduced or irreduced form \quad (in dependence of fixed
values of dimensions $n$ and $m$ ):

$$
(\sigma _{<\widehat{\alpha }>})_{\underline{\beta }}^{\cdot \underline{%
\gamma }},~(\sigma ^{<\widehat{\alpha }>})^{\underline{\beta }\underline{%
\gamma }},~(\sigma ^{<\widehat{\alpha }>})_{\underline{\beta }\underline{%
\gamma }},...,$$ $$(\sigma _{<\widehat{a}>})^{\underline{b}\underline{c}%
},...,(\sigma _{\widehat{i}})_{\underline{j}\underline{k}},...,(\sigma _{<%
\widehat{a}>})^{AA^{\prime }},...,(\sigma ^{\widehat{i}})_{II^{\prime }},....%
\eqno(58)
$$
It is obvious that contracting with corresponding $\sigma $-objects (58)
we can introduce instead of d-tensors indices the d-spinor ones, for
instance,%
$$
\omega ^{\underline{\beta }\underline{\gamma }}=(\sigma ^{<\widehat{\alpha }%
>})^{\underline{\beta }\underline{\gamma }}\omega _{<\widehat{\alpha }%
>},\quad \omega _{AB^{\prime }}=(\sigma ^{<\widehat{a}>})_{AB^{\prime
}}\omega _{<\widehat{a}>},\quad ...,\zeta _{\cdot \underline{j}}^{\underline{%
i}}=(\sigma ^{\widehat{k}})_{\cdot \underline{j}}^{\underline{i}}\zeta _{%
\widehat{k}},....
$$
For d-tensors containing groups of antisymmetric indices there is a more
simple procedure of theirs transforming into d-spinors because the objects
$$
(\sigma _{\widehat{\alpha }\widehat{\beta }...\widehat{\gamma }})^{%
\underline{\delta }\underline{\nu }},\quad (\sigma ^{\widehat{a}\widehat{b}%
...\widehat{c}})^{\underline{d}\underline{e}},\quad ...,(\sigma ^{\widehat{i}%
\widehat{j}...\widehat{k}})_{II^{\prime }},\quad ...\eqno(59)
$$
can be used for sets of such indices into pairs of d-spinor indices. Let us
enumerate some properties of $\sigma $-objects of type (59) (for
simplicity we consider only h-components having q indices $\widehat{i},%
\widehat{j},\widehat{k},...$ taking values from 1 to $n;$ the properties of v%
$_p$-components can be written in a similar manner with respect to indices $%
\widehat{a}_p,\widehat{b}_p,\widehat{c}_p...$ taking values from 1 to $m$):%
$$
(\sigma _{\widehat{i}...\widehat{j}})^{\underline{k}\underline{l}}%
\mbox{
 is\ }\left\{ \
\begin{array}{c}
\mbox{symmetric on }\underline{k},\underline{l}\mbox{ for }n-2q\equiv
1,7~(mod~8); \\ \mbox{antisymmetric on }\underline{k},\underline{l}%
\mbox{
for }n-2q\equiv 3,5~(mod~8)
\end{array}
\right\} \eqno(60)
$$
for odd values of $n,$ and an object
$$
(\sigma _{\widehat{i}...\widehat{j}})^{IJ}~\left( (\sigma _{\widehat{i}...%
\widehat{j}})^{I^{\prime }J^{\prime }}\right)
$$
$$
\mbox{ is\ }\left\{
\begin{array}{c}
\mbox{symmetric on }I,J~(I^{\prime },J^{\prime })\mbox{ for }n-2q\equiv
0~(mod~8); \\ \mbox{antisymmetric on }I,J~(I^{\prime },J^{\prime })%
\mbox{
for }n-2q\equiv 4~(mod~8)
\end{array}
\right\} \eqno(61)
$$
or%
$$
(\sigma _{\widehat{i}...\widehat{j}})^{IJ^{\prime }}=\pm (\sigma _{\widehat{i%
}...\widehat{j}})^{J^{\prime }I}\{
\begin{array}{c}
n+2q\equiv 6(mod8); \\
n+2q\equiv 2(mod8),
\end{array}
\eqno(62)
$$
with vanishing of the rest of reduced components of the d-tensor $(\sigma _{%
\widehat{i}...\widehat{j}})^{\underline{k}\underline{l}}$ with prime/unprime
sets of indices.

\subsection{ Transformation of d-spinors into d-tensors; fundamental
d-spinors}

We can transform every d-spinor $\xi ^{\underline{\alpha }}=\left( \xi ^{%
\underline{i}},\xi ^{\underline{a}_1},...,\xi ^{\underline{a}_z}\right) $
into a corresponding d-tensor. For simplicity, we consider this construction
only for a h-component $\xi ^{\underline{i}}$ on a h-space being of
dimension $n$. The values%
$$
\xi ^{\underline{\alpha }}\xi ^{\underline{\beta }}(\sigma ^{\widehat{i}...%
\widehat{j}})_{\underline{\alpha }\underline{\beta }}\quad \left( n%
\mbox{ is
odd}\right) \eqno(63)
$$
or
$$
\xi ^I\xi ^J(\sigma ^{\widehat{i}...\widehat{j}})_{IJ}~\left( \mbox{or }\xi
^{I^{\prime }}\xi ^{J^{\prime }}(\sigma ^{\widehat{i}...\widehat{j}%
})_{I^{\prime }J^{\prime }}\right) ~\left( n\mbox{ is even}\right)
\eqno(64)
$$
with a different number of indices $\widehat{i}...\widehat{j},$ taken
together, defines the h-spinor $\xi ^{\underline{i}}\,$ to an accuracy to
the sign. We emphasize that it is necessary to choose only those
h-components of d-tensors (63) (or (64)) which are symmetric on pairs of
indices $\underline{\alpha }\underline{\beta }$ (or $IJ\,$ (or $I^{\prime
}J^{\prime }$ )) and the number $q$ of indices $\widehat{i}...\widehat{j}$
satisfies the condition (as a respective consequence of the properties
(60) and/or (61), (62))%
$$
n-2q\equiv 0,1,7~(mod~8).\eqno(65)
$$
Of special interest is the case when
$$
q=\frac 12\left( n\pm 1\right) ~\left( n\mbox{ is odd}\right) \eqno(66)
$$
or
$$
q=\frac 12n~\left( n\mbox{ is even}\right) .\eqno(67)
$$
If all expressions (63) and/or (64) are zero for all values of $q\,$
with the exception of one or two ones defined by the conditions (65),
(66) (or
(67)), the value $\xi ^{\widehat{i}}$ (or $\xi ^I$ ($\xi ^{I^{\prime }}))$
is called a fundamental h-spinor. Defining in a similar manner the
fundamental v-spinors we can introduce fundamental d-spinors as pairs of
fundamental h- and v-spinors. Here we remark that a h(v$_p$)-spinor $\xi ^{%
\widehat{i}}~(\xi ^{\widehat{a}_p})\,$ (we can also consider reduced
components) is always a fundamental one for $n(m)<7,$ which is a consequence
of (67)).

Finally, in this subsection, we note that the geometry of fundamental h- and
v-spinors is similar to that of usual fundamental spinors (see Appendix to
the monograph \cite{penr2}). We omit such details in this work, but
emphasize that constructions with fundamental d-spinors, for a la-space,
must be adapted to the corresponding global splitting by N-connection of the
space.

\section{ The Differential Geometry of D--Spinors}

This subsection is devoted to the differential geometry of d--spinors in higher
order anisotropic spaces.
We shall use denotations of type
$$
v^{<\alpha >}=(v^i,v^{<a>})\in \sigma ^{<\alpha >}=(\sigma ^i,\sigma ^{<a>})
$$
and%
$$
\zeta ^{\underline{\alpha }_p}=(\zeta ^{\underline{i}_p},\zeta ^{\underline{a%
}_p})\in \sigma ^{\alpha _p}=(\sigma ^{i_p},\sigma ^{a_p})\,
$$
for, respectively, elements of modules of d-vector and irreduced d-spinor
fields (see details in \cite{vjmp}). D-tensors and d-spinor tensors
(irreduced or reduced) will be interpreted as elements of corresponding $%
{\cal \sigma }$--modules, for instance,
$$
q_{~<\beta >...}^{<\alpha >}\in {\cal \sigma ^{<\alpha >}~_{<\beta >....}}%
,\psi _{~\underline{\beta }_p\quad ...}^{\underline{\alpha }_p\quad
\underline{\gamma }_p}\in {\cal \sigma }_{~\underline{\beta _p}\quad ...}^{%
\underline{\alpha }_p\quad \underline{\gamma }_p}~,\xi _{\quad
J_pK_p^{\prime }N_p^{\prime }}^{I_pI_p^{\prime }}\in {\cal \sigma }_{\quad
J_pK_p^{\prime }N_p^{\prime }}^{I_pI_p^{\prime }}~,...
$$

We can establish a correspondence between the la-adapted metric $g_{\alpha
\beta }$ (12) and d-spinor metric $\epsilon _{\underline{\alpha }%
\underline{\beta }}$ ( $\epsilon $-objects (50) for both h- and v$_p$%
-subspaces of ${\cal E}^{<z>}{\cal \,}$ ) of a ha-space ${\cal E}^{<z>}$ by
using the relation%
$$
g_{<\alpha ><\beta >}=-\frac 1{N(n)+N(m_1)+...+N(m_z)}\times \eqno(68)
$$
$$
((\sigma _{(<\alpha >}(u))^{\underline{\alpha }\underline{\beta }}(\sigma
_{<\beta >)}(u))^{\underline{\delta }\underline{\gamma }})\epsilon _{%
\underline{\alpha }\underline{\gamma }}\epsilon _{\underline{\beta }%
\underline{\delta }},
$$
where%
$$
(\sigma _{<\alpha >}(u))^{\underline{\nu }\underline{\gamma }}=l_{<\alpha
>}^{<\widehat{\alpha }>}(u)(\sigma _{<\widehat{\alpha }>})^{<\underline{\nu }%
><\underline{\gamma }>},\eqno(69)
$$
which is a consequence of formulas (43)-(50). In brief we can write
(68) as
$$
g_{<\alpha ><\beta >}=\epsilon _{\underline{\alpha }_1\underline{\alpha }%
_2}\epsilon _{\underline{\beta }_1\underline{\beta }_2}\eqno(70)
$$
if the $\sigma $-objects are considered as a fixed structure, whereas $%
\epsilon $-objects are treated as caring the metric ''dynamics '' , on
la-space. This variant is used, for instance, in the so-called 2-spinor
geometry \cite{penr1,penr2} and should be preferred if we have to make
explicit the algebraic symmetry properties of d-spinor objects by using
 metric decomposition (70). An
alternative way is to consider as fixed the algebraic structure of $\epsilon
$-objects and to use variable components of $\sigma $-objects of type (69)
for developing a variational d-spinor approach to gravitational and matter
field interactions on ha-spaces ( the spinor Ashtekar variables \cite{ash}
are introduced in this manner).

We note that a d--spinor metric
$$
\epsilon _{\underline{\nu }\underline{\tau }}=\left(
\begin{array}{cccc}
\epsilon _{\underline{i}\underline{j}} & 0 & ... & 0 \\
0 & \epsilon _{\underline{a}_1\underline{b}_1} & ... & 0 \\
... & ... & ... & ... \\
0 & 0 & ... & \epsilon _{\underline{a}_z\underline{b}_z}
\end{array}
\right)
$$
on the d-spinor space ${\cal S}=({\cal S}_{(h)},{\cal S}_{(v_1)},...,{\cal S}%
_{(v_z)})$ can have symmetric or antisymmetric h (v$_p$) -components $%
\epsilon _{\underline{i}\underline{j}}$ ($\epsilon _{\underline{a}_p%
\underline{b}_p})$ , see $\epsilon $-objects (50). For simplicity,
in order to avoid cumbersome calculations connected with eight-fold
periodicity on dimensions $n$ and $m_p$ of a ha-space ${\cal E}^{<z>},$
 we shall develop a general d-spinor formalism only by using irreduced
spinor spaces ${\cal S}_{(h)}$ and ${\cal S}_{(v_p)}.$

\subsection{ D-covariant derivation on ha--spaces}

Let ${\cal E}^{<z>}$ be a ha-space. We define the action on a d-spinor of a
d-covariant operator%
$$
\nabla _{<\alpha >}=\left( \nabla _i,\nabla _{<a>}\right) =(\sigma _{<\alpha
>})^{\underline{\alpha }_1\underline{\alpha }_2}\nabla _{^{\underline{\alpha
}_1\underline{\alpha }_2}}=\left( (\sigma _i)^{\underline{i}_1\underline{i}%
_2}\nabla _{^{\underline{i}_1\underline{i}_2}},~(\sigma _{<a>})^{\underline{a%
}_1\underline{a}_2}\nabla _{^{\underline{a}_1\underline{a}_2}}\right)
$$
$$
 =
\left( (\sigma _i)^{\underline{i}_1\underline{i}_2}\nabla _{^{\underline{i}_1%
\underline{i}_2}},~(\sigma _{a_1})^{\underline{a}_1\underline{a}_2}\nabla
_{(1)^{\underline{a}_1\underline{a}_2}},...,(\sigma _{a_p})^{\underline{a}_1%
\underline{a}_2}\nabla _{(p)^{\underline{a}_1\underline{a}_2}},...,(\sigma
_{a_z})^{\underline{a}_1\underline{a}_2}\nabla _{(z)^{\underline{a}_1%
\underline{a}_2}}\right)
$$
(in brief, we shall write
$$
\nabla _{<\alpha >}=\nabla _{^{\underline{\alpha }_1\underline{\alpha }%
_2}}=\left( \nabla _{^{\underline{i}_1\underline{i}_2}},~\nabla _{(1)^{%
\underline{a}_1\underline{a}_2}},...,\nabla _{(p)^{\underline{a}_1\underline{%
a}_2}},...,\nabla _{(z)^{\underline{a}_1\underline{a}_2}}\right) )
$$
as maps
$$
\nabla _{{\underline{\alpha }}_1{\underline{\alpha }}_2}\ :\ {\cal \sigma }^{%
\underline{\beta }}\rightarrow \sigma _{<\alpha >}^{\underline{\beta }%
}=\sigma _{{\underline{\alpha }}_1{\underline{\alpha }}_2}^{\underline{\beta
}}=
$$
$$
\left( \sigma _i^{\underline{\beta }}=\sigma _{{\underline{i}}_1{\underline{i%
}}_2}^{\underline{\beta }},\sigma _{(1)a_1}^{\underline{\beta }}=\sigma _{(1)%
{\underline{\alpha }}_1{\underline{\alpha }}_2}^{\underline{\beta }%
},...,\sigma _{(p)a_p}^{\underline{\beta }}=\sigma _{(p){\underline{\alpha }}%
_1{\underline{\alpha }}_2}^{\underline{\beta }},...,\sigma _{(z)a_z}^{%
\underline{\beta }}=\sigma _{(z){\underline{\alpha }}_1{\underline{\alpha }}%
_2}^{\underline{\beta }}\right)
$$
satisfying conditions%
$$
\nabla _{<\alpha >}(\xi ^{\underline{\beta }}+\eta ^{\underline{\beta }%
})=\nabla _{<\alpha >}\xi ^{\underline{\beta }}+\nabla _{<\alpha >}\eta ^{%
\underline{\beta }},
$$
and%
$$
\nabla _{<\alpha >}(f\xi ^{\underline{\beta }})=f\nabla _{<\alpha >}\xi ^{%
\underline{\beta }}+\xi ^{\underline{\beta }}\nabla _{<\alpha >}f
$$
for every $\xi ^{\underline{\beta }},\eta ^{\underline{\beta }}\in {\cal %
\sigma ^{\underline{\beta }}}$ and $f$ being a scalar field on ${\cal E}%
^{<z>}{\cal .\ }$ It is also required that one holds the Leibnitz rule%
$$
(\nabla _{<\alpha >}\zeta _{\underline{\beta }})\eta ^{\underline{\beta }%
}=\nabla _{<\alpha >}(\zeta _{\underline{\beta }}\eta ^{\underline{\beta }%
})-\zeta _{\underline{\beta }}\nabla _{<\alpha >}\eta ^{\underline{\beta }}
$$
and that $\nabla _{<\alpha >}\,$ is a real operator, i.e. it commuters with
the operation of complex conjugation:%
$$
\overline{\nabla _{<\alpha >}\psi _{\underline{\alpha }\underline{\beta }%
\underline{\gamma }...}}=\nabla _{<\alpha >}(\overline{\psi }_{\underline{%
\alpha }\underline{\beta }\underline{\gamma }...}).
$$

Let now analyze the question on uniqueness of action on d-spinors of an
operator $\nabla _{<\alpha >}$ satisfying necessary conditions . Denoting by
$\nabla _{<\alpha >}^{(1)}$ and $\nabla _{<\alpha >}$ two such d-covariant
operators we consider the map%
$$
(\nabla _{<\alpha >}^{(1)}-\nabla _{<\alpha >}):{\cal \sigma ^{\underline{%
\beta }}\rightarrow \sigma _{\underline{\alpha }_1\underline{\alpha }_2}^{%
\underline{\beta }}}.\eqno(71)
$$
Because the action on a scalar $f$ of both operators $\nabla _\alpha ^{(1)}$
and $\nabla _\alpha $ must be identical, i.e.%
$$
\nabla _{<\alpha >}^{(1)}f=\nabla _{<\alpha >}f,
$$
the action (71) on $f=\omega _{\underline{\beta }}\xi ^{\underline{\beta }%
} $ must be written as
$$
(\nabla _{<\alpha >}^{(1)}-\nabla _{<\alpha >})(\omega _{\underline{\beta }%
}\xi ^{\underline{\beta }})=0.
$$
In consequence we conclude that there is an element $\Theta _{\underline{%
\alpha }_1\underline{\alpha }_2\underline{\beta }}^{\quad \quad \underline{%
\gamma }}\in {\cal \sigma }_{\underline{\alpha }_1\underline{\alpha }_2%
\underline{\beta }}^{\quad \quad \underline{\gamma }}$ for which%
$$
\nabla _{\underline{\alpha }_1\underline{\alpha }_2}^{(1)}\xi ^{\underline{%
\gamma }}=\nabla _{\underline{\alpha }_1\underline{\alpha }_2}\xi ^{%
\underline{\gamma }}+\Theta _{\underline{\alpha }_1\underline{\alpha }_2%
\underline{\beta }}^{\quad \quad \underline{\gamma }}\xi ^{\underline{\beta }%
}\eqno(72)
$$
and%
$$
\nabla _{\underline{\alpha }_1\underline{\alpha }_2}^{(1)}\omega _{%
\underline{\beta }}=\nabla _{\underline{\alpha }_1\underline{\alpha }%
_2}\omega _{\underline{\beta }}-\Theta _{\underline{\alpha }_1\underline{%
\alpha }_2\underline{\beta }}^{\quad \quad \underline{\gamma }}\omega _{%
\underline{\gamma }}~.
$$
The action of the operator (71) on a d-vector $v^{<\beta >}=v^{\underline{%
\beta }_1\underline{\beta }_2}$ can be written by using formula (72) for
both indices $\underline{\beta }_1$ and $\underline{\beta }_2$ :%
$$
(\nabla _{<\alpha >}^{(1)}-\nabla _{<\alpha >})v^{\underline{\beta }_1%
\underline{\beta }_2}=\Theta _{<\alpha >\underline{\gamma }}^{\quad
\underline{\beta }_1}v^{\underline{\gamma }\underline{\beta }_2}+\Theta
_{<\alpha >\underline{\gamma }}^{\quad \underline{\beta }_2}v^{\underline{%
\beta }_1\underline{\gamma }}=
$$
$$
(\Theta _{<\alpha >\underline{\gamma }_1}^{\quad \underline{\beta }_1}\delta
_{\underline{\gamma }_2}^{\quad \underline{\beta }_2}+\Theta _{<\alpha >%
\underline{\gamma }_1}^{\quad \underline{\beta }_2}\delta _{\underline{%
\gamma }_2}^{\quad \underline{\beta }_1})v^{\underline{\gamma }_1\underline{%
\gamma }_2}=Q_{\ <\alpha ><\gamma >}^{<\beta >}v^{<\gamma >},
$$
where%
$$
Q_{\ <\alpha ><\gamma >}^{<\beta >}=Q_{\qquad \underline{\alpha }_1%
\underline{\alpha }_2~\underline{\gamma }_1\underline{\gamma }_2}^{%
\underline{\beta }_1\underline{\beta }_2}=\Theta _{<\alpha >\underline{%
\gamma }_1}^{\quad \underline{\beta }_1}\delta _{\underline{\gamma }%
_2}^{\quad \underline{\beta }_2}+\Theta _{<\alpha >\underline{\gamma }%
_1}^{\quad \underline{\beta }_2}\delta _{\underline{\gamma }_2}^{\quad
\underline{\beta }_1}.\eqno(73)
$$
The d-commutator $\nabla _{[<\alpha >}\nabla _{<\beta >]}$ defines the
d-torsion (see (23)-(25) and (30)). So, applying operators $\nabla
_{[<\alpha >}^{(1)}\nabla _{<\beta >]}^{(1)}$ and $\nabla _{[<\alpha
>}\nabla _{<\beta >]}$ on $f=\omega _{\underline{\beta }}\xi ^{\underline{%
\beta }}$ we can write
$$
T_{\quad <\alpha ><\beta >}^{(1)<\gamma >}-T_{~<\alpha ><\beta >}^{<\gamma
>}=Q_{~<\beta ><\alpha >}^{<\gamma >}-Q_{~<\alpha ><\beta >}^{<\gamma >}
$$
with $Q_{~<\alpha ><\beta >}^{<\gamma >}$ from (73).

The action of operator $\nabla _{<\alpha >}^{(1)}$ on d-spinor tensors of
type $\chi _{\underline{\alpha }_1\underline{\alpha }_2\underline{\alpha }%
_3...}^{\qquad \quad \underline{\beta }_1\underline{\beta }_2...}$ must be
constructed by using formula (72) for every upper index $\underline{\beta }%
_1\underline{\beta }_2...$ and formula (73) for every lower index $%
\underline{\alpha }_1\underline{\alpha }_2\underline{\alpha }_3...$ .

\subsection{Infeld--van der Waerden co\-ef\-fi\-ci\-ents and d-con\-nec\-ti\-ons}

Let
$$
\delta _{\underline{{\bf \alpha }}}^{\quad \underline{\alpha }}=\left(
\delta _{\underline{{\bf 1}}}^{\quad \underline{i}},\delta _{\underline{{\bf %
2}}}^{\quad \underline{i}},...,\delta _{\underline{{\bf N(n)}}}^{\quad
\underline{i}},\delta _{\underline{{\bf 1}}}^{\quad \underline{a}},\delta _{%
\underline{{\bf 2}}}^{\quad \underline{a}},...,\delta _{\underline{{\bf N(m)}%
}}^{\quad \underline{i}}\right)
$$
be a d--spinor basis. The dual to it basis is denoted as
$$
\delta _{\underline{\alpha }}^{\quad \underline{{\bf \alpha }}}=\left(
\delta _{\underline{i}}^{\quad \underline{{\bf 1}}},\delta _{\underline{i}%
}^{\quad \underline{{\bf 2}}},...,\delta _{\underline{i}}^{\quad \underline{%
{\bf N(n)}}},\delta _{\underline{i}}^{\quad \underline{{\bf 1}}},\delta _{%
\underline{i}}^{\quad \underline{{\bf 2}}},...,\delta _{\underline{i}%
}^{\quad \underline{{\bf N(m)}}}\right) .
$$
A d-spinor $\kappa ^{\underline{\alpha }}\in {\cal \sigma }$ $^{\underline{%
\alpha }}$ has components $\kappa ^{\underline{{\bf \alpha }}}=\kappa ^{%
\underline{\alpha }}\delta _{\underline{\alpha }}^{\quad \underline{{\bf %
\alpha }}}.$ Taking into account that
$$
\delta _{\underline{{\bf \alpha }}}^{\quad \underline{\alpha }}\delta _{%
\underline{{\bf \beta }}}^{\quad \underline{\beta }}\nabla _{\underline{%
\alpha }\underline{\beta }}=\nabla _{\underline{{\bf \alpha }}\underline{%
{\bf \beta }}},
$$
we write out the components $\nabla _{\underline{\alpha }\underline{\beta }}$
$\kappa ^{\underline{\gamma }}$ as%
$$
\delta _{\underline{{\bf \alpha }}}^{\quad \underline{\alpha }}~\delta _{%
\underline{{\bf \beta }}}^{\quad \underline{\beta }}~\delta _{\underline{%
\gamma }}^{\quad \underline{{\bf \gamma }}}~\nabla _{\underline{\alpha }%
\underline{\beta }}\kappa ^{\underline{\gamma }}=            \eqno(74)$$
$$\delta _{\underline{{\bf %
\epsilon }}}^{\quad \underline{\tau }}~\delta _{\underline{\tau }}^{\quad
\underline{{\bf \gamma }}}~\nabla _{\underline{{\bf \alpha }}\underline{{\bf %
\beta }}}\kappa ^{\underline{{\bf \epsilon }}}+\kappa ^{\underline{{\bf %
\epsilon }}}~\delta _{\underline{\epsilon }}^{\quad \underline{{\bf \gamma }}%
}~\nabla _{\underline{{\bf \alpha }}\underline{{\bf \beta }}}\delta _{%
\underline{{\bf \epsilon }}}^{\quad \underline{\epsilon }}=
\nabla _{\underline{{\bf \alpha }}\underline{{\bf \beta }}}\kappa ^{%
\underline{{\bf \gamma }}}+\kappa ^{\underline{{\bf \epsilon }}}\gamma _{~%
\underline{{\bf \alpha }}\underline{{\bf \beta }}\underline{{\bf \epsilon }}%
}^{\underline{{\bf \gamma }}},
$$
where the coordinate components of the d--spinor connection $\gamma _{~%
\underline{{\bf \alpha }}\underline{{\bf \beta }}\underline{{\bf \epsilon }}%
}^{\underline{{\bf \gamma }}}$ are defined as
$$
\gamma _{~\underline{{\bf \alpha }}\underline{{\bf \beta }}\underline{{\bf %
\epsilon }}}^{\underline{{\bf \gamma }}}\doteq \delta _{\underline{\tau }%
}^{\quad \underline{{\bf \gamma }}}~\nabla _{\underline{{\bf \alpha }}%
\underline{{\bf \beta }}}\delta _{\underline{{\bf \epsilon }}}^{\quad
\underline{\tau }}.\eqno(75)
$$
We call the Infeld - van der Waerden d-symbols a set of $\sigma $-objects ($%
\sigma _{{\bf \alpha }})^{\underline{{\bf \alpha }}\underline{{\bf \beta }}}$
paramet\-ri\-zed with respect to a coordinate d-spinor basis. Defining
$$
\nabla _{<{\bf \alpha >}}=(\sigma _{<{\bf \alpha >}})^{\underline{{\bf %
\alpha }}\underline{{\bf \beta }}}~\nabla _{\underline{{\bf \alpha }}%
\underline{{\bf \beta }}},
$$
introducing denotations
$$
\gamma ^{\underline{{\bf \gamma }}}{}_{<{\bf \alpha >\underline{\tau }}%
}\doteq \gamma ^{\underline{{\bf \gamma }}}{}_{{\bf \underline{\alpha }%
\underline{\beta }\underline{\tau }}}~(\sigma _{<{\bf \alpha >}})^{%
\underline{{\bf \alpha }}\underline{{\bf \beta }}}
$$
and using properties (74) we can write relations%
$$
l_{<{\bf \alpha >}}^{<\alpha >}~\delta _{\underline{\beta }}^{\quad
\underline{{\bf \beta }}}~\nabla _{<\alpha >}\kappa ^{\underline{\beta }%
}=\nabla _{<{\bf \alpha >}}\kappa ^{\underline{{\bf \beta }}}+\kappa ^{%
\underline{{\bf \delta }}}~\gamma _{~<{\bf \alpha >}\underline{{\bf \delta }}%
}^{\underline{{\bf \beta }}}\eqno(76)
$$
and%
$$
l_{<{\bf \alpha >}}^{<\alpha >}~\delta _{\underline{{\bf \beta }}}^{\quad
\underline{\beta }}~\nabla _{<\alpha >}~\mu _{\underline{\beta }}=\nabla _{<%
{\bf \alpha >}}~\mu _{\underline{{\bf \beta }}}-\mu _{\underline{{\bf \delta
}}}\gamma _{~<{\bf \alpha >}\underline{{\bf \beta }}}^{\underline{{\bf %
\delta }}}\eqno(77)
$$
for d-covariant derivations $~\nabla _{\underline{\alpha }}\kappa ^{%
\underline{\beta }}$ and $\nabla _{\underline{\alpha }}~\mu _{\underline{%
\beta }}.$

We can consider expressions similar to (76) and (77) for values having
both types of d-spinor and d-tensor indices, for instance,%
$$
l_{<{\bf \alpha >}}^{<\alpha >}~l_{<\gamma >}^{<{\bf \gamma >}}~\delta _{%
\underline{{\bf \delta }}}^{\quad \underline{\delta }}~\nabla _{<\alpha
>}\theta _{\underline{\delta }}^{~<\gamma >}=\nabla _{<{\bf \alpha >}}\theta
_{\underline{{\bf \delta }}}^{~<{\bf \gamma >}}-\theta _{\underline{{\bf %
\epsilon }}}^{~<{\bf \gamma >}}\gamma _{~<{\bf \alpha >}\underline{{\bf %
\delta }}}^{\underline{{\bf \epsilon }}}+\theta _{\underline{{\bf \delta }}%
}^{~<{\bf \tau >}}~\Gamma _{\quad <{\bf \alpha ><\tau >}}^{~<{\bf \gamma >}}
$$
(we can prove this by a straightforward calculation).

Now we shall consider some possible relations between components of
d-connec\-ti\-ons $\gamma _{~<{\bf \alpha >}\underline{{\bf \delta }}}^{%
\underline{{\bf \epsilon }}}$ and $\Gamma _{\quad <{\bf \alpha ><\tau >}}^{~<%
{\bf \gamma >}}$ and derivations of $(\sigma _{<{\bf \alpha >}})^{\underline{%
{\bf \alpha }}\underline{{\bf \beta }}}$ . According to definitions (12)
we can write%
$$
\Gamma _{~<{\bf \beta ><\gamma >}}^{<{\bf \alpha >}}=l_{<\alpha >}^{<{\bf %
\alpha >}}\nabla _{<{\bf \gamma >}}l_{<{\bf \beta >}}^{<\alpha >}=
$$
$$
l_{<\alpha >}^{<{\bf \alpha >}}\nabla _{<{\bf \gamma >}}(\sigma _{<{\bf %
\beta >}})^{\underline{\epsilon }\underline{\tau }}=l_{<\alpha >}^{<{\bf %
\alpha >}}\nabla _{<{\bf \gamma >}}((\sigma _{<{\bf \beta >}})^{\underline{%
{\bf \epsilon }}\underline{{\bf \tau }}}\delta _{\underline{{\bf \epsilon }}%
}^{~\underline{\epsilon }}\delta _{\underline{{\bf \tau }}}^{~\underline{%
\tau }})=
$$
$$
l_{<\alpha >}^{<{\bf \alpha >}}\delta _{\underline{{\bf \alpha }}}^{~%
\underline{\alpha }}\delta _{\underline{{\bf \epsilon }}}^{~\underline{%
\epsilon }}\nabla _{<{\bf \gamma >}}(\sigma _{<{\bf \beta >}})^{\underline{%
{\bf \alpha }}\underline{{\bf \epsilon }}}+l_{<\alpha >}^{<{\bf \alpha >}%
}(\sigma _{<{\bf \beta >}})^{\underline{{\bf \epsilon }}\underline{{\bf \tau
}}}(\delta _{\underline{{\bf \tau }}}^{~\underline{\tau }}\nabla _{<{\bf %
\gamma >}}\delta _{\underline{{\bf \epsilon }}}^{~\underline{\epsilon }%
}+\delta _{\underline{{\bf \epsilon }}}^{~\underline{\epsilon }}\nabla _{<%
{\bf \gamma >}}\delta _{\underline{{\bf \tau }}}^{~\underline{\tau }})=
$$
$$
l_{\underline{{\bf \epsilon }}\underline{{\bf \tau }}}^{<{\bf \alpha >}%
}~\nabla _{<{\bf \gamma >}}(\sigma _{<{\bf \beta >}})^{\underline{{\bf %
\epsilon }}\underline{{\bf \tau }}}+l_{\underline{{\bf \mu }}\underline{{\bf %
\nu }}}^{<{\bf \alpha >}}\delta _{\underline{\epsilon }}^{~\underline{{\bf %
\mu }}}\delta _{\underline{\tau }}^{~\underline{{\bf \nu }}}(\sigma _{<{\bf %
\beta >}})^{\underline{\epsilon }\underline{\tau }}(\delta _{\underline{{\bf %
\tau }}}^{~\underline{\tau }}\nabla _{<{\bf \gamma >}}\delta _{\underline{%
{\bf \epsilon }}}^{~\underline{\epsilon }}+\delta _{\underline{{\bf \epsilon
}}}^{~\underline{\epsilon }}\nabla _{<{\bf \gamma >}}\delta _{\underline{%
{\bf \tau }}}^{~\underline{\tau }}),
$$
where $l_{<\alpha >}^{<{\bf \alpha >}}=(\sigma _{\underline{{\bf \epsilon }}%
\underline{{\bf \tau }}})^{<{\bf \alpha >}}$ , from which it follows%
$$
(\sigma _{<{\bf \alpha >}})^{\underline{{\bf \mu }}\underline{{\bf \nu }}%
}(\sigma _{\underline{{\bf \alpha }}\underline{{\bf \beta }}})^{<{\bf \beta >%
}}\Gamma _{~<{\bf \gamma ><\beta >}}^{<{\bf \alpha >}}=$$
 $$(\sigma _{\underline{%
{\bf \alpha }}\underline{{\bf \beta }}})^{<{\bf \beta >}}\nabla _{<{\bf %
\gamma >}}(\sigma _{<{\bf \alpha >}})^{\underline{{\bf \mu }}\underline{{\bf %
\nu }}}+\delta _{\underline{{\bf \beta }}}^{~\underline{{\bf \nu }}}\gamma
_{~<{\bf \gamma >\underline{\alpha }}}^{\underline{{\bf \mu }}}+\delta _{%
\underline{{\bf \alpha }}}^{~\underline{{\bf \mu }}}\gamma _{~<{\bf \gamma >%
\underline{\beta }}}^{\underline{{\bf \nu }}}.
$$
Connecting the last expression on \underline{${\bf \beta }$} and \underline{$%
{\bf \nu }$} and using an orthonormalized d-spinor basis when $\gamma _{~<%
{\bf \gamma >\underline{\beta }}}^{\underline{{\bf \beta }}}=0$ (a
consequence from (75)) we have
$$
\gamma _{~<{\bf \gamma >\underline{\alpha }}}^{\underline{{\bf \mu }}}=\frac
1{N(n)+N(m_1)+...+N(m_z)}\times $$
$$(\Gamma _{\quad <{\bf \gamma >~\underline{\alpha }%
\underline{\beta }}}^{\underline{{\bf \mu }}\underline{{\bf \beta }}%
}-(\sigma _{\underline{{\bf \alpha }}\underline{{\bf \beta }}})^{<{\bf \beta
>}}\nabla _{<{\bf \gamma >}}(\sigma _{<{\bf \beta >}})^{\underline{{\bf \mu }%
}\underline{{\bf \beta }}}),\eqno(78)
$$
where
$$
\Gamma _{\quad <{\bf \gamma >~\underline{\alpha }\underline{\beta }}}^{%
\underline{{\bf \mu }}\underline{{\bf \beta }}}=(\sigma _{<{\bf \alpha >}})^{%
\underline{{\bf \mu }}\underline{{\bf \beta }}}(\sigma _{\underline{{\bf %
\alpha }}\underline{{\bf \beta }}})^{{\bf \beta }}\Gamma _{~<{\bf \gamma
><\beta >}}^{<{\bf \alpha >}}.\eqno(79)
$$
We also note here that, for instance, for the canonical (see (18) and
(19)) and Berwald (see (20)) connections, Christoffel d-symbols (see
(21)) we can express d-spinor connection (79) through corresponding
locally adapted derivations of components of metric and N-connection by
introducing corresponding coefficients instead of $\Gamma _{~<{\bf \gamma
><\beta >}}^{<{\bf \alpha >}}$ in (79) and than in (78).

\subsection{ D--spinors of ha--space curvature and torsion}

The d-tensor indices of the commutator (29), $\Delta _{<\alpha ><\beta >},$
can be transformed into d-spinor ones:%
$$
\Box _{\underline{\alpha }\underline{\beta }}=(\sigma ^{<\alpha ><\beta >})_{%
\underline{\alpha }\underline{\beta }}\Delta _{\alpha \beta }=(\Box _{%
\underline{i}\underline{j}},\Box _{\underline{a}\underline{b}})=\eqno(80)
$$
$$
(\Box _{\underline{i}\underline{j}},\Box _{\underline{a}_1\underline{b}%
_1},...,\Box _{\underline{a}_p\underline{b}_p},...,\Box _{\underline{a}_z%
\underline{b}_z}),
$$
with h- and v$_p$-components,
$$
\Box _{\underline{i}\underline{j}}=(\sigma ^{<\alpha ><\beta >})_{\underline{%
i}\underline{j}}\Delta _{<\alpha ><\beta >}\mbox{ and }\Box _{\underline{a}%
\underline{b}}=(\sigma ^{<\alpha ><\beta >})_{\underline{a}\underline{b}%
}\Delta _{<\alpha ><\beta >},
$$
being symmetric or antisymmetric in dependence of corresponding values of
dimensions $n\,$ and $m_p$ (see eight-fold parametrizations (50) and
(51)). Considering the actions of operator (80) on d-spinors $\pi ^{%
\underline{\gamma }}$ and $\mu _{\underline{\gamma }}$ we introduce the
d-spinor curvature $X_{\underline{\delta }\quad \underline{\alpha }%
\underline{\beta }}^{\quad \underline{\gamma }}\,$ as to satisfy equations%
$$
\Box _{\underline{\alpha }\underline{\beta }}\ \pi ^{\underline{\gamma }}=X_{%
\underline{\delta }\quad \underline{\alpha }\underline{\beta }}^{\quad
\underline{\gamma }}\pi ^{\underline{\delta }}\eqno(81)
$$
and%
$$
\Box _{\underline{\alpha }\underline{\beta }}\ \mu _{\underline{\gamma }}=X_{%
\underline{\gamma }\quad \underline{\alpha }\underline{\beta }}^{\quad
\underline{\delta }}\mu _{\underline{\delta }}.
$$
The gravitational d-spinor $\Psi _{\underline{\alpha }\underline{\beta }%
\underline{\gamma }\underline{\delta }}$ is defined by a corresponding
symmetrization of d-spinor indices:%
$$
\Psi _{\underline{\alpha }\underline{\beta }\underline{\gamma }\underline{%
\delta }}=X_{(\underline{\alpha }|\underline{\beta }|\underline{\gamma }%
\underline{\delta })}.
$$
We note that d-spinor tensors $X_{\underline{\delta }\quad \underline{\alpha
}\underline{\beta }}^{\quad \underline{\gamma }}$ and $\Psi _{\underline{%
\alpha }\underline{\beta }\underline{\gamma }\underline{\delta }}\,$ are
transformed into similar 2-spinor objects on locally isotropic spaces \cite
{penr1,penr2} if we consider vanishing of the N-connection structure and a
limit to a locally isotropic space.

Putting $\delta _{\underline{\gamma }}^{\quad {\bf \underline{\gamma }}}$
instead of $\mu _{\underline{\gamma }}$ in (81)  we can
express respectively the curvature and gravitational d-spinors as
$$
X_{\underline{\gamma }\underline{\delta }\underline{\alpha }\underline{\beta
}}=\delta _{\underline{\delta }\underline{{\bf \tau }}}\Box _{\underline{%
\alpha }\underline{\beta }}\delta _{\underline{\gamma }}^{\quad {\bf
\underline{\tau }}}
$$
and%
$$
\Psi _{\underline{\gamma }\underline{\delta }\underline{\alpha }\underline{%
\beta }}=\delta _{\underline{\delta }\underline{{\bf \tau }}}\Box _{(%
\underline{\alpha }\underline{\beta }}\delta _{\underline{\gamma })}^{\quad
{\bf \underline{\tau }}}.
$$

The d-spinor torsion $T_{\qquad \underline{\alpha }\underline{\beta }}^{%
\underline{\gamma }_1\underline{\gamma }_2}$ is defined similarly as for
d-tensors (see (30)) by using the d-spinor commutator (80) and equations
$$
\Box _{\underline{\alpha }\underline{\beta }}f=T_{\qquad \underline{\alpha }%
\underline{\beta }}^{\underline{\gamma }_1\underline{\gamma }_2}\nabla _{%
\underline{\gamma }_1\underline{\gamma }_2}f.\eqno(82)
$$

The d-spinor components $R_{\underline{\gamma }_1\underline{\gamma }_2\qquad
\underline{\alpha }\underline{\beta }}^{\qquad \underline{\delta }_1%
\underline{\delta }_2}$ of the curvature d-tensor $R_{\gamma \quad \alpha
\beta }^{\quad \delta }$ can be computed by using relations (79), and
(80) and (82) as to satisfy the equations (the d-spinor analogous of
equations (31) )%
$$
(\Box _{\underline{\alpha }\underline{\beta }}-T_{\qquad \underline{\alpha }%
\underline{\beta }}^{\underline{\gamma }_1\underline{\gamma }_2}\nabla _{%
\underline{\gamma }_1\underline{\gamma }_2})V^{\underline{\delta }_1%
\underline{\delta }_2}=R_{\underline{\gamma }_1\underline{\gamma }_2\qquad
\underline{\alpha }\underline{\beta }}^{\qquad \underline{\delta }_1%
\underline{\delta }_2}V^{\underline{\gamma }_1\underline{\gamma }_2},%
$$
here d-vector $V^{\underline{\gamma }_1\underline{\gamma }_2}$ is considered
as a product of d-spinors, i.e. $V^{\underline{\gamma }_1\underline{\gamma }%
_2}=\nu ^{\underline{\gamma }_1}\mu ^{\underline{\gamma }_2}$. We find

$$
R_{\underline{\gamma }_1\underline{\gamma }_2\qquad \underline{\alpha }%
\underline{\beta }}^{\qquad \underline{\delta }_1\underline{\delta }%
_2}=\left( X_{\underline{\gamma }_1~\underline{\alpha }\underline{\beta }%
}^{\quad \underline{\delta }_1}+T_{\qquad \underline{\alpha }\underline{%
\beta }}^{\underline{\tau }_1\underline{\tau }_2}\quad \gamma _{\quad
\underline{\tau }_1\underline{\tau }_2\underline{\gamma }_1}^{\underline{%
\delta }_1}\right) \delta _{\underline{\gamma }_2}^{\quad \underline{\delta }%
_2}+       \eqno(83)
$$
$$
\left( X_{\underline{\gamma }_2~\underline{\alpha }\underline{\beta }%
}^{\quad \underline{\delta }_2}+T_{\qquad \underline{\alpha }\underline{%
\beta }}^{\underline{\tau }_1\underline{\tau }_2}\quad \gamma _{\quad
\underline{\tau }_1\underline{\tau }_2\underline{\gamma }_2}^{\underline{%
\delta }_2}\right) \delta _{\underline{\gamma }_1}^{\quad \underline{\delta }%
_1}.
$$

It is convenient to use this d-spinor expression for the curvature d-tensor
$$
R_{\underline{\gamma }_1\underline{\gamma }_2\qquad \underline{\alpha }_1%
\underline{\alpha }_2\underline{\beta }_1\underline{\beta }_2}^{\qquad
\underline{\delta }_1\underline{\delta }_2}=\left( X_{\underline{\gamma }_1~%
\underline{\alpha }_1\underline{\alpha }_2\underline{\beta }_1\underline{%
\beta }_2}^{\quad \underline{\delta }_1}+T_{\qquad \underline{\alpha }_1%
\underline{\alpha }_2\underline{\beta }_1\underline{\beta }_2}^{\underline{%
\tau }_1\underline{\tau }_2}~\gamma _{\quad \underline{\tau }_1\underline{%
\tau }_2\underline{\gamma }_1}^{\underline{\delta }_1}\right) \delta _{%
\underline{\gamma }_2}^{\quad \underline{\delta }_2}+
$$
$$
\left( X_{\underline{\gamma }_2~\underline{\alpha }_1\underline{\alpha }_2%
\underline{\beta }_1\underline{\beta }_2}^{\quad \underline{\delta }%
_2}+T_{\qquad \underline{\alpha }_1\underline{\alpha }_2\underline{\beta }_1%
\underline{\beta }_2~}^{\underline{\tau }_1\underline{\tau }_2}\gamma
_{\quad \underline{\tau }_1\underline{\tau }_2\underline{\gamma }_2}^{%
\underline{\delta }_2}\right) \delta _{\underline{\gamma }_1}^{\quad
\underline{\delta }_1}
$$
in order to get the d--spinor components of the Ricci d-tensor%
$$
R_{\underline{\gamma }_1\underline{\gamma }_2\underline{\alpha }_1\underline{%
\alpha }_2}=R_{\underline{\gamma }_1\underline{\gamma }_2\qquad \underline{%
\alpha }_1\underline{\alpha }_2\underline{\delta }_1\underline{\delta }%
_2}^{\qquad \underline{\delta }_1\underline{\delta }_2}= \eqno(84)
$$
$$
X_{\underline{\gamma }_1~\underline{\alpha }_1\underline{\alpha }_2%
\underline{\delta }_1\underline{\gamma }_2}^{\quad \underline{\delta }%
_1}+T_{\qquad \underline{\alpha }_1\underline{\alpha }_2\underline{\delta }_1%
\underline{\gamma }_2}^{\underline{\tau }_1\underline{\tau }_2}~\gamma
_{\quad \underline{\tau }_1\underline{\tau }_2\underline{\gamma }_1}^{%
\underline{\delta }_1}+X_{\underline{\gamma }_2~\underline{\alpha }_1%
\underline{\alpha }_2\underline{\delta }_1\underline{\gamma }_2}^{\quad
\underline{\delta }_2}+T_{\qquad \underline{\alpha }_1\underline{\alpha }_2%
\underline{\gamma }_1\underline{\delta }_2~}^{\underline{\tau }_1\underline{%
\tau }_2}\gamma _{\quad \underline{\tau }_1\underline{\tau }_2\underline{%
\gamma }_2}^{\underline{\delta }_2}
$$
and this d-spinor decomposition of the scalar curvature:%
$$
q\overleftarrow{R}=R_{\qquad \underline{\alpha }_1\underline{\alpha }_2}^{%
\underline{\alpha }_1\underline{\alpha }_2}=X_{\quad ~\underline{~\alpha }%
_1\quad \underline{\delta }_1\underline{\alpha }_2}^{\underline{\alpha }_1%
\underline{\delta }_1~~\underline{\alpha }_2}+T_{\qquad ~~\underline{\alpha }%
_2\underline{\delta }_1}^{\underline{\tau }_1\underline{\tau }_2\underline{%
\alpha }_1\quad ~\underline{\alpha }_2}~\gamma _{\quad \underline{\tau }_1%
\underline{\tau }_2\underline{\alpha }_1}^{\underline{\delta }_1}+
\eqno(85)$$
$$
X_{\qquad \quad \underline{\alpha }_2\underline{\delta }_2\underline{\alpha }%
_1}^{\underline{\alpha }_2\underline{\delta }_2\underline{\alpha }%
_1}+T_{\qquad \underline{\alpha }_1\quad ~\underline{\delta }_2~}^{%
\underline{\tau }_1\underline{\tau }_2~~\underline{\alpha }_2\underline{%
\alpha }_1}\gamma _{\quad \underline{\tau }_1\underline{\tau }_2\underline{%
\alpha }_2}^{\underline{\delta }_2}.
$$

Putting (84) and (85) into (34) and, correspondingly, (35) we find
the d--spinor components of the Einstein and $\Phi _{<\alpha ><\beta >}$
d--tensors:%
$$
\overleftarrow{G}_{<\gamma ><\alpha >}=\overleftarrow{G}_{\underline{\gamma }%
_1\underline{\gamma }_2\underline{\alpha }_1\underline{\alpha }_2}=X_{%
\underline{\gamma }_1~\underline{\alpha }_1\underline{\alpha }_2\underline{%
\delta }_1\underline{\gamma }_2}^{\quad \underline{\delta }_1}+T_{\qquad
\underline{\alpha }_1\underline{\alpha }_2\underline{\delta }_1\underline{%
\gamma }_2}^{\underline{\tau }_1\underline{\tau }_2}~\gamma _{\quad
\underline{\tau }_1\underline{\tau }_2\underline{\gamma }_1}^{\underline{%
\delta }_1}+ \eqno(86)
$$
$$
X_{\underline{\gamma }_2~\underline{\alpha }_1\underline{\alpha }_2%
\underline{\delta }_1\underline{\gamma }_2}^{\quad \underline{\delta }%
_2}+T_{\qquad \underline{\alpha }_1\underline{\alpha }_2\underline{\gamma }_1%
\underline{\delta }_2~}^{\underline{\tau }_1\underline{\tau }_2}\gamma
_{\quad \underline{\tau }_1\underline{\tau }_2\underline{\gamma }_2}^{%
\underline{\delta }_2}-
$$
$$
\frac 12\varepsilon _{\underline{\gamma }_1\underline{\alpha }_1}\varepsilon
_{\underline{\gamma }_2\underline{\alpha }_2}[X_{\quad ~\underline{~\beta }%
_1\quad \underline{\mu }_1\underline{\beta }_2}^{\underline{\beta }_1%
\underline{\mu }_1~~\underline{\beta }_2}+T_{\qquad ~~\underline{\beta }_2%
\underline{\mu }_1}^{\underline{\tau }_1\underline{\tau }_2\underline{\beta }%
_1\quad ~\underline{\beta }_2}~\gamma _{\quad \underline{\tau }_1\underline{%
\tau }_2\underline{\beta }_1}^{\underline{\mu }_1}+
$$
$$
X_{\qquad \quad \underline{\beta }_2\underline{\mu }_2\underline{\nu }_1}^{%
\underline{\beta }_2\underline{\mu }_2\underline{\nu }_1}+T_{\qquad
\underline{\beta }_1\quad ~\underline{\delta }_2~}^{\underline{\tau }_1%
\underline{\tau }_2~~\underline{\beta }_2\underline{\beta }_1}\gamma _{\quad
\underline{\tau }_1\underline{\tau }_2\underline{\beta }_2}^{\underline{%
\delta }_2}]
$$
and%
$$
\Phi _{<\gamma ><\alpha >}=\Phi _{\underline{\gamma }_1\underline{\gamma }_2%
\underline{\alpha }_1\underline{\alpha }_2}=\frac
1{2(n+m_1+...+m_z)}\varepsilon _{\underline{\gamma }_1\underline{\alpha }%
_1}\varepsilon _{\underline{\gamma }_2\underline{\alpha }_2}[X_{\quad ~%
\underline{~\beta }_1\quad \underline{\mu }_1\underline{\beta }_2}^{%
\underline{\beta }_1\underline{\mu }_1~~\underline{\beta }_2}+\eqno(87)
$$
$$
T_{\qquad ~~\underline{\beta }_2\underline{\mu }_1}^{\underline{\tau }_1%
\underline{\tau }_2\underline{\beta }_1\quad ~\underline{\beta }_2}~\gamma
_{\quad \underline{\tau }_1\underline{\tau }_2\underline{\beta }_1}^{%
\underline{\mu }_1}+X_{\qquad \quad \underline{\beta }_2\underline{\mu }_2%
\underline{\nu }_1}^{\underline{\beta }_2\underline{\mu }_2\underline{\nu }%
_1}+T_{\qquad \underline{\beta }_1\quad ~\underline{\delta }_2~}^{\underline{%
\tau }_1\underline{\tau }_2~~\underline{\beta }_2\underline{\beta }_1}\gamma
_{\quad \underline{\tau }_1\underline{\tau }_2\underline{\beta }_2}^{%
\underline{\delta }_2}]-
$$
$$
\frac 12[X_{\underline{\gamma }_1~\underline{\alpha }_1\underline{\alpha }_2%
\underline{\delta }_1\underline{\gamma }_2}^{\quad \underline{\delta }%
_1}+T_{\qquad \underline{\alpha }_1\underline{\alpha }_2\underline{\delta }_1%
\underline{\gamma }_2}^{\underline{\tau }_1\underline{\tau }_2}~\gamma
_{\quad \underline{\tau }_1\underline{\tau }_2\underline{\gamma }_1}^{%
\underline{\delta }_1}+
$$
$$
X_{\underline{\gamma }_2~\underline{\alpha }_1\underline{\alpha }_2%
\underline{\delta }_1\underline{\gamma }_2}^{\quad \underline{\delta }%
_2}+T_{\qquad \underline{\alpha }_1\underline{\alpha }_2\underline{\gamma }_1%
\underline{\delta }_2~}^{\underline{\tau }_1\underline{\tau }_2}\gamma
_{\quad \underline{\tau }_1\underline{\tau }_2\underline{\gamma }_2}^{%
\underline{\delta }_2}].
$$

The components of the conformal Weyl d-spinor can be computed by putting
d-spinor values of the curvature (83) and Ricci (84) d-tensors into
corresponding expression for the d-tensor (33). We omit this calculus in
this work.

\section{ Field Equations on Ha-Spaces}

The problem of formulation gravitational and gauge field equations on
different types of la-spaces is considered, for instance, in \cite
{ma94,bej,asa88} and \cite{vg}. In this subsection we shall introduce the basic
field equations for gravitational and matter field la-interactions in a
generalized form for generic higher order anisotropic spaces.

\subsection{ Locally anisotropic scalar field interactions}

Let $\varphi \left( u\right) =(\varphi _1\left( u\right) ,\varphi _2\left(
u\right) \dot ,...,\varphi _k\left( u\right) )$ be a complex k-component
scalar field of mass $\mu $ on ha-space ${\cal E}^{<z>}.$ The
d-covariant generalization of the conformally invariant (in the massless
case) scalar field equation \cite{penr1,penr2} can be defined by using the
d'Alambert locally anisotropic operator \cite{ana94,vst96} $\Box =D^{<\alpha
>}D_{<\alpha >}$, where $D_{<\alpha >}$ is a d-covariant derivation on $%
{\cal E}^{<z>}$ satisfying conditions (14) and (15) and constructed, for
simplicity, by using Christoffel d--symbols (21) (all formulas for field
equations and conservation values can be deformed by using corresponding
deformations d--tensors $P_{<\beta ><\gamma >}^{<\alpha >}$ from the
Christoffel d--symbols, or the canonical d--connection to a general
d-connection into consideration):

$$
(\Box +\frac{n_E-2}{4(n_E-1)}\overleftarrow{R}+\mu ^2)\varphi \left(
u\right) =0,\eqno(88)
$$
where $n_E=n+m_1+...+m_z.$We must change d-covariant derivation $D_{<\alpha
>}$ into $^{\diamond }D_{<\alpha >}=D_{<\alpha >}+ieA_{<\alpha >}$ and take
into account the d-vector current
$$
J_{<\alpha >}^{(0)}\left( u\right) =i(\left( \overline{\varphi }\left(
u\right) D_{<\alpha >}\varphi \left( u\right) -D_{<\alpha >}\overline{%
\varphi }\left( u\right) )\varphi \left( u\right) \right)
$$
if interactions between locally anisotropic electromagnetic field ( d-vector
potential $A_{<\alpha >}$ ), where $e$ is the electromagnetic constant, and
charged scalar field $\varphi $ are considered. The equations (88) are
(locally adapted to the N-connection structure) Euler equations for the
Lagrangian%
$$
{\cal L}^{(0)}\left( u\right) = \eqno(89)$$
$$\sqrt{|g|}\left[ g^{<\alpha ><\beta >}\delta
_{<\alpha >}\overline{\varphi }\left( u\right) \delta _{<\beta >}\varphi
\left( u\right) -\left( \mu ^2+\frac{n_E-2}{4(n_E-1)}\right) \overline{%
\varphi }\left( u\right) \varphi \left( u\right) \right] ,
$$
where $|g|=detg_{<\alpha ><\beta >}.$

The locally adapted variations of the action with Lagrangian (89) on
variables $\varphi \left( u\right) $ and $\overline{\varphi }\left( u\right)
$ leads to the locally anisotropic generalization of the energy-momentum
tensor,%
$$
E_{<\alpha ><\beta >}^{(0,can)}\left( u\right) =\delta _{<\alpha >}\overline{%
\varphi }\left( u\right) \delta _{<\beta >}\varphi \left( u\right) +
\eqno(90)  $$
$$ \delta
_{<\beta >}\overline{\varphi }\left( u\right) \delta _{<\alpha >}\varphi
\left( u\right) -\frac 1{\sqrt{|g|}}g_{<\alpha ><\beta >}{\cal L}%
^{(0)}\left( u\right) ,
$$
and a similar variation on the components of a d-metric (12) leads to a
symmetric metric energy-momentum d-tensor,%
$$
E_{<\alpha ><\beta >}^{(0)}\left( u\right) =E_{(<\alpha ><\beta
>)}^{(0,can)}\left( u\right) -\eqno(91)
$$
$$
\frac{n_E-2}{2(n_E-1)}\left[ R_{(<\alpha ><\beta >)}+D_{(<\alpha >}D_{<\beta
>)}-g_{<\alpha ><\beta >}\Box \right] \overline{\varphi }\left( u\right)
\varphi \left( u\right) .
$$
Here we note that we can obtain a nonsymmetric energy-momentum d-tensor if
we firstly vary on $G_{<\alpha ><\beta >}$ and than impose constraints of
type (10) in order to have a compatibility with the N-connection
structure. We also conclude that the existence of a N-connection in
dv-bundle ${\cal E}^{<z>}$ results in a nonequivalence of energy-momentum
d-tensors (90) and (91), nonsymmetry of the Ricci tensor (see (29)),
nonvanishing of the d-covariant derivation of the Einstein d-tensor (34), $%
D_{<\alpha >}\overleftarrow{G}^{<\alpha ><\beta >}\neq 0$ and, in
consequence, a corresponding breaking of conservation laws on ha-spaces when
$D_{<\alpha >}E^{<\alpha ><\beta >}\neq 0\,$ . The problem of formulation of
conservation laws on la-spaces is discussed in details and two variants of
its solution (by using nearly autoparallel maps and tensor integral
formalism on locally anisotropic and higher order multispaces) where proposed
in \cite{vst96}. In this subsection we shall present only
straightforward generalizations of field equations and necessary formulas
for energy-momentum d-tensors of matter fields on ${\cal E}^{<z>}$
considering that it is naturally that the conservation laws (usually being
consequences of global, local and/or intrinsic symmetries of the fundamental
space-time and of the type of field interactions) have to be broken on
locally anisotropic spaces.

\subsection{ Proca equations on ha--spaces}

Let consider a d-vector field $\varphi _{<\alpha >}\left( u\right) $ with
mass $\mu ^2$ (locally anisotropic Proca field ) interacting with exterior
la-gravitational field. From the Lagrangian
$$
{\cal L}^{(1)}\left( u\right) =\sqrt{\left| g\right| }\left[ -\frac 12{%
\overline{f}}_{<\alpha ><\beta >}\left( u\right) f^{<\alpha ><\beta >}\left(
u\right) +\mu ^2{\overline{\varphi }}_{<\alpha >}\left( u\right) \varphi
^{<\alpha >}\left( u\right) \right] ,\eqno(92)
$$
where $f_{<\alpha ><\beta >}=D_{<\alpha >}\varphi _{<\beta >}-D_{<\beta
>}\varphi _{<\alpha >},$ one follows the Proca equations on higher order
anisotropic spaces
$$
D_{<\alpha >}f^{<\alpha ><\beta >}\left( u\right) +\mu ^2\varphi ^{<\beta
>}\left( u\right) =0.\eqno(93)
$$
Equations (93) are a first type constraints for $\beta =0.$ Acting with $%
D_{<\alpha >}$ on (93), for $\mu \neq 0$ we obtain second type constraints%
$$
D_{<\alpha >}\varphi ^{<\alpha >}\left( u\right) =0.\eqno(94)
$$

Putting (94) into (93) we obtain second order field equations with
respect to $\varphi _{<\alpha >}$ :%
$$
\Box \varphi _{<\alpha >}\left( u\right) +R_{<\alpha ><\beta >}\varphi
^{<\beta >}\left( u\right) +\mu ^2\varphi _{<\alpha >}\left( u\right) =0.%
\eqno(95)
$$
The energy-momentum d-tensor and d-vector current following from the (95)
can be written as
$$
E_{<\alpha ><\beta >}^{(1)}\left( u\right) =-g^{<\varepsilon ><\tau >}\left(
{\overline{f}}_{<\beta ><\tau >}f_{<\alpha ><\varepsilon >}+{\overline{f}}%
_{<\alpha ><\varepsilon >}f_{<\beta ><\tau >}\right) +
$$
$$
\mu ^2\left( {\overline{\varphi }}_{<\alpha >}\varphi _{<\beta >}+{\overline{%
\varphi }}_{<\beta >}\varphi _{<\alpha >}\right) -\frac{g_{<\alpha ><\beta >}%
}{\sqrt{\left| g\right| }}{\cal L}^{(1)}\left( u\right) .
$$
and%
$$
J_{<\alpha >}^{\left( 1\right) }\left( u\right) =i\left( {\overline{f}}%
_{<\alpha ><\beta >}\left( u\right) \varphi ^{<\beta >}\left( u\right) -{%
\overline{\varphi }}^{<\beta >}\left( u\right) f_{<\alpha ><\beta >}\left(
u\right) \right) .
$$

For $\mu =0$ the d-tensor $f_{<\alpha ><\beta >}$ and the Lagrangian (92)
are invariant with respect to locally anisotropic gauge transforms of type
$$
\varphi _{<\alpha >}\left( u\right) \rightarrow \varphi _{<\alpha >}\left(
u\right) +\delta _{<\alpha >}\Lambda \left( u\right) ,
$$
where $\Lambda \left( u\right) $ is a d-differentiable scalar function, and
we obtain a locally anisot\-rop\-ic variant of Maxwell theory.

\subsection{ Higher order an\-i\-sot\-rop\-ic gravitons}

Let a massless d-tensor field $h_{<\alpha ><\beta >}\left( u\right) $ is
interpreted as a small perturbation of the locally anisotropic background
metric d-field $g_{<\alpha ><\beta >}\left( u\right) .$ Considering, for
simplicity, a torsionless background we have locally anisotropic Fierz--Pauli
equations%
$$\Box h_{<\alpha ><\beta >}\left( u\right) +2R_{<\tau ><\alpha ><\beta ><\nu
>}\left( u\right) ~h^{<\tau ><\nu >}\left( u\right) =0
$$
and d--gauge conditions%
$$
D_{<\alpha >}h_{<\beta >}^{<\alpha >}\left( u\right) =0,\quad h\left(
u\right) \equiv h_{<\beta >}^{<\alpha >}(u)=0,
$$
where $R_{<\tau ><\alpha ><\beta ><\nu >}\left( u\right) $ is curvature
d-tensor of the la-background space (these formulae can be obtained by using
a perturbation formalism with respect to $h_{<\alpha ><\beta >}\left(
u\right) $ developed in \cite{gri}; in our case we must take into account
the distinguishing of geometrical objects and operators on ha--spaces).

\subsection{ Higher order anisotropic Dirac equations}

Let denote the Dirac d--spinor field on ${\cal E}^{<z>}$ as $\psi \left(
u\right) =\left( \psi ^{\underline{\alpha }}\left( u\right) \right) $ and
consider as the generalized Lorentz transforms the group of automorphysm of
the metric $G_{<\widehat{\alpha }><\widehat{\beta }>}$ (see the ha-frame
decomposition of d-metric (12), (68) and (69) ).The d-covariant
derivation of field $\psi \left( u\right) $ is written as
$$
\overrightarrow{\nabla _{<\alpha >}}\psi =\left[ \delta _{<\alpha >}+\frac
14C_{\widehat{\alpha }\widehat{\beta }\widehat{\gamma }}\left( u\right)
~l_{<\alpha >}^{\widehat{\alpha }}\left( u\right) \sigma ^{\widehat{\beta }%
}\sigma ^{\widehat{\gamma }}\right] \psi ,\eqno(96)
$$
where coefficients $C_{\widehat{\alpha }\widehat{\beta }\widehat{\gamma }%
}=\left( D_{<\gamma >}l_{\widehat{\alpha }}^{<\alpha >}\right) l_{\widehat{%
\beta }<\alpha >}l_{\widehat{\gamma }}^{<\gamma >}$ generalize for ha-spaces
the corresponding Ricci coefficients on Riemannian spaces \cite{foc}. Using $%
\sigma $-objects $\sigma ^{<\alpha >}\left( u\right) $ (see (44) and
(60)--(62)) we define the Dirac equations on ha--spaces:
$$
(i\sigma ^{<\alpha >}\left( u\right) \overrightarrow{\nabla _{<\alpha >}}%
-\mu )\psi =0,
$$
which are the Euler equations for the Lagrangian%
$$
{\cal L}^{(1/2)}\left( u\right) =\sqrt{\left| g\right| }\{[\psi ^{+}\left(
u\right) \sigma ^{<\alpha >}\left( u\right) \overrightarrow{\nabla _{<\alpha
>}}\psi \left( u\right) - \eqno(97)
$$
$$
(\overrightarrow{\nabla _{<\alpha >}}\psi ^{+}\left( u\right) )\sigma
^{<\alpha >}\left( u\right) \psi \left( u\right) ]-\mu \psi ^{+}\left(
u\right) \psi \left( u\right) \},
$$
where $\psi ^{+}\left( u\right) $ is the complex conjugation and
transposition of the column $~\psi \left( u\right) .$

From (97) we obtain the d-metric energy-momentum d-tensor%
$$
E_{<\alpha ><\beta >}^{(1/2)}\left( u\right) =\frac i4[\psi ^{+}\left(
u\right) \sigma _{<\alpha >}\left( u\right) \overrightarrow{\nabla _{<\beta
>}}\psi \left( u\right) +\psi ^{+}\left( u\right) \sigma _{<\beta >}\left(
u\right) \overrightarrow{\nabla _{<\alpha >}}\psi \left( u\right) -
$$
$$
(\overrightarrow{\nabla _{<\alpha >}}\psi ^{+}\left( u\right) )\sigma
_{<\beta >}\left( u\right) \psi \left( u\right) -(\overrightarrow{\nabla
_{<\beta >}}\psi ^{+}\left( u\right) )\sigma _{<\alpha >}\left( u\right)
\psi \left( u\right) ]
$$
and the d-vector source%
$$
J_{<\alpha >}^{(1/2)}\left( u\right) =\psi ^{+}\left( u\right) \sigma
_{<\alpha >}\left( u\right) \psi \left( u\right) .
$$
We emphasize that la-interactions with exterior gauge fields can be
introduced by changing the higher order anisotropic partial derivation from
(96) in this manner:%
$$
\delta _\alpha \rightarrow \delta _\alpha +ie^{\star }B_\alpha ,
$$
where $e^{\star }$ and $B_\alpha $ are respectively the constant d-vector
potential of locally anisotropic gauge interactions on  higher order
anisotropic spaces (see \cite{vg}).

\subsection{ D--spinor locally anisotropic Yang--Mills fields}

We consider a dv-bundle ${\cal B}_E,~\pi _B:{\cal B\rightarrow E}^{<z>}{\cal %
,}$ on ha-space ${\cal E}^{<z>}{\cal .\,}$ Additionally to d-tensor and
d-spinor indices we shall use capital Greek letters, $\Phi ,\Upsilon ,\Xi
,\Psi ,...$ for fibre (of this bundle) indices (see details in \cite
{penr1,penr2} for the case when the base space of the v-bundle $\pi _B$ is a
locally isotropic space-time). Let $\underline{\nabla }_{<\alpha >}$ be, for
simplicity, a torsionless, linear connection in ${\cal B}_E$ satisfying
conditions:
$$
\underline{\nabla }_{<\alpha >}:{\em \Upsilon }^\Theta \rightarrow {\em %
\Upsilon }_{<\alpha >}^\Theta \quad \left[ \mbox{or }{\em \Xi }^\Theta
\rightarrow {\em \Xi }_{<\alpha >}^\Theta \right] ,
$$
$$
\underline{\nabla }_{<\alpha >}\left( \lambda ^\Theta +\nu ^\Theta \right) =%
\underline{\nabla }_{<\alpha >}\lambda ^\Theta +\underline{\nabla }_{<\alpha
>}\nu ^\Theta ,
$$
$$
\underline{\nabla }_{<\alpha >}~(f\lambda ^\Theta )=\lambda ^\Theta
\underline{\nabla }_{<\alpha >}f+f\underline{\nabla }_{<\alpha >}\lambda
^\Theta ,\quad f\in {\em \Upsilon }^\Theta ~[\mbox{or }{\em \Xi }^\Theta ],
$$
where by ${\em \Upsilon }^\Theta ~\left( {\em \Xi }^\Theta \right) $ we
denote the module of sections of the real (complex) v-bundle ${\cal B}_E$
provided with the abstract index $\Theta .$ The curvature of connection $%
\underline{\nabla }_{<\alpha >}$ is defined as
$$
K_{<\alpha ><\beta >\Omega }^{\qquad \Theta }\lambda ^\Omega =\left(
\underline{\nabla }_{<\alpha >}\underline{\nabla }_{<\beta >}-\underline{%
\nabla }_{<\beta >}\underline{\nabla }_{<\alpha >}\right) \lambda ^\Theta .
$$

For Yang-Mills fields as a rule one considers that ${\cal B}_E$ is enabled
with a unitary (complex) structure (complex conjugation changes mutually the
upper and lower Greek indices). It is useful to introduce instead of $%
K_{<\alpha ><\beta >\Omega }^{\qquad \Theta }$ a Hermitian matrix $%
F_{<\alpha ><\beta >\Omega }^{\qquad \Theta }=i$ $K_{<\alpha ><\beta >\Omega
}^{\qquad \Theta }$ connected with components of the Yang-Mills d-vector
potential $B_{<\alpha >\Xi }^{\quad \Phi }$ according the formula:

$$
\frac 12F_{<\alpha ><\beta >\Xi }^{\qquad \Phi }=\underline{\nabla }%
_{[<\alpha >}B_{<\beta >]\Xi }^{\quad \Phi }-iB_{[<\alpha >|\Lambda
|}^{\quad \Phi }B_{<\beta >]\Xi }^{\quad \Lambda },\eqno(98)
$$
where the la-space indices commute with capital Greek indices. The gauge
transforms are written in the form:

$$
B_{<\alpha >\Theta }^{\quad \Phi }\mapsto B_{<\alpha >\widehat{\Theta }%
}^{\quad \widehat{\Phi }}=B_{<\alpha >\Theta }^{\quad \Phi }~s_\Phi ^{\quad
\widehat{\Phi }}~q_{\widehat{\Theta }}^{\quad \Theta }+is_\Theta ^{\quad
\widehat{\Phi }}\underline{\nabla }_{<\alpha >}~q_{\widehat{\Theta }}^{\quad
\Theta },
$$
$$
F_{<\alpha ><\beta >\Xi }^{\qquad \Phi }\mapsto F_{<\alpha ><\beta >\widehat{%
\Xi }}^{\qquad \widehat{\Phi }}=F_{<\alpha ><\beta >\Xi }^{\qquad \Phi
}s_\Phi ^{\quad \widehat{\Phi }}q_{\widehat{\Xi }}^{\quad \Xi },
$$
where matrices $s_\Phi ^{\quad \widehat{\Phi }}$ and $q_{\widehat{\Xi }%
}^{\quad \Xi }$ are mutually inverse (Hermitian conjugated in the unitary
case). The Yang-Mills equations on torsionless la-spaces \cite{vg}
 are written in this form:%
$$
\underline{\nabla }^{<\alpha >}F_{<\alpha ><\beta >\Theta }^{\qquad \Psi
}=J_{<\beta >\ \Theta }^{\qquad \Psi },\eqno(99)
$$
$$
\underline{\nabla }_{[<\alpha >}F_{<\beta ><\gamma >]\Theta }^{\qquad \Xi
}=0.\eqno(100)
$$
We must introduce deformations of connection of type  $\underline{\nabla }%
_\alpha ^{\star }~\longrightarrow \underline{\nabla }_\alpha +P_\alpha ,$
(the deformation d-tensor $P_\alpha $ is induced by the torsion in dv-bundle
${\cal B}_E)$ into the definition of the curvature of ha-gauge fields
(98) and motion equations (99) and (100) if interactions are modeled
on a generic ha-space.

\subsection{D--spinor Einstein--Cartan equ\-a\-ti\-ons }

Now we can write out the field equations of the Einstein-Cartan theory in
the d-spinor form. So, for the Einstein equations (34) we have

$$
\overleftarrow{G}_{\underline{\gamma }_1\underline{\gamma }_2\underline{%
\alpha }_1\underline{\alpha }_2}+\lambda \varepsilon _{\underline{\gamma }_1%
\underline{\alpha }_1}\varepsilon _{\underline{\gamma }_2\underline{\alpha }%
_2}=\kappa E_{\underline{\gamma }_1\underline{\gamma }_2\underline{\alpha }_1%
\underline{\alpha }_2},
$$
with $\overleftarrow{G}_{\underline{\gamma }_1\underline{\gamma }_2%
\underline{\alpha }_1\underline{\alpha }_2}$ from (86), or, by using the
d-tensor (87),

$$
\Phi _{\underline{\gamma }_1\underline{\gamma }_2\underline{\alpha }_1%
\underline{\alpha }_2}+(\frac{\overleftarrow{R}}4-\frac \lambda
2)\varepsilon _{\underline{\gamma }_1\underline{\alpha }_1}\varepsilon _{%
\underline{\gamma }_2\underline{\alpha }_2}=-\frac \kappa 2E_{\underline{%
\gamma }_1\underline{\gamma }_2\underline{\alpha }_1\underline{\alpha }_2},
$$
which are the d-spinor equivalent of the equations (35). These equations
must be completed by the algebraic equations (36) for the d-torsion and
d-spin density with d-tensor indices changed into corresponding d-spinor
ones.

\section{Summary and outlook}

We have developed the spinor differential geometry of distinguished vector
bundles provided with nonlinear and distinguished connections and metric
structures and shown in detail the way of formulation the theory of
fundamental field (gravitational, gauge and spinor) interactions on generic
higher order an\-isot\-rop\-ic spaces.

We investigated the problem of definition of spinors on spaces with higher
order anisotropy. Our approach is based on the formalism of Clifford
distinguished (by a nonlinear connection structure) algebras. We introduced
spinor structures on higher order anisotropic spaces as Clifford
 distinguished module structures on distinguished vector bundles. We also
 proposed the second definition,
as distinguished spinor structures, by using Clifford fibrations. It was
shown that almost Hermitian models of generalized Lagrange spaces,
$H^{2n}$--spaces admit as a proper characteristic the almost
complex spinor structures.

It should be noted that we introduced \cite{vjmp,vb295,vod} distinguished
spinor structures in an algebraic topological manner, and that in our
considerations the compatibility of distinguished connection and metric,
 adapted to a given nonlinear connection, plays a crucial role. The Yano and
 Ishihara method of
lifting of geometrical objects in the total spaces of tangent bundles \cite
{yan} and the general formalism for vector bundles of Miron and Anastasiei
\cite{ma87,ma94} and for higher order Lagrange spaces
 of Miron and Atanasiu \cite{mirata}
 clearing up the possibility and way of definition of
spinors on higher order anisotropic spaces. Even a straightforward
definition of spinors on Finsler and Lagrange spaces, and, of course, on
various theirs extensions, with general noncompatible connection and metric
structures, is practically impossible (if spinors are introduced locally
with respect to a given metric quadratic form, the spinor constructions will
not be invariant on parallel transports), we can avoid this difficulty by
lifting in a convenient manner the geometric objects and physical values
from the base of a locally anisotropic space on the tangent bundles of
 vector and tangent bundles under
consideration. We shall introduce corresponding discordance laws and values
and define nonstandard spinor structures by using nonmetrical distinguished
tensors (see such constructions for locally isotropic curved spaces with
torsion and nonmetricity in \cite{lue}).

The distinguishing by a nonlinear connection structure of the multidimensional space
into horizontal and vertical subbundles points out to the necessity to start
up the spinor constructions for locally anisotropic
 spaces with a study of distinguished
Clifford algebras for vector spaces split into horizontal and vector subspaces.
 The distinguished spinor objects exhibit a eight-fold periodicity on
dimensions of the mentioned subspaces. As it was shown in \cite{vjmp,vb295},
 see also
 Sections 3--5 of this work,
a corresponding distinguished spinor technique can be
developed, which is a generalization for higher dimensional with
nonlinear connection structure of that proposed by Penrose and Rindler \cite
{pen,penr1,penr2} for locally isotropic curved spaces, if the locally
adapted to the nonlinear connection structures
 distinguished spinor and distinguished vector frames are
used. It is clear the distinguished spinor calculus is more tedious than
the two--spinor
one for Einstein spaces because of multidimensional and multiconnection
character of generic higher order anisotropic spaces.
The distinguished spinor differential geometry formulated in Section 7 can be
considered as a branch of the geometry of Clifford fibrations for
vector bundles provided with nonlinear connection, distinguished connection
 and metric structures. We have
emphasized only the features containing distinguished spinor torsions and
 curvatures
which are necessary for a distinguished spinor formulation of
 locally anisotropic gravity. To develop a
conformally invariant distinguished
spinor calculus is possible only for a particular
class of higher order anisotropic spaces when the Weyl
 distinguished tensor (33) is defined by the
nonlinear connection and distinguished metric structures.
In general, we have to extend the
class of conformal transforms to that of nearly autoparallel maps of
higher order anisotropic spaces (see \cite{vod,voa,vodg,vcl96}).

Having fixed compatible nonlinear connection,
 distinguished connection and metric structures on
a distinguished vector bundle  ${\cal E}^{<z>}$ we can develop physical
models on this space
by using a covariant variational distinguished tensor calculus as on
Riemann--Cartan
spaces (really there are specific complexities because, in general, the
Ricci distinguished tensor is not symmetric and the locally anisotropic
frames are
nonholonomic). The systems of basic field equations for fundamental matter
(scalar, Proca and Dirac) fields and gauge and gravitational fields have
been introduced in a geometric manner by using distinguished
covariant operators and
locally anisotropic frame decompositions of
distinguished metric. These equations and expressions for
energy--momentum distinguished tensors and
 distinguished vector currents can be established by using
the standard variational procedure, but correspondingly adapted to the
nonlinear connection structure if we work by using locally adapted bases.

Let us try to summarize our results, discuss their possible implications and
make the basic conclusions. Firstly, we have shown that the Einstein--Cartan
theory has a natural extension for a various class of
higher order anisotropic spaces. Following
the R. Miron,  M. Anastasiei and Gh. Atanasiu approach
\cite{ma87,ma94,mirata} to the geometry of
locally anisotropic and higher order anisotropic
spaces it becomes evident the possibility and manner of
 formulation of
classical and quantum field theories on such spaces. Here we note that in
locally anisotropic theories we have an additional geometric structure,
the nonlinear connection.
From physical point of view it can be interpreted, for instance, as a
fundamental field managing the dynamics of splitting of higher--dimensional
space--time into the four--dimensional and compactified ones. We can also
consider the nonlinear connection as a generalized type of gauge field which
reflects some specifics of higher order anisotropic
field interactions and possible intrinsic
structures of higher order anisotropic spaces.
It was convenient to analyze the geometric structure
of different variants of higher order anisotropic spaces
 (for instance, Finsler, Lagrange and
generalized Lagrange spaces) in order to make obvious physical properties
and compare theirs perspectives in developing of new  models of
 higher order anisotropic gravity,
 locally anisotropic strings and
 higher order anisotropic superstrings \cite{vlags, v96jpa}.

\end{document}